\documentclass[10pt,onecolumn,openright]{book}

\usepackage{multicol}
\usepackage{bm}
\usepackage{braket}

\usepackage[acronym]{glossaries}   
\makenoidxglossaries
\newacronym{ldmx}{LDMX}{Light Dark Matter eXperiment}
\newacronym{sm}{SM}{Standard Model}
\newacronym{dm}{DM}{Dark Matter}
\newacronym{faser}{FASER}{ForwArd Search ExpeRiment}
\newacronym{lhc}{LHC}{Large Hadron Collider}
\newacronym{sps}{SPS}{Super Proton Synchrotron}
\newacronym{ship}{SHiP}{Search for Hidden Particles}
\newacronym{lsnd}{LSND}{Liquid Scintillator Neutrino Detector}
\newacronym{snd}{SND}{Scattering and Neutrino Detector}
\newacronym{nova}{NO$\nu$A}{NuMI Off-Axis $\nu_e$ Appearance}
\newacronym{ufo}{UFO}{Universal Feynman Output}
\newacronym{WW}{WW}{Weizs\"acker-Williams}
\newacronym{stt}{STT}{Silicon Tagging Tracker}
\newacronym{srt}{SRT}{Silicon Recoil Tracker}
\newacronym{ts}{TS}{Trigger Scintillator}
\newacronym{ecal}{ECal}{Electromagnetic Calorimeter}
\newacronym{hcal}{HCal}{Hadronic Calorimeter}
\newacronym{eot}{EOT}{Electrons On Target}
\newacronym{cl}{C.L.}{Confidence Level}
\newacronym{KM}{KM}{Kinetic Mixing}
\newacronym{M}{M}{Magnetic dipole}
\newacronym{E}{E}{Electric dipole}
\newacronym{C}{C}{Charge radius}
\newacronym{A}{A}{Anapole moment}
\newacronym{uv}{UV}{Ultraviolet}
\newacronym[firstplural={Vacuum Expectation Values (vevs)}]{vev}{vev}{Vacuum Expectation Value}
\newacronym{bbn}{BBN}{Big Bang Nucleosynthesis}
\newacronym{WIMP}{WIMP}{Weakly Interacting Massive Particle}
\newacronym{SIMP}{SIMP}{Strongly Interacting Massive Particle}
\newacronym{cmb}{CMB}{Cosmic Microwave Background}
\newacronym{mond}{MOND}{Modified Newtonian Dynamics}
\newacronym{qcd}{QCD}{Quantum Chromodynamics}
\newacronym{bsm}{BSM}{Beyond the Standard Model}
\newacronym{macho}{MACHO}{Massive Astrophysical Compact Halo Object}
\newacronym{dDSCGF}{dDSCGF}{deformed Dyson Self-Consistent Green’s function}

\def\phdThesis{1}

\newcommand{\currentyear}{2026}
\newcommand{\authorname}{Taylor Rose Gray}
\newcommand{\mytitle}{Illuminating Light Dark Matter with Fixed-Target Experiments}

\newcommand{\division}{Subatomic, High Energy, and Plasma Physics}

\newcommand{\chalmers}{Chalmers University of Technology}
\newcommand{\gotuni}{University of Gothenburg}
\newcommand{\mydepartment}{Department of Physics and Astronomy}
\newcommand{\chalIgu}{\chalmers~\IfItalicsTF{$|\!$}{$|$}~\gotuni} 

\newcommand{\degreetitle}{Doctor of Philosophy}

\ifx\phdThesis\undefined

\else

\fi

\usepackage{geometry}

\ifx\inlagaPage\undefined
\else
\fi

\usepackage{fancyhdr}
\usepackage[labelformat=simple]{subcaption}

\usepackage{graphicx}
\usepackage{url}
\usepackage{booktabs}
\usepackage{bm}

\usepackage[british]{babel} 
\usepackage[backend=biber,style=numeric-comp,maxcitenames=3,backref=true,language=british]{biblatex}

\usepackage{amsmath,amssymb,amscd,latexsym,dsfont}
\usepackage{textcomp}
\usepackage{psfrag}
\usepackage{rotating}
\usepackage{enumitem}
\usepackage{pdflscape}
\usepackage{zref}
\usepackage{hyperref}
\usepackage[final]{pdfpages}
\usepackage{xspace}
\usepackage{relsize}

\usepackage{url}
\usepackage{IEEEtrantools}

\usepackage{microtype}

\usepackage[textsize=tiny,textwidth=20mm,backgroundcolor=orange!70]{todonotes}
\newcommand{\figref}[1]{Fig.~\ref{#1}}
\newcommand{\tabref}[1]{Table~\ref{#1}}
\newcommand{\chapref}[1]{Chapter~\ref{#1}}
\newcommand{\secref}[1]{Section~\ref{#1}}
\newcommand{\eqnref}[1]{Eq.~\ref{#1}}
\newcommand{\appref}[1]{Appendix~\ref{#1}}
\newcommand{\partref}[1]{Part~\ref{#1}}

\makeatletter
\newcommand*{\IfItalicsTF}{%
  \ifx\f@shape\my@test@it
    \expandafter\@firstoftwo
  \else
    \expandafter\@secondoftwo
  \fi
}
\newcommand*{\my@test@it}{it}
\makeatother

\newcommand{\romannum}[1]{{\uppercase\expandafter{\romannumeral #1\relax}}}

\usepackage{hyperref}
\usepackage{amsmath}
\usepackage{mathrsfs}

\def\Ap{{A^\prime}}
\def\Zp{{Z^\prime}}
\def\pp{{p^\prime}}

\newcommand{\BdNMC}{\textsc{BdNMC}\xspace}

\newcommand{\GAMBIT}{\textsc{GAMBIT}\xspace}

\newcommand{\madgraph}{\textsc{MadGraph}\xspace}
\newcommand{\feynrules}{\textsc{FeynRules}\xspace}
\newcommand{\maddump}{\textsc{MadDump}\xspace}
\newcommand{\pythia}{\textsc{Pythia}\xspace}

\usepackage{slashed}

\begin{document}
\frontmatter


\thispagestyle{empty} 
\begin{center}
  \textsc{Thesis for The Degree of \degreetitle}\\
\end{center}

\vspace{4cm}

\begin{center} {\LARGE \textbf{\mytitle}}\\
\ifx\mysubtitle\undefined
\else
  \vspace{4mm}
  \textit{\large\mysubtitle}
\fi
\end{center}

\ifx\mysubtitle\undefined
\vspace{5mm}
\else
\vspace{2mm}
\fi

\begin{center}
\textsc{\large\authorname} \\
\end{center}


\vfill

\begin{center}
\mydepartment\\
\textsc{\chalmers}\\
Gothenburg, Sweden, 2026 \\
\end{center}

\newpage
\thispagestyle{empty}

\vspace{2cm} \noindent \textbf
\mytitle\\
\ifx\mysubtitle\undefined
\else
  \textit{\small\mysubtitle}
  \\
\fi

\noindent
\textsc{\authorname}\\

\noindent
ISBN 978-91-8103-463-9

\vspace{1.5cm}
\noindent
Acknowledgements, dedications, and similar personal statements in this thesis, reflect the author’s own views.

\vskip 0.5cm\noindent
\copyright \space \authorname,~\currentyear  \\
\noindent

\noindent
Doktorsavhandlingar vid Chalmers tekniska högskola \\ Ny serie nr 5920

\noindent
ISSN 0346-718X

\noindent
https://doi.org/10.63959/chalmers.dt/5920

\vspace{1.25cm}


\noindent 
\mydepartment\\
Division of \division\\
\ifx\researchgroup\undefined\else\researchgroup\\\fi
\chalmers\\
SE-412 96 Gothenburg,\\
Sweden\\
Phone: +46(0)31 772 1000

\vskip 1.25cm

\noindent
This thesis has been prepared using \LaTeX.
\\
\vskip 0.25 cm
\noindent
The author acknowledges the use of Claude (Anthropic) and Gemini (Google) as editing assistants during the drafting of this thesis. Written statements presented in this work remain the sole responsibility of the author.

\vskip 1.5cm

\noindent
Cover:
Artistic rendition of galaxies and the unseen structure binding them, by Xinyue Wang.
\\

\vskip 1.25cm

\noindent
Printed by Chalmers Digitaltryck\\
Gothenburg, Sweden 2026.

\setcounter{page}{3}
\thispagestyle{plain}

\begin{table}
  \begin{flushright}
	\large{\emph{~ In loving memory of my aunt, Anna Maria Romano.}}  
\end{flushright}
\end{table}

\cleardoublepage \addcontentsline{toc}{chapter}{Abstract}
\thispagestyle{plain}
\begin{raggedright}
\noindent 
{\large\textbf{\mytitle}}\\
\ifx\mysubtitle\undefined\else\textit{\small\mysubtitle}\fi
\end{raggedright}

\vskip 1mm
\noindent
{\sc\authorname}
\vskip 1mm
\noindent
\mydepartment\\
Chalmers University of Technology

\vskip 8mm

\section*{Abstract}

\noindent

Astronomical and cosmological observations provide overwhelming evidence for a non-luminous matter component — dark matter (DM) — whose particle nature remains unknown. 
Extensive experimental searches for non-gravitational DM interactions have largely been geared for DM with masses $\mathcal{O}(\text{GeV-TeV})$ via nuclear recoils, yielding no discovery so far. DM in the MeV–GeV mass range, however, is theoretically motivated, reproducing the observed relic abundance while remaining invisible to nuclear recoil searches. Such light DM is better probed via electron recoils or at fixed-target experiments.
This thesis advances the theoretical and statistical toolkit for light DM searches at fixed-target experiments, spanning model building, statistical inference, global fits, and signal simulation.
We survey the landscape of fixed-target searches for light DM, with a central focus on Light Dark Matter eXperiment (LDMX). We also consider the next-generation experiment SHiP, where we simulate DM signatures to assess its sensitivity to light DM.

Within the standard paradigm of minimal DM models, we present a likelihood-based statistical framework for LDMX using the full two-dimensional recoil electron distribution, spanning exclusion, discovery, parameter estimation, and model comparison in frequentist and Bayesian formulations. This framework establishes the discovery potential of LDMX, and we further quantify the direct detection exposure needed to independently confirm a potential signal.

Moving beyond this standard paradigm, we perform global fits of scalar and fermionic light DM, combining cosmological, astrophysical, and laboratory constraints.
We then extend the theoretical landscape of fixed-target searches to spin-1 DM and higher electromagnetic moment dark photons, yielding rich signatures with strong complementarity to other probes.
Finally, we go beyond standard nuclear modelling of dark photon production, incorporating elastic nuclear form factors and spectral functions from many-body \textit{ab initio} methods in place of standard analytic treatments.

\vspace{10 mm}
\noindent{\textbf{Keywords}}

\vspace{3 mm}
\noindent{Light Dark Matter, Dark Photon, Fixed-Target Experiment, LDMX}

\cleardoublepage \addcontentsline{toc}{chapter}{List of Publications}
\thispagestyle{plain}
\chapter*{List of Publications}
\label{chap:papers}

\section*{Publications and Manuscripts}
This thesis is based on the following papers (alphabetical author name ordering in high energy physics):

\newcommand{\pubitem}[3]{\item #1,
\newblock {\textit{#2}}\\
\newblock {\em #3}.
}

\begin{enumerate}[align=left,label={[\textbf{Paper \Roman*}]}]

    \pubitem{R. Catena, T. R. Gray}{Spin-1 Thermal Targets for Dark Matter Searches at Beam Dump and Fixed Target Experiments}{JCAP vol. 11, p. 058 2023. DOI: 10.1088/1475-7516/2023/11/058. arXiv: 2307.02207 [hep-ph] }

    \pubitem{S. Balan, C. Balázs, T. Bringmann, C. Cappiello, R. Catena, T. Emken, T. E. Gonzalo, T. R. Gray, W. Handley, Q. Huynh, F. Kahlhoefer, A. C. Vincent}{Resonant or Asymmetric: The Status of Sub-GeV Dark Matter}{JCAP vol. 1, p. 053, 2025. DOI: 10.1088/1475-7516/2025/01/053. arXiv: 2405.17548 [hep-ph]}

\pubitem{R. Catena, T. R. Gray, A. Lund}
{On the Dark Matter Origin of an LDMX Signal}
{JCAP vol. 7, p. 020, 2025. DOI: 10.1088/1475-7516/2025/07/020. arXiv: 2411.10216 [hep-ph]}

\pubitem{R. Catena, T. R. Gray, T. Jerkvall}
{Production of Dark Photons through Higher Electromagnetic Moments at LDMX: Simulations and Model Discrimination}
{JHEP vol. 9, p. 040, 2025. DOI: 10.1007/JHEP09(2025)040. arXiv: 2502.13635 [hep-ph]}

\pubitem{T. Åkesson et al. (LDMX Collaboration)}
{LDMX -- The Light Dark Matter eXperiment}
{arXiv: 2508.11833 [hep-ex] \textit{Included in thesis: Section 2.4.1.1 [Spin-1 Dark Matter]}}

\pubitem{A. Banerjee, R. Catena, T. R. Gray}
{Light Vector Dark Matter via a Magnetic Dipole Portal: Bridging Direct Detection and Fixed-Target Searches}
{accepted for publication in JHEP. arXiv: 2511.23259 [hep-ph]}

\pubitem{A. Berger, R.Catena, J. Conrad, T. Gray}
{Light Dark Matter Discovery Potential and Model Selection at LDMX}
{submitted to JHEP. arXiv: 2607.24524 [hep-ph]}

\pubitem{T. Gray, A. Scalesi}
{First-Principles Nuclear Modeling for Light Dark Matter Experiments at the Intensity Frontier}
{manuscript}

\end{enumerate}


\section*{Other publications}

\begin{enumerate}[align=left,label={[\textbf{\alph*}]}]

     \pubitem{C. Cosme, M. Dutra, S. Godfrey, T. R. Gray}{Testing freeze-in with axial and vector Z' bosons}{JHEP vol. 09 p. 056 2021. DOI: 10.1007/JHEP09(2021)056. arXiv: 2307.02207 [hep-ph] }

\end{enumerate}

\addcontentsline{toc}{chapter}{Statement of Contributions}
\thispagestyle{plain}
\chapter*{Statement of Contributions}
\vspace{5 mm}

\begin{itemize}
    \item \textbf{Paper I:} In this first publication, I was responsible for the analytical derivation and numerical implementation of the majority of the results. Specifically, I derived the helicity amplitudes for unitarity considerations, calculated the dark matter annihilation cross sections, and computed the relic density. I extended the existing Monte Carlo simulation software to incorporate additional particle physics models and executed these production simulations for beam dump environments. Furthermore, I integrated data from all other relevant experiments and generated all of the final figures. The interpretation of these results was carried out in collaboration with my supervisor, Riccardo Catena.
    \item \textbf{Paper II:}  As one of the four corresponding authors in this paper, my main responsibility was the fixed target experiment likelihoods. This task involved extending the Monte Carlo simulation software to include our models, running simulations of relevant experiments for each model, tabulating the results, and implementing this data into the Colliderbit GAMBIT module. I drafted the section on accelerated-based searches.
     \item \textbf{Paper III:}
     My primary role was the technical guidance and co-supervision of the Master's student, Andreas Lund, who generated the core results. I developed and provided the Universal FeynRules Output (UFO) files and the specialized computational code framework required to execute the MadGraph simulations of LDMX dark matter signatures. In addition to provisioning these software tools, I guided the simulation workflow, managed the technical implementation, and contributed to the final interpretation of the results.
    \item \textbf{Paper IV:}
    I wrote the manuscript for this paper, and I co-supervised the Master's student, Thomas Jerkvall, providing him with the computational code framework required to execute the MadGraph simulations for dark photon production and guiding him through the analysis process. To ensure the accuracy of the project, I developed an independent implementation to fully validate the student's simulation results.
    \item \textbf{Paper V:} As this is a large collaboration paper by the LDMX Collaboration, my contribution was focused on a specific physics case. My supervisor Riccardo Catena and I were invited to coauthor a subsection for the spin-1 dark matter scenario. I drafted the text for this dedicated subsection and produced the corresponding sensitivity plot featured within it.
     \item \textbf{Paper VI:} For this paper, the design of the phenomenological aspect of this study was carried out by me. I was responsible for the complete numerical and computational implementation of the framework, generating all results and figures presented in the paper except for the direct detection analysis done by Riccardo Catena. This includes implementation in \feynrules, computation of the relic density, simulation of signatures at experiments, and inclusion of relevant experimental bounds. Additionally, I wrote the majority of the manuscript. The theoretical formulation of the light vector dark matter model and its corresponding theory section, were designed and contributed by Avik Benerjee.
     \item \textbf{Paper VII:} For this manuscript, I wrote the text and implemented the frequentist statistical framework. I implemented the full pseudo-experiment pipeline to ensure robust statistical inference. Additionally, I co-supervised the Master's student, Adam Berger, guiding his development and implementation of the Bayesian analysis and model comparison components of the framework and providing simulation data.
     \item \textbf{Paper VIII:} For this work, I originated the concept for this study and designed the methodology governing the analysis steps that followed the nuclear input. I integrated the nuclear physics inputs provided by Alberto Scalesi into the simulation framework to evaluate light dark matter fixed-target signatures. This involved computing the nuclear structure functions from the spectroscopic amplitudes for inelastic scattering, as well as implementing the numerical form factors into the Monte Carlo simulation pipeline to generate the final results. Regarding the writing of the manuscript, I drafted all parts excluding the dedicated nuclear theory text.
\end{itemize}

\addcontentsline{toc}{chapter}{Acknowledgements}
\thispagestyle{plain}
\chapter*{Acknowledgments}

This research has been funded by the Knut and Alice Wallenberg project grant Light Dark Matter (Dnr. KAW 2019.0080).
\vskip 0.5cm

Throughout this PhD experience, I have had the privilege of working with and learning from many wonderful people to whom I owe my sincere gratitude.

First and foremost, a PhD supervisor shapes the entire doctoral journey, and I am profoundly grateful to my supervisor, Riccardo Catena, who gave me this opportunity and inspired me throughout. 
Grazie per tutto ciò che mi hai insegnato, per la tua costante guida e per il tuo prezioso supporto nel mio passaggio alla fase successiva della mia carriera accademica. Oltre alla ricerca, ti sono immensamente grata per il tuo calore, la tua simpatia e l'ambiente stimolante che crei all'interno del nostro gruppo. È stata una vera fortuna poter lavorare con te e non vedo l'ora di continuare a collaborare in futuro.

Turning to my research collaborations, I wish to convey my sincere appreciation to Felix Kahlhöfer, for not only our collaboration but his mentorship and support throughout my postdoc search, my wonderful visit to KIT, and our many fruitful and pleasant discussions.
My warmest thanks go to my collaborators, Sowmiya Balan, Csaba Balazs, Torsten Bringmann, Christopher Cappiello, Tomas Gonzalo, Will Handley, Quan Huynh, Aaron Vincent, Avik Banerjee, Jan Conrad, Alberto Scalesi, and a special thanks to Timon Emken for a great time during our overlap at Chalmers.
I am grateful to my collaborators at Lund University for their valuable feedback during our meetings and for offering essential insights into the experimental side of our work, and particularly to Torsten Åkesson for his support with my academic career.
I would like to sincerely thank my examiner, Christian Forssén, my manager, Gabriele Ferretti, and Christophe Demaziere who organized the study plan meetings. 
I extend my appreciation to the opponent of my Licentiate thesis defence, Roman Pasechnik, for his feedback on my research and our engaging discussions.

Beyond direct research, my time at Chalmers was enriched by teaching and outreach.
I was fortunate to work with several wonderful Master's students: Andreas Lund, Thomas Jerkvall, Elias Svensson, Albin Ledell, and Adam Berger, who all made impactful contributions to this thesis.
I have had the opportunity to participate in outreach activities with the Onsala Space Observatory. I express my sincere gratitude to Robert Cumming for this fantastic experience, and to the rest of the outreach team from Astronomy.

To past and present group members:
I had the privilege of sharing an office with Einar Urdshals. Thank you for the great memories from daily office life to celebrating your wedding in Albania.
To Michał Iglicki, thank you for the jokes, coffee breaks, climbing sessions, and your warmth throughout your time here. To Theresa Backes, thank you for being the best dance partner and belayer; I wish you the best in your PhD journey and beyond.

I am incredibly proud to call such brilliant, supportive, and inspiring individuals my friends. To Iva, Chiara, and Xinyue, as well as everyone else who made this journey brighter: thank you for everything.
A special thanks to Xinyue for creating the beautiful artwork for the thesis cover.
To my family -- Jesse, Mom, Dad, RockC and Zeppy -- as well as my extended family, thank you for being a constant anchor throughout this journey and for always believing in me.
To Alberto, grazie per essere la mia più grande fonte d'ispirazione ogni giorno. Il tuo sostegno ha significato per me più di quanto le parole possano esprimere.

Finally, I would like to extend my gratitude to my younger self, for persevering through early uncertainty and staying true to a vision that was entirely my own.

\vspace{2mm}
\noindent

\clearpage

\addcontentsline{toc}{chapter}{List of Acronyms}
\renewcommand*{\glossarysection}[2][]{}
\chapter*{List of Acronyms}
\begin{multicols}{2}
    \printnoidxglossary[type=\acronymtype]
\end{multicols}

\cleardoublepage
\thispagestyle{plain}
\tableofcontents 
\chapter*{Preface}
\addcontentsline{toc}{chapter}{Preface}

It often occurs that we are faced with observations that cannot be explained by current scientific theories. Throughout history, the quest to understand the nature of these observations has triggered new theoretical developments and ultimately new knowledge.
Today, we are faced with cosmological and astrophysical evidence, exclusively gravitational in nature, for the existence of an elusive and invisible new form of matter, \textit{dark matter}. This evidence spans a wide range of scales and epochs, from the anomalous rotation curves of individual galaxies and the gravitational lensing of galaxy clusters, to the acoustic peaks of the cosmic microwave background and the large scale structure of the Universe. These observations call for theoretical advancement to understand the underlying nature of dark matter.
This problem, which follows from years of scientific advancement in fields such as astronomy, cosmology, classical mechanics, and quantum mechanics, is one of the greatest of our time. Let us take a step back in time to the 1500s, prior to the current understanding of the Universe that we have today, and journey forward from there to the present, reviewing past examples in which observations have led to new scientific advancements and introducing useful concepts used throughout this thesis.

It took more than 100 years for Copernicus' initially controversial heliocentric model of the solar system to become accepted in the scientific community. Prior to 1543 when Copernicus' \textit{On the Revolutions of the Heavenly Spheres} \cite{Copernicus} was published, the Earth was thought to be the centre of the Universe. Later, using Tycho Brahe's measurements, Johannes Kepler formulated the laws of planetary motion. Galileo used a telescope for the first time to observe planets and stars, finding the moons of Jupiter thus proving Copernicus' hypothesis of a non-geocentric solar system. He also made vital contributions to the understanding of the laws of motion, which were important for Isaac Newton to formulate the laws of gravitation. In 1687, the phenomenon of gravitational attraction experienced between astronomical bodies was connected with the tendency for objects here on Earth to fall to the floor, through Newton's law of gravitation \cite{Newton}. This fundamentally connects the seemingly separate worlds of astronomical bodies in outer space and our experienced lives on Earth. Newton also that white light could be decomposed into a spectrum of colours, laying the grounds for spectroscopy, the foundation of observational astronomy \cite{longair_2006}.

All while, and long before,  this important progress in our understanding of physics and astronomy was being made, astronomy was being used as a means of navigation and timekeeping. Navigating the seas became easier using accurate measurements of the positions of stars from understanding planetary motion.  Uncovering the laws of nature governing the Universe has not only been to feed our curiosity, but also for practical applications leading to invaluable improvements in society.

After the discovery of Uranus in 1781, it was noticed that its orbit around the solar system's barycenter was slightly different than predicted by Newtonian gravity. One viable explanation for these perturbations in the orbit was that there exists a new planet beyond it. In 1846, Neptune was discovered after predicting its position using these perturbations from the expected orbit. This serves as an example of how astronomical objects can be inferred from their gravitational influence on known and measurable objects. When deviations from the predicted motion of objects under gravity are discovered, it can be explained by one of two scenarios: 1. there is an undiscovered object which is influencing other objects gravitationally or 2. the current theory of gravity is incomplete.

Newtonian gravity works well for explaining objects falling on Earth or many planetary orbits, although it was discovered that it fails to describe the observed precession of Mercury's perihelion. It was hypothesized at the time that there could exist another planet between the sun and Mercury, namely Vulcan, although no new planet was ever observed and instead a new theory of gravitation was born. Albert Einstein published his theories of special and general relativity \cite{einstein,einstein2} in 1905 and 1915, respectively, where special relativity postulates that the speed of light is the same in all inertial reference frames, and general relativity establishes a connection between gravity and geometry, describing gravity not as a force in the Newtonian sense, but as the curvature of spacetime. Newtonian gravity can only handle scenarios with small gravitational fields and where velocities are small compared to the speed of light; outside of these realms general relativity is needed. For example, it is required for explaining phenomena such as Mercury's orbit (where the gravitational field from the sun is large) and for the accuracy of Global Positioning Systems (GPSs).
There is even yet another proposed extension of the current framework of gravitation, namely \gls{mond} \cite{Milgrom:1983ca}, which attempts to provide an explanation for anomalous stellar and galactic dynamics. However, it is debated whether or not MOND is substantial. Later we will discuss the existence of this unknown mass, dark matter, in much more detail.

Following the formulation of special and general relativity, Einstein developed a static and closed model for the Universe \cite{Einstein:1917ce} -- setting the foundation for modern cosmology.
Einstein introduced an additional term to his field equations, the cosmological constant, where this term counteracts the inward pull of gravity in the Universe. In 1929, Edwin Hubble measured the velocities of galaxies and discovered that they are receding away from us \cite{Hubble:1929ig}, thus the Universe is expanding. Einstein then assumed that the cosmological constant term is no longer needed. Even prior to these observations of the expanding Universe, Georges Lemaître worked on an expanding universe solution to Einstein's equations, independently of  Alexander Friedmann.  Lemaître had the idea that the Universe began from one point and then expanded radially, being now known as the Big Bang theory \cite{Lemaitre:1931zzb,Lemaitre:1933gd}.
Later on, in 1998, the acceleration of the expansion of the Universe was measured using distant supernovae independently by two groups, and it was concluded that the expansion of the Universe is speeding up \cite{SupernovaSearchTeam:1998fmf,SupernovaCosmologyProject:1998vns}. With this discovery, Einstein's cosmological constant regained validity as it represents this unknown repulsive force now called dark energy.

\begin{figure}
    \centering
    \includegraphics[width=0.7\linewidth]{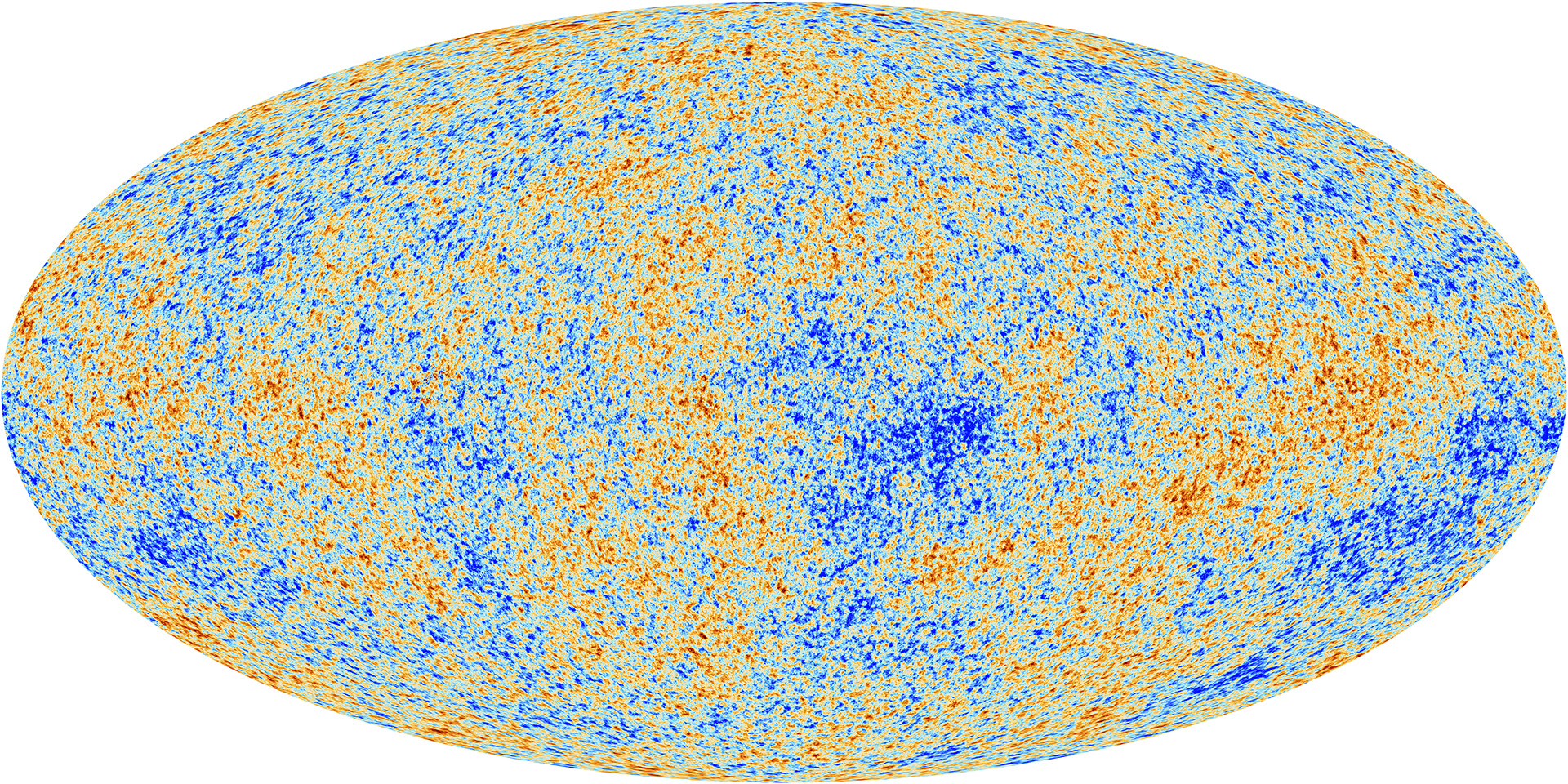}
    \caption{Sky map of the CMB photons observed by Planck \cite{Planck:2018vyg}. Colours show the temperature of these photons, with red being slightly hotter and blue slightly colder. The average temperature is 2.7 Kelvin, and the difference between the blue and the red is on the order of $10^{-5}$ Kelvin.}
    \label{fig:CMB}
\end{figure}
If the Universe began from a ``single point", the Universe would have been very dense and hot in its earliest times. Just 1 second after the Big Bang, its temperature would still have been around $10^{10}$ Kelvin -- far too hot for atoms to even form.
In this dense cosmic soup of free electrons and nuclei, photons could not travel far without interacting, similar to fog. After 380 thousand years, when the Universe expanded and cooled to $\sim 3000$ Kelvin, electrons combined with hydrogen and helium nuclei and photons could now travel freely, thus the Universe became transparent. These first photons, from when the Universe first became transparent, are the \gls{cmb}, our earliest probe of the Universe using light measurements. The CMB was first predicted by George Gamow, Ralph Alpher, and Robert Herman in 1948
\cite{Alpher:1948srz,Alpher:1949sef,Alpher:1988kr}, and was first measured in 1965 unintentionally as excess noise in a radio receiver \cite{Penzias:1965wn,Dicke:1965zz}. \figref{fig:CMB} is a skymap of the CMB temperature measured by the Planck satellite \cite{Planck:2018vyg}. The anisotropies of the CMB embed information on the abundances of baryons and other content (which will be discussed in the following sections) in the Universe.

The \textit{ultraviolet catastrophe} was realized at the end of the $19^{th}$ century, when experimental results of blackbody radiation (spectra of intensity as a function of wavelength of emitted light of almost perfect absorbers) did not agree with theoretical predictions. At the time, theory predicted that the energy emitted at smaller wavelengths grew to infinity -- something clearly unphysical. In 1900, Max Planck solved this problem by introducing \textit{quanta}, discrete energy levels that the atoms in a blackbody can have. These discrete energy levels are integer multiples of a base energy unit proportional to Planck's constant, $h$. At the time, the deeper consequences of these discretized energies were not yet realized; it was viewed merely as a mathematical trick to fit observations. Einstein also used Planck's quanta to explain the \textit{photoelectric effect}, where light incident on a metal ejects electrons. A wave description of light was not consistent with the observations of the photoelectric effect, although other phenomena such as those in thermodynamics required both a wave and a quantized description of light -- two inconsistent frameworks.

In 1913, Niels Bohr studied the Rutherford nuclear model, one of the models of the atom at the time.
In the Rutherford picture, heavy positively charged nuclei are orbited by much lighter negatively charged electrons, similarly to a mini solar system. There were a couple of problems with this picture. For one, the emission and absorption spectra of hydrogen consisting of discrete lines could not be explained. In addition, this atom could not be stable: an orbiting electron is constantly accelerating, and by classical electrodynamics an accelerating charge must radiate electromagnetic energy. As the electron continuously loses energy to this radiation, it would spiral in towards the nucleus, causing the atom to collapse.

In order to fix this description of the atom, Bohr proposed a new model. In this model, the electron orbits the nucleus in discrete orbits -- quantized angular momentum, and when the electron is jumping between these orbits, light is emitted or absorbed at a specific energy.
Here, Planck's quantum theory was being applied to the model of the atom.
Despite the inevitable realization that this light of specific energy being absorbed or emitted in the atom is indeed the same light quanta that Einstein introduced, Bohr was famously and persistently resistant to accepting them as the same, not openly embracing the photon concept until 1925.
Later, Einstein connected Bohr's atomic model to blackbody radiation, using Bohr's discrete energy levels to re-derive Planck's blackbody spectrum from first principles \cite{Einstein1917}. In doing so, he showed that transitions between atomic energy levels could only be predicted probabilistically, anticipating the fundamentally statistical character of quantum mechanics. Einstein himself was uneasy about this conclusion: in closing the paper, he identified the probabilistic character of individual transitions as a weakness of the theory, foreshadowing the objections to quantum indeterminism he would maintain for the rest of his career.
However, beyond these interpretational questions, there were more concrete gaps in Bohr's atomic theory: atoms with more than one electron could not be predicted well within this framework.

In 1925, Werner Heisenberg developed a mathematical formalism for quantum mechanics \cite{Heisenberg:1925zz}, based on matrix mechanics. One year later, Erwin Schrödinger presented his formalism involving matter waves represented as wavefunctions \cite{Schrodinger:1926gei}. Many technologies rely on the developments of quantum mechanics, such as magnetic resonance imaging used in medicine, lasers, solar cells, atomic clocks, and quantum computing.

The development of special relativity and quantum mechanics made way for their combination, quantum field theory, the theoretical framework used in describing the physics of the smallest known constituents of matter. We rely on the methods of quantum field theory in this thesis to describe these constituents of matter and their forces, describing them by fields. The dynamics of these fields (and the particles arising as their quantized excitations) are encoded in a Lagrangian density, $\mathscr{L}$. The current best theory for describing these fundamental particles and forces is the Standard Model of particle physics.

Analogous to the examples from history discussed above, precision cosmological and astrophysical measurements over the past decades have provided mounting evidence for the existence of dark matter in the Universe. What is now sought is a particle physics theory capable of explaining the origin of this anomalous observation, that is, a microscopic description underlying this purely gravitational phenomenon. This thesis explores the hypothesis that this cosmological component is composed of new, hypothetical particles beyond the Standard Model, described in terms of a Lagrangian, as an alternative to the modified gravity approach of MOND discussed above. Concretely, this amounts to extending the Standard Model Lagrangian, $\mathscr{L}_{\rm SM}$, with new fields and interaction terms. Specifically, this work hypothesizes viable theoretical scenarios for the nature of dark matter, describing its fundamental properties and its interactions with other species, and computes the expected signals and sensitivity at experiments searching for it.

 
\cleardoublepage

\mainmatter

\thispagestyle{empty} 
\cleardoublepage
\part{Background and Motivation}
\label{part:overview}
 \newpage
 \thispagestyle{empty} 
 \cleardoublepage

 \chapter{Introduction}
\label{chap:Introduction}

Constituting approximately 85\% of the matter content of the Universe, \gls{dm} is one of the most significant open problems in fundamental physics. Gravitational evidence for DM spans an enormous range of scales, from galactic rotation curves and cluster dynamics to the precision measurements of the \gls{cmb} and large scale 
structure~\cite{Planck:2018vyg}. Despite this overwhelming evidence, the 
particle properties of DM -- its mass,  interactions, and
connection (if any) to the \gls{sm} -- are unknown.
This thesis focuses on the hypothesis that DM is a particle in the sub-GeV mass range, and explores its signatures at fixed-target accelerator-based experiments.

This thesis is organized into five parts. Part I, \textit{Background and Motivation}, lays out the broader context for this thesis and the appended papers. \chapref{chap:Introduction} reviews the Standard Model of particle physics, its shortcomings, and the evidence for the existence of DM. \chapref{chap:ParticleDarkMatter} then surveys DM candidates and their early Universe production, outlines ongoing DM search strategies, and motivates the focus of this thesis: sub-GeV DM and the dark photon mediator.
Part II, \textit{The Standard Paradigm: Minimal Dark Matter Models at Fixed-Target Experiments}, develops fixed-target searches within a minimal and most widely studied scenario, a kinetically mixed dark photon mediator coupled to complex scalar or fermionic DM. While our focus here is on this benchmark case,
the methods and experimental treatment developed in this part apply equally well to the broader scenarios considered in Part III.
\chapref{chap:AcceleratorExperiments} introduces fixed-target experiments and their light DM signatures, \chapref{chap:LDMX} presents the upcoming electron fixed-target experiment LDMX in detail, and \chapref{chap:Likelihood} develops a
likelihood-based statistical framework for LDMX light DM searches and evaluates the discovery parameter inference potential.
Part III, \textit{Beyond the Standard Paradigm: Global Fits, Dark Sector Theories, and Nuclear Modelling}, advances this picture along three complementary frontiers.
\chapref{chap:GlobalFits} combines cosmological, astrophysical, and laboratory constraints within global fits of minimal light DM models, providing a status update for light DM. \chapref{chap:Models} looks beyond the standard predictive models to richer DM theories, spin-1 DM scenarios and dark photons with higher-order electromagnetic moments, in the context of fixed-target searches and complementary probes such as direct detection through electron recoils. \chapref{chap:NuclearPhysics} implements state-of-the-art \textit{ab initio} many-body nuclear theory input for modelling the target nucleus in these searches.
The thesis concludes with Part IV, summarizing this thesis and outlining future and ongoing directions.
Part V collects the appended papers underlying this thesis.

\section{The Standard Model}

\begin{figure}[h]
    \centering
   \includegraphics[width=\linewidth]{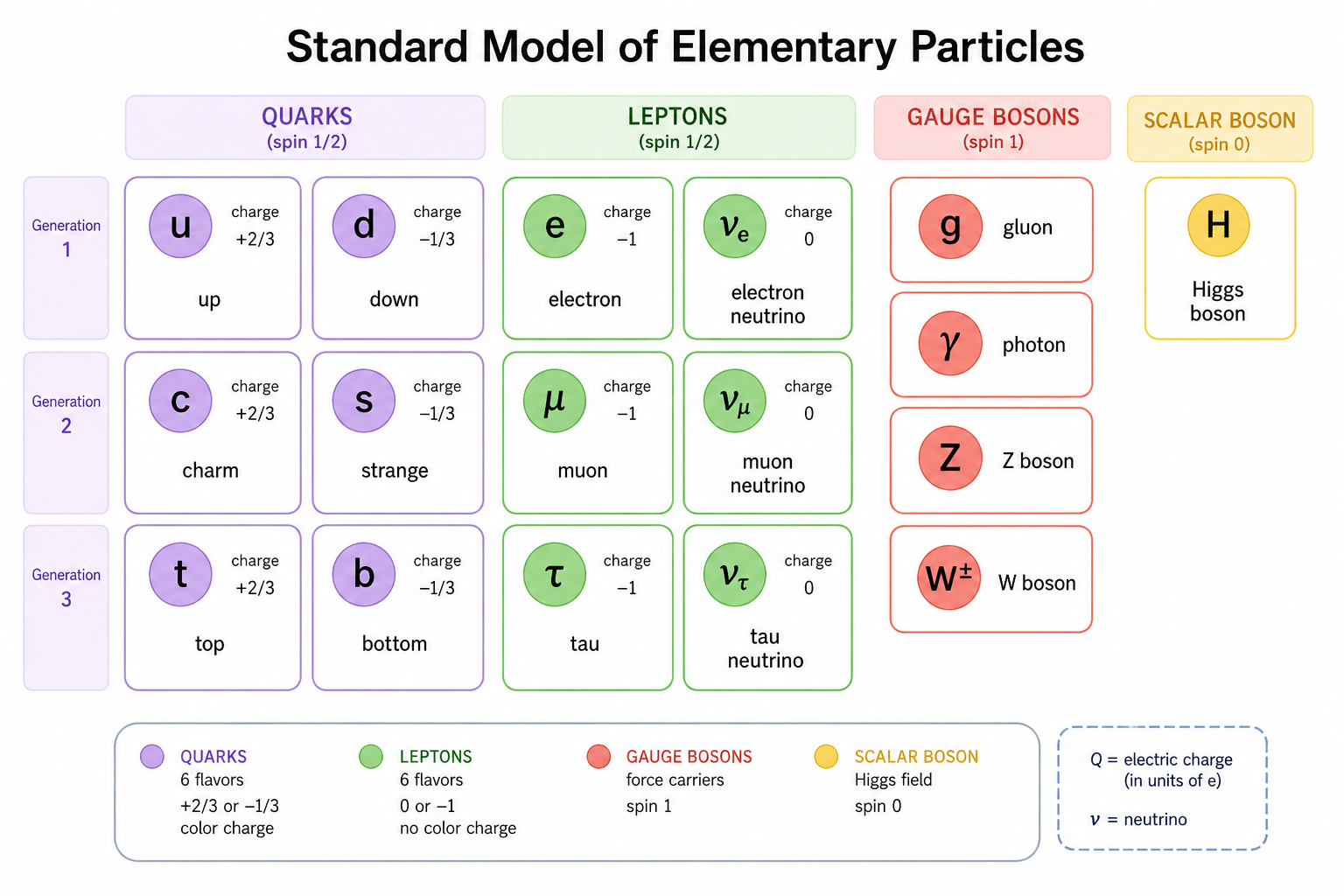}
    \caption{Overview of the SM particle content. The three generations of quarks and leptons are shown alongside the gauge bosons, which mediate the electromagnetic, weak, and strong interactions, and the Higgs boson, which is responsible for electroweak symmetry breaking and the generation of particle masses.}
    \label{fig:SM}
\end{figure}
 The particles that make up the \gls{sm} of particle physics, the current best theory describing the constituents of matter and their interactions, are shown in \figref{fig:SM}. These particles, grouped into quarks, leptons, force carrier  bosons, and the Higgs, are involved in electroweak theory describing electromagnetism and the weak interaction, and \gls{qcd} describing the strong interaction. 
 
 The SM has been a highly successful theory, describing a number of phenomena.
 For instance, the Higgs boson was first predicted in 1964, and then in 2012 it was discovered experimentally by the ATLAS and the CMS collaborations \cite{ATLAS:2012yve,CMS:2012qbp}. Other particles such as the top quark \cite{Campagnari:1996ai}, and the W and Z bosons \cite{Watkins:1986va} were discovered after being predicted by the SM.
As another incredible example of the SM's success, the electron magnetic dipole moment measurement agrees with SM theory predictions to one part in $10^{12}$ \cite{Odom:2006zz}.

Although the SM works well in predicting many observations, the SM is not (yet) a complete theory, lacking the ability to explain many phenomena. These shortcomings have opened up a new field of particle physics, \gls{bsm} physics, where these missing pieces are addressed with new physics. The following is a non-exhaustive list of some of the most studied phenomena lacking in the SM:

\noindent
{\bf Gravity:}
As of yet, there is no known way to unify our current best theory of gravitation, namely general relativity, with the other three forces (electromagnetic, weak, and strong) explained by the SM, since general relativity breaks down in the quantum realm. In addition, there has not been a force carrier particle discovered for gravity, namely a graviton \cite{Maldacena:1999fi}. There are ongoing efforts to develop a \textit{theory of everything}, such as string theory \cite{Marchesano}, which would include the unification of gravity and the SM.

\noindent
{\bf Dark Energy:}
As mentioned above, dark energy, or the origin of cosmic accelerated expansion, is unknown and unexplained by the SM \cite{Mortonson:2013zfa}. One explanation is that dark energy is a cosmological vacuum energy, although this leads to results for the theoretically calculated energy of this quantum vacuum far too large \cite{Binetruy:2009zz}. Another possibility is that dark energy comes from new physics, such as from a light scalar field \cite{Wang_1998,PhysRevLett.82.896,Frieman:1995pm,Ratra:1987rm,Wetterich:1987fm}.

\noindent
{\bf Baryon Asymmetry:}
The mechanism explaining the asymmetry in the amount of matter to anti-matter, baryogenesis, is an open question \cite{Sather:1996cz,Canetti:2012zc}. There must exist an asymmetry in the amount of matter compared to anti-matter, since if they both existed in equal amounts, all matter would have annihilated away in the early Universe. Interestingly, you and I both exist as beings made of matter, gravitationally bound to a planet made of matter, within a galaxy of matter, etc.

\noindent
{\bf Hierarchy Problem:}
The hierarchy problem refers to the large hierarchy between the Higgs mass and the Planck mass (the scale of gravity) \cite{Peskin:2025lsg}. Supersymmetry has been one well-studied option to account for this problem.

\noindent
{\bf Neutrino Masses:}
The SM does not predict that neutrinos have any mass, which is in conflict with measurements of neutrino oscillations indicating that they do indeed have mass \cite{Super-Kamiokande:1998kpq,SNO:2001kpb,SNO:2002tuh}.

\noindent
{\bf Dark Matter:} Finally we have DM, the topic of this thesis. DM is the term used to call the elusive form of matter which interacts minimally via electromagnetism, that we have evidence of through gravitational implications in astrophysics and cosmology, as we discuss in the following section. DM cannot be accounted for by the SM, so there must exist new physics to explain it.
The following sections will focus on the following questions: How do we know about the existence of DM? What are some ways it could have been produced in the early Universe? What could be its particle nature? What are the ways in which we are trying to detect it?

\section{The Unknown Mass}

It is not trivial to pinpoint the first discussions and scientific work considering the existence of unobserved (or dark) astronomical objects \cite{Bertone:2016nfn}. For example, in 1844 Friedrich Bessel realized that there must exist unobserved stars influencing the motion of stars Sirius and Procyon \cite{10.1093/mnras/6.11.136a}. Similarly to the discovery of the planet Neptune, this marks a significant occasion in history where an astronomical object was anticipated solely based on its gravitational interactions.

It turns out that the amount of mass observed in galaxies is not consistent with theoretical predictions. In 1904, Lord Kelvin noticed that the mass in our galaxy, the Milky Way, estimated from measurements of velocities of stars, must include dark bodies which we cannot see \cite{Kelvin}.

Often you will hear this story begin in 1933 with the astronomer Fritz Zwicky and his comparison between measured versus theoretically predicted velocities of galaxies within the Coma Cluster \cite{Andernach:2017vpm}. Using the virial theorem, Zwicky calculated what the velocities of 800 galaxies each $\sim 10^9$ solar masses within a radius of one million light years should be theoretically and found a value of approximately $80\: km/s$.
However, he found that the measurements of velocities using redshifts are approximately $1000\: km/s$. This discrepancy of over an order of magnitude called for the existence of some unknown missing matter, dubbed \textit{dark matter}. 

\begin{figure}
    \centering
    \includegraphics[width=0.7\linewidth]{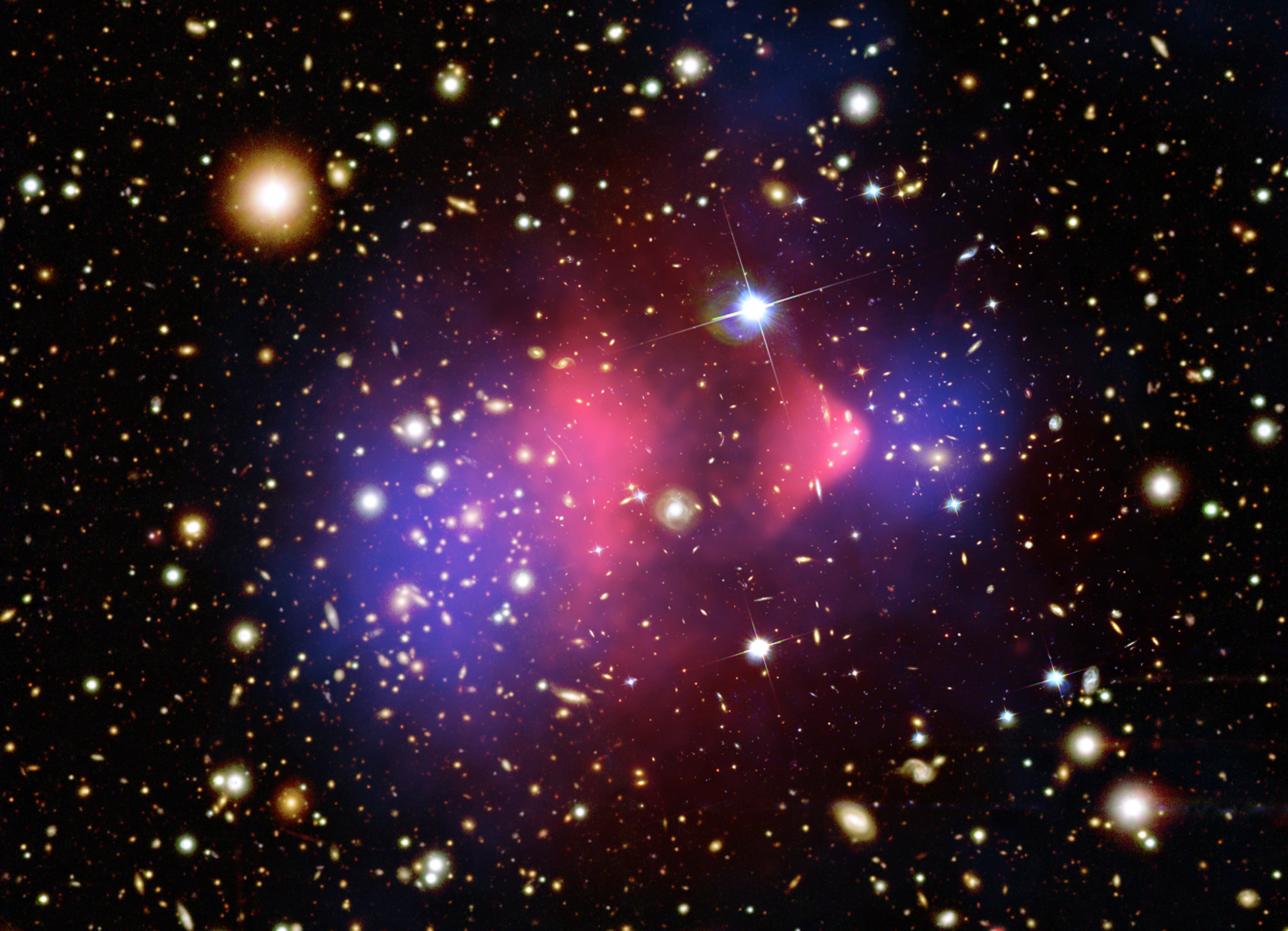}
    
    \caption{Composite image of the Bullet Cluster (1E~0657$-$56), a system of two colliding galaxy clusters. The optical component (white/orange) shows the galaxy populations of the two subclusters. The X-ray emission (pink/red) traces the hot gas, which constitutes the bulk of the baryonic mass and has been slowed and displaced by ram pressure during the collision. The blue contours show the projected mass distribution reconstructed from weak gravitational lensing, which instead traces the collisionless DM and remains spatially coincident with the galaxies rather than the gas. Image from \cite{Clowe:2003tk}.}
    \label{fig:BulletCluster}
\end{figure}
The Bullet Cluster (also known as 1E 0657-56), shown in \figref{fig:BulletCluster}, magnificently demonstrates the existence of DM and the necessity for it to be a new type of matter distinct from ordinary matter. This false colour image is of the collision of two galaxy clusters, where each cluster has an ordinary matter component and a DM component. The white and yellow spots are other galaxies in the distance, taken in the optical by the Magellan and the Hubble Space telescopes. The red is the X-rays detected by the Chandra telescope, which shows where the baryonic/ordinary matter (hydrogen gas) exists within the clusters. Finally, the blue is where most of the mass is concentrated from gravitational lensing\footnote{Gravitational lensing is the effect where gravity from a massive object causes light passing near it to bend. By measuring the distorted shapes of luminous objects under this effect by performing gravitational lensing surveys, the distribution of mass can be determined.} measurements. As evident in \figref{fig:BulletCluster}, there is a mismatch between the red and blue regions, namely the expected and the measured locations of the mass in the clusters. Notice also that the blue region has moved further from the collision center than the red regions, showing the slowing of the gas from frictional forces and the blue matter's lack of friction. With this, we have evidence that the majority of the mass in these galaxy clusters demonstrating this particular behaviour is some additional invisible matter, DM, which interacts very weakly. This is the most famous example, although there are several other examples of galaxy clusters demonstrating the existence of DM.

Zooming into individual galaxies, the shapes of the rotation curves of galaxies also demonstrate the existence of DM \cite{Begeman:1991iy,Sofue:2000jx}. It is expected that after a sufficient distance from the centre of the galaxy, the radial velocities, $v(r)$, will decrease with increasing radius, $r$, with the relationship,
\begin{equation}
    v(r) = \sqrt{\frac{GM(r)}{r}},
\end{equation}
where $G$ is the gravitational constant and $M(r) = 4\pi \int_0^r \rho(r')r'^2 dr'$, with $\rho(r)$ as the mass density profile of ordinary matter.
This is in contradiction to measurements where the radial velocity curves instead flatten out \cite{Begeman:1991iy,Sofue:2000jx}.
\begin{figure}
    \centering
    \includegraphics[width=0.5\linewidth]{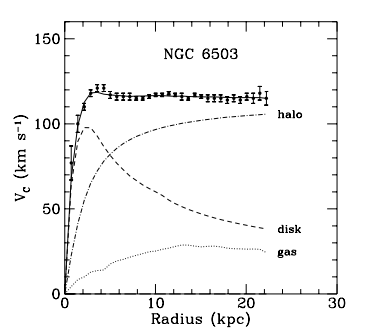}
    \caption{Rotational velocities of stars in the spiral galaxy NGC 6503 \cite{Begeman:1991iy}. The points with error bars are measurements using red shifts. The dashed curve labeled ``disk", the dotted curve labeled ``gas", the dash dotted line labeled ``halo", and the solid curve going through the data are theoretical curves considering only ordinary matter, gas, a DM halo, and ordinary matter with a DM halo added in quadrature, respectively.} 
    \label{fig:RotationCurves}
\end{figure}
As an example, we take the rotation curve of spiral galaxy NGC 6503, as shown in \figref{fig:RotationCurves}. Here it is evident that only considering the ordinary matter component leads to predictions that do not match the data. There must be a DM halo component in order to reproduce the flattened rotation curves. The DM halos that are hypothesized to exist as the major component of galaxies are predicted to have a spherical shape\footnote{It is plausible that the DM halo shapes are not exactly spherical \cite{Kamionkowski:1997xg}.} distributed according to rotation curve observations with a mass density following $\rho \sim r^{-2}$. However, the density profiles of DM in the center regions of galaxies differ whether considering simulations or observational data \cite{Dutton:2018nop}. N-body simulations of galaxy formation lead to central profiles that are \textit{cuspy} following $\rho \sim 1/r$ at small $r$ \cite{Navarro:1995iw}, while the data is instead consistent with a constant density \textit{core} where $\rho \sim r^0$ \cite{Moore:1994yx}. This disagreement is called the ``core-cusp problem", which has had many proposed solutions \cite{Deng:2018jjz}.

During the 1970s, it became more accepted that galaxies are enveloped inside of these massive DM halos \cite{Frenk:2012ph,Ostriker:1973uit}. These halos provide the galaxies with large gravitational potential wells in which ordinary matter can cool and form into galaxies \cite{White:1977jf}. From simulations of structure formation, the Universe as we know it today cannot exist without a significant DM component. Without DM present, structure cannot begin to grow until after recombination\footnote{Recombination is the cosmological epoch where the Universe is cooled sufficiently for electrons and protons to form neutral hydrogen atoms.} -- which is inconsistent with the galaxy we observe today. Density fluctuations of baryonic matter do not increase with time before matter-radiation decoupling due to the photon pressure from frequent Compton scattering. Therefore, DM is necessary to have the galactic structure in which we live.

\begin{figure}
    \centering
    \includegraphics[width=0.7\linewidth]{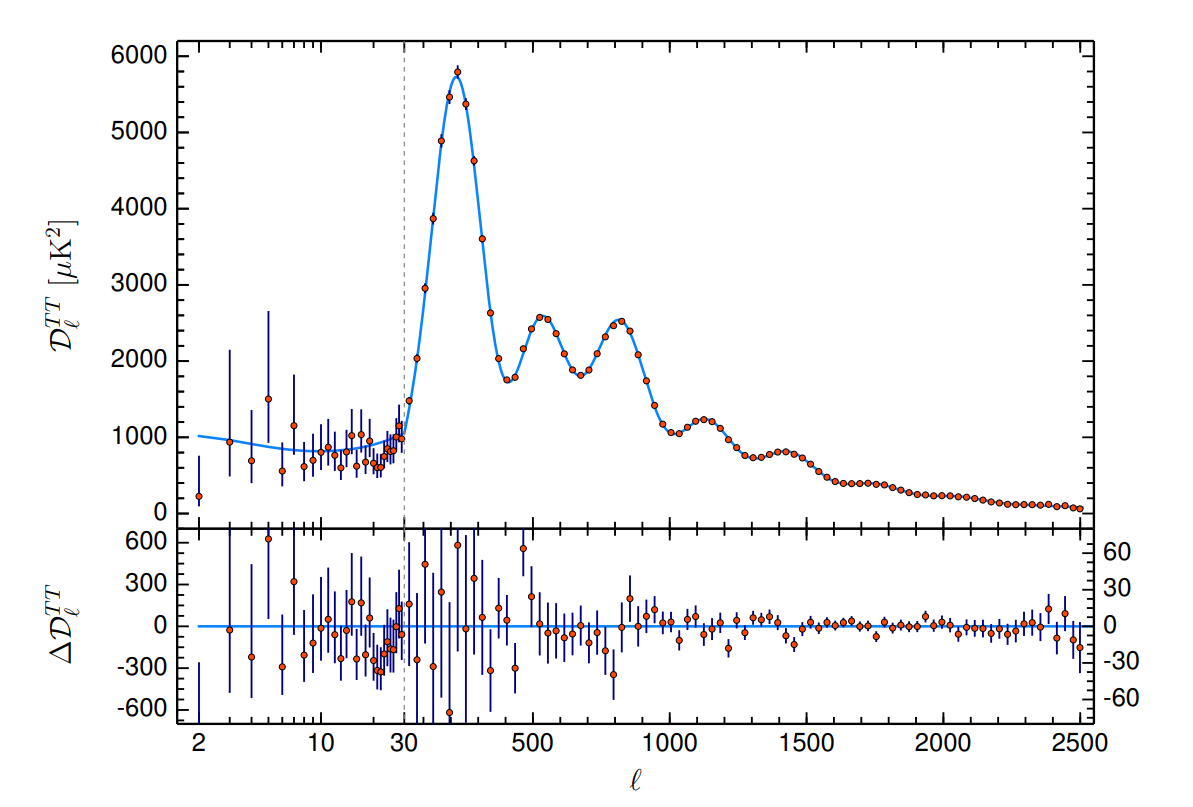}
    \caption{The temperature power spectrum of the CMB \cite{Planck:2018vyg}, where the temperature fluctuations are plotted versus multipole $l$. The measurements made by the Planck satellite are plotted in red along with their error bars, and the $\Lambda$CDM theoretical best fit is plotted in light blue.}
    \label{fig:CMBpower}
\end{figure}

The final piece of evidence for DM to mention: the CMB anisotropies. The oscillations of the photon and baryon plasma in the early Universe cause the angular fluctuations of the CMB temperature. These oscillations are caused by the competing forces from gravity and photon pressure. 
In order to quantify the temperature anisotropies, we compute the power spectrum -- the angular correlation function of the temperature differences expanded in spherical harmonics, as shown in \figref{fig:CMBpower}.
The angular scales of the CMB are depicted by the peaks and troughs, which change with the abundance of DM, ordinary matter, and dark energy showing their individual contributions \cite{Hu:2001bc}. The blue curve in \figref{fig:CMBpower} is the best fit of the $\Lambda$CDM (Cold DM) model, the current best cosmological model \cite{Bull:2015stt}. The cosmological parameters that are fit include the abundance  of baryons and DM in the Universe, $\Omega_\text{B}$ and $\Omega_\text{DM}$\footnote{Often (and later in this thesis) cosmological abundances are expressed in units of $\Omega h^2$, in which $H_0$ is factored out.}, where $\Omega_i \equiv \rho_i / \rho_c$, and the critical density is given by,
\begin{equation}
    \rho_c \equiv \frac{3H_0^2}{8\pi G}.
\end{equation}
The Universe's energy budget is 69\% dark energy, 26\% DM, and 5\% ordinary matter \cite{Planck:2018vyg}.

With this undeniable evidence for DM, existing at many scales and times in the Universe, we arrive at attempting to solve one of the greatest scientific mysteries of our time.

 \chapter{Particle Dark Matter}
\label{chap:ParticleDarkMatter}

\section{Its Nature and Cosmological Origin}
\label{section:ParticleDM}

What could make up this mysterious matter causing measurable gravitational effects in astrophysical and cosmological objects? An obvious explanation is that DM is made up of ordinary matter compact objects that are simply less luminous. These objects include planets, brown/red/white dwarfs, neutron stars, and black holes: categorized as \gls{macho} \cite{Hegyi:1983bj}. MACHOs have mostly been ruled out as a plausible DM candidate because of the lack of them found in microlensing surveys \cite{EROS-2:2006ryy,Lasserre:2001ygk} and measurements of the baryon density (such as from the CMB \cite{Planck:2018vyg}), leaving a small mass window of primordial black holes as technically viable DM candidates \cite{Villanueva-Domingo:2021spv,Carr:2020xqk}.

Another explanation is that it is in fact not matter at all, but instead evidence for the necessity for a more complete theory of gravity such as MOND \cite{Milgrom:1983ca,Milgrom:1983pn,Milgrom:1983zz} as introduced earlier. MOND is able to explain some of the shapes of rotation curves, however MOND has trouble when it comes to observations of galaxy clusters and the CMB \cite{Aguirre:2001fj,Clowe:2006eq}.

In order to explain the total amount of DM and the entirety of its gravitational evidence, the idea that DM is made up of particles (similarly to ordinary matter) became the leading hypothesis in the field~\cite{Bertone:2016nfn}. Only interacting through the weak and gravitational forces, SM neutrinos were hypothesized to be DM. One of the problems with neutrinos is that they were hot (relativistic) during structure formation of galaxies \cite{Peebles:1982ib}, which is inconsistent with simulations of structure formation showing the need for cold (non-relativistic) DM \cite{White:1983fcs}. Another problem is that there is a lower bound on the mass of galactic halo leptons called the ``Tremaine-Gunn bound" \cite{Tremaine:1979we}, since the density of fermions cannot exceed that of a degenerate Fermi gas. Since neutrinos are the only stable and electromagnetically neutral particle in the SM, this leaves us with the need for a new type of BSM particle for DM. Despite the lack of information we have on the nature of this mysterious matter, this new hypothetical particle must fulfill some criteria: stable on cosmological scales, sufficiently weak interactions with ordinary matter consistent with null experimental results, has an early universe production mechanism giving rise to the observed relic abundance (see \secref{section:FreezeOut}), non-relativistic during structure formation, and small enough self interactions consistent with cluster observations.

The \gls{WIMP}~\cite{Steigman:1984ac,Saikawa:2020swg,Schumann:2019eaa,Queiroz:2017kxt,Arcadi:2017kky} has been one of the most popular classes of DM candidates\footnote{The definition of WIMP in the literature is not fully consistent - some consider them to interact via the $SU(2)_L$ gauge bosons in the SM (via the weak force), while others consider them to interact through new interactions via a new mediator. Here we take them to be of mass $O(GeV)$ to $\approx O(100 TeV)$ and produced through the freeze-out production mechanism, interacting via SM or BSM forces.}, arising from BSM theories such as Supersymmetry or from theories that attempt to solve the hierarchy problem. Interestingly for this candidate, the observed relic abundance by Planck \cite{Planck:2018vyg} is obtained through freeze-out while having a cross section and mass of the weak scale. This realization (or coincidence) has been coined the ``WIMP miracle". WIMPs have a mass in the $\sim \mathcal{O}(\rm GeV)$ to $\sim \mathcal{O}(\rm 100 \: TeV)$ range, bounded below by the relic density \cite{Lee:1977ua} and above by unitarity \cite{Griest:1989wd}. This lower bound exists if the WIMP interacts through the SM weak force, where the Lee-Weinberg bound sets a lower bound on the mass due to avoiding overproduction of the DM density. However, this lower bound can be circumvented by the addition of a new gauge mediator, allowing for masses below a GeV.

\subsection{Thermal Production: Freeze-Out}
\label{section:FreezeOut}
The cosmological abundance of DM is known from measurements of the CMB \cite{Planck:2018vyg}. However, the precise history of how this abundance came to be is not known. If we consider that DM was once in thermal equilibrium with the SM bath, the DM history will follow that of the \textit{freeze-out} mechanism \cite{Gondolo:1990dk,Srednicki:1988ce}. There are other possible mechanisms that can produce the observed abundance of DM such as \textit{freeze-in} \cite{Cosme:2021baj}, where DM was never in thermal equilibrium with the SM bath, but here we focus on DM produced through freeze-out.

\begin{figure}
    \centering
    \includegraphics[width=0.8\linewidth]{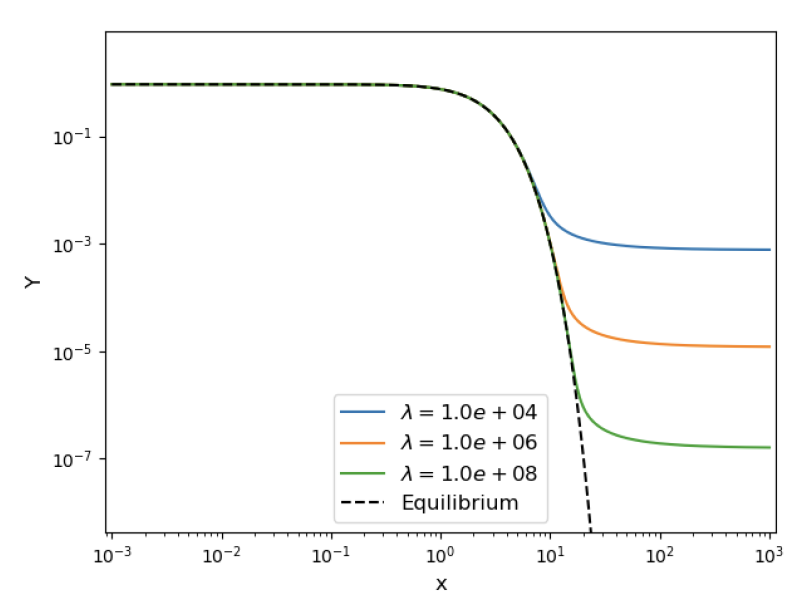}
    \caption{The evolution of the DM yield, or comoving number density, $Y$ plotted vs $x\equiv m/T$ (a variable representing something similar to time) where $m$ is a relevant mass scale and $T$ is temperature. Coloured curves are the DM yield in the freeze-out scenario at various coupling strengths, and the dashed curve is the equilibrium yield assuming Maxwell-Boltzmann statistics.}
    \label{fig:freeze_out}
\end{figure}

The assumption that DM was once in thermal equilibrium with the SM bath is not implausible, since most other species were once in thermal equilibrium, sharing a common temperature $T$. Figure~\ref{fig:freeze_out} plots the evolution of the DM density, or yield (described below) as a function of time. In the freeze-out mechanism, the DM abundance (coloured solid curves), follows the equilibrium curve (dashed curve) until chemical decoupling -- when the Hubble rate exceeds the production and annihilation rate of DM. Before freeze-out occurs, the average kinetic energy of the SM bath species will not be sufficient to efficiently produce DM. This phenomenon, called \textit{Boltzmann suppression}, causes the equilibrium yield curve to exponentially fall to zero, from the exponential in the Maxwell-Boltzmann distribution that decreases rapidly as temperature decreases.
The temperature at which Boltzmann suppression occurs is whichever happens first: either when the temperature drops below the mass of the DM (if the DM is more massive than the initial species it is produced from) or when the abundance of the initial species becomes small due to their own Boltzmann suppression (if DM is lighter than the other species). The interaction strength, or coupling, between the DM and SM (or other dark sector particles), governed by the DM model, influences how long the yield can follow the equilibrium curve until decoupling. Larger couplings lead to a smaller final DM yield.

The number density of a species of mass $m$ and $\gamma$ internal degrees of freedom in thermal equilibrium, assuming Maxwell Boltzmann statistics, follows \cite{Kolb:1990vq},
\begin{equation}
    n^{eq} (T)=
    \frac{\gamma m^2 T K_2\left(\frac{m}{T}\right)}{2\pi^2},
    \label{eq:neq}
\end{equation}
where $K_i(x)$ is the modified Bessel function of the second kind of order $i$.
Departures from thermal equilibrium were vital in establishing a relic, such as a DM relic and of course you, the reader. These departures from thermal equilibrium, namely the process of \textit{decoupling} as mentioned already above, occur when the interaction rate, $\Gamma$, of the given species with the other bath species drops below the Hubble expansion rate, $H$.

In order to calculate the abundance of DM for a given theoretical framework for DM, the decoupling must be treated by using the Boltzmann equation to compute the evolution of the particle's phase space distribution \cite{Gondolo:1990dk,Kolb:1990vq}. You may refer to \cite{Gray:2020yzn, Kolb:1990vq,Gondolo:1990dk} for a review of equilibrium thermodynamics and more details on the Boltzmann equation.
The number density $n$ is governed by the Boltzmann fluid equation,
\begin{equation}
    \dot{n}+3Hn = R(t),
    \label{eq:Boltzmann_n}
\end{equation}
where $R$ is the interaction rate density, the number of interactions per unit time and volume,
\begin{equation}
    R(t) = \frac{\gamma}{(2\pi)^3}\int C[f]\frac{d^3p}{E},
\end{equation}
where $C[f]$ is the collision operator\footnote{The collision operator is defined as the rate of change of the distribution function of the species of interest due to collision processes such as scattering and annihilation.} as a function of the distribution function $f$ containing all particle number changing interactions, $E$ is the energy and $p$ is the momentum of the particle of interest and $H$ is the Hubble rate. 
The Hubble rate describes the expansion rate of the universe,
\begin{equation}
    H\equiv\frac{\dot{a}}{a}
\end{equation}
where $a$ is the scale factor.
During the radiation dominated era the Hubble parameter is given by,
\begin{equation}
    H(T) = \sqrt{\frac{8\pi^3}{90}}\frac{\sqrt{g_{*e}}T^2}{m_{Pl}},
    \label{eq:H}
\end{equation}
where $m_{Pl}$ is the Planck mass and $g_{*e}(T)$ is the effective number of degrees of freedom associated with energy \cite{Husdal:2016haj} given by,
\begin{equation}
    g_{*e} = \sum_{\text{bosons}\:i}\gamma_i\left(\frac{T_i}{T}\right)^4 + \frac{7}{8}\sum_{\text{fermions}\:j} \gamma_j \left( \frac{T_j}{T}\right)^4
\end{equation}
and $T_{i(j)}$ is the temperature of species $i$ ($j$), while $T$ is the thermal bath temperature.

Considering the process $i+j+\cdots \leftrightarrow k+l+\cdots$, if we are interested in the particle $i$,
\begin{equation}
\begin{aligned}
    R_i(t) &= \frac{\gamma_i}{(2\pi)^3}\int C[f] \frac{d^3p_i}{E_i}\\
     &= \int d\Pi_i d\Pi_j \cdots d\Pi_k d\Pi_l \cdots (2\pi)^4 \delta^4(p_i+p_j+\cdots - p_k - p_l - \cdots)\\ &\left[|\bar{M}|_{k+l+\cdots \to i+j+\cdots}^2 f_k f_l \cdots (1\pm f_i)(1\pm f_j)\cdots \right. \\
    &\left. - |\bar{M}|_{i+j+\cdots \to k + l+\cdots} f_i f_j \cdots (1 \pm f_k)(1 \pm f_l) \cdots \right]
\end{aligned}
\end{equation}
where $|\bar{M}|^2$ is the squared amplitude averaged over initial and final spins\footnote{In cosmology we average over initial and final spins, rather than averaging over initial and summing over final spins, since we do not know the initial or final states.}, $d\Pi_i$ are the phase space differential elements,
\begin{equation}
    d\Pi_i=\frac{\gamma_i d^3p_i}{(2\pi)^3 2E_i},
\end{equation}
for every species index $i$,
and the $(1 \pm f)$ factors are the stimulated ($+$) and blocking ($-$) factors for final state bosons and fermions respectively.

Let us now define two useful parameters. To scale out the universe's expansion, we use the comoving number density or yield, $Y$, defined by,
\begin{equation}
    Y \equiv \frac{n}{s},
\end{equation}
where $s$ is the entropy density. We use $x$ as a parameter representing time in a convenient way,
\begin{equation}
    x \equiv \frac{m}{T},
\end{equation}
where $m$ is a relevant mass scale, usually set to be of the particle of interest, $m_i$.

Let us consider the process $34\leftrightarrow 12$\footnote{All other number changing processes such as $1 \to 2$ and $3 \to 2$ (where the latter is relevant for Paper I) must be considered when solving the Bolzmann equation.} where we are interested in calculating the final yield of particle 3, $Y_\infty$. Let us also make a few assumptions that simplify the problem: (1) charge-parity (time) invariance hence $|\bar{M}|^2_{12\to34}=|\bar{M}|^2_{34\to12}$, (2) Maxwell Boltzmann statistics thus emitting the stimulated and blocking $(1\pm f)$ factors, and (3) that particle 3 is in thermal equilibrium until it decouples and moreover is in kinetic equilibrium during chemical decoupling.
We write the Boltzmann equation, \eqnref{eq:Boltzmann_n}, in terms of $Y$ and $x$ under these assumptions,
\begin{equation}
    \frac{dY_3}{dx}= \sqrt{\frac{\pi}{45}}\frac{m_{Pl}m_3g_*}{x^2}\langle \sigma v\rangle_{34\to12} \left( \left(Y_3^{eq}\right)^2 - Y_3^2 \right)
    \label{eq:2to2Boltz}
\end{equation}
where $Y^{eq}$ is the equilibrium yield given by,
\begin{equation}
    Y_i^{eq} = \frac{n_i^{eq}}{s}= \frac{45 \gamma_i x^2K_2(x)}{4\pi^4 g_{*s}},
    \label{eq:Yeq}
\end{equation}
and the thermally averaged cross section is,
\begin{equation}
    \langle \sigma v\rangle_{34\to12} = \frac{\int\sigma_{34\to12}(s_m-4m_3^2)\sqrt{s_m}K_1\left( \frac{\sqrt{s_m}}{T}\right)ds_m}{8m_3^4TK_2\left(\frac{m_3}{T}\right)^2},
\end{equation}
with the Mandelstam variable $s_m \equiv (p_3+p_4)^2 = (p_1+p_2)^2$ and the cross section of the process $\sigma_{34\to12}$. 
\eqnref{eq:2to2Boltz} is a differential equation with no analytic solution, therefore we must rely on numerical methods to solve it. However, in the case of freeze-out, if we can determine the freeze-out temperature $T_F$, the final present day yield can be approximated. After freeze-out, $Y^{eq} \ll Y$, so $Y^{eq}$ in \eqnref{eq:2to2Boltz} can be safely neglected.

We take the result of \cite{Gondolo:1990dk} (also see \cite{Gray:2020yzn} for more details) for the yield of particle three at $t=\infty$, or the present day yield,
\begin{equation}
    \frac{1}{Y_{\infty}} \approx \frac{1}{Y_f} + \int_{0}^{T_f} \sqrt{\frac{\pi}{45}}m_{Pl}m_3g_*\langle \sigma v\rangle_{34\to12}dT.
    \label{eq:FOapprox2to2}
\end{equation}
The freeze-out temperature can be calculated in two main ways. 
For the first method, we can numerically solve for $T_f$, or $x_f \equiv \frac{m_3}{T_f}$, using \cite{Gondolo:1990dk},
\begin{equation}
    \left.\frac{dY_3^{eq}}{dx}\right|_{x_f} \approx -\sqrt{\frac{\pi}{45}}\frac{m_{Pl}m_3g_*(x_f)}{x_f^2}\langle \sigma v\rangle _{34\to12}(Y_3^{eq}(x_f))^2 \left(\delta^2+2\delta \right)
\end{equation}
where $\left.\frac{dY_3^{eq}}{dx}\right|_{x_f}$ is the derivative of $Y_3^{eq}(x)$ with respect to $x$ evaluated at $x_f$, and we can take $\delta=1.5$ \cite{Gondolo:1990dk}.
The second method for calculating $T_f$ is to take the temperature at which the Hubble rate exceeds the annihilation rate of particle 3. We use the first method throughout this thesis.

These methods of calculating the relic abundance of DM are used throughout this thesis for determining the parameters consistent with the measured abundance, the so called \textit{relic targets}.

\section{Dark Matter Searches}
In this section, we briefly outline the types of efforts that exist to search for signals from DM particles.
\begin{figure}[h]
    \centering
    \includegraphics[width=0.5\linewidth]{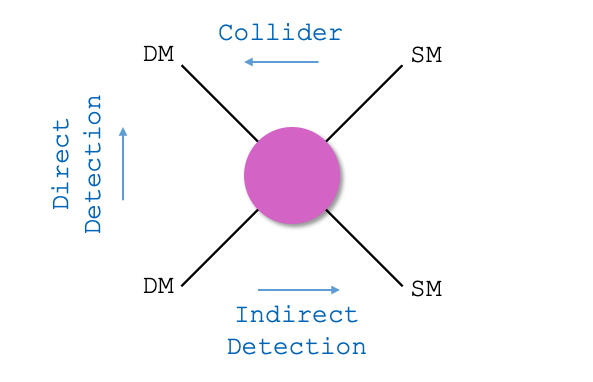}
    \caption{The DM search channels at collider, direct detection, and indirect detection searches.}
    \label{fig:DMsearches}
\end{figure}

\subsection*{Indirect Detection}
Going from left to right in \figref{fig:DMsearches}, we have DM annihilations into SM species. If DM particles annihilate into ordinary matter, a measurable signal of photons, neutrinos \cite{deDiosZornoza:2021rgw}, or cosmic rays can be produced and observed \cite{PerezdelosHeros:2020qyt}. DM annihilations can also have an impact on the CMB, \gls{bbn}, and the 21-cm Hydrogen line \cite{Slatyer:2015jla,Slatyer:2015kla}. This detection channel plays less of a prominent role in this thesis compared to the others, however we consider it in Paper II in combination with other channels.

\subsection*{Direct Detection}
\label{sec:DirectDetection}

Now starting at the bottom and going upwards in \figref{fig:DMsearches}, we have DM scattering with SM particles in direct detection searches \cite{Baudis:2025yva,Schumann:2019eaa,Catena:2022fnk,Akerib:2022ort}. DM particles from the galaxy can scatter off target particles in detector materials on Earth, with the expected signal rate dependent on the local DM velocity distribution, typically taken to follow the Standard Halo Model \cite{Baxter:2021pqo}. Depending on its kinetic energy and the momentum transfer, DM can scatter with electrons, partons, nucleons, or nuclei, and cause a measurable signal. Traditional searches target nuclear recoils from WIMP-scale DM, such as XENONnT \cite{XENON:2023cxc}, LZ \cite{LZ:2024zvo}, and PandaX \cite{PandaX:2024qfu}, but this sensitivity degrades sharply for sub-GeV DM masses and quickly falls below realistic detector thresholds, motivating the electron scattering channel \cite{Mitridate:2022tnv}.

For sub-GeV DM, a central element of this thesis which is introduced below, the dominant detectable process is instead DM-electron scattering \cite{Essig:2011nj}. Because the electron mass is far smaller than the nucleon mass, DM as light as $\mathcal{O}(\mathrm{MeV})$ can transfer enough momentum to ionize an atom or excite an electron across a semiconductor band gap, producing a detectable signal even though the corresponding nuclear recoil would be invisible. Experimentally, this channel has been pursued by SENSEI \cite{SENSEI:2023zdf}, DAMIC-M \cite{DAMIC-M:2025luv}, PandaX-4T \cite{PandaX:2025rrz}, SuperCDMS \cite{SuperCDMS:2024yiv}, and XENONnT \cite{XENON:2026qow}, among others. 
Calculating the DM-electron scattering rate requires the electronic structure of the target material, described either through atomic or dielectric response methods, and more generally within an effective framework for DM-electron interactions \cite{Catena:2019gfa,Catena:2024rym}.

Direct detection experiments can extend their sensitivity to light DM via the \textit{Migdal effect}, the ionization of atomic electrons from the instantaneous accelerations of the atom, a mechanism whose existence was confirmed in early 2026 through the direct observation of Migdal-induced electron emission in neutron–nucleus collisions \cite{Yi:2026fmf}.

\subsection*{Collider}
Finally from right to left in \figref{fig:DMsearches}, we have searches where the collision of SM particles produces DM or dark sector particles \cite{Graham:2021ggy}. These SM collisions can be between two high energy beams (e.g., electrons, protons, heavy ions) or a beam and a fixed-target. The largest particle accelerator in the world, the \gls{lhc}, among other physics, searches for BSM physics such as DM \cite{PerezAdan:2023rsl}. However, since LHC searches are more sensitive to $\gtrsim \textrm{GeV}$ mass DM, this thesis focuses on lower energy, higher intensity accelerator experiments better suited as sub-GeV DM searches. 
These searches can be grouped into invisible and visible searches: where the former searches for ``invisible" or dark signatures such as missing energy/momentum or DM scattering, and the latter searches for the visible decay products of dark sector particles (such as from the dark photon).
In \chapref{chap:AcceleratorExperiments}, we discuss in detail a class of accelerator experiments: fixed-target experiments -- a core aspect of this thesis -- and consider these experiments in the context of light DM.

\section{Below a GeV: Beyond the WIMP}

If DM is produced thermally in the early universe, i.e. it was once in thermal equilibrium with the SM and produced through freeze-out, its mass can be from $\sim \mathcal{O}(\text{MeV})$ to $\sim \mathcal{O}(100\:\text{TeV})$ \cite{Lee:1977ua}. The upper mass bound arises because of unitarity violations, and the lower bound from cosmological observations\footnote{Energy injection into the SM during BBN is tightly constrained \cite{Chu:2022xuh}.}. 
The experimental efforts in search of nuclear recoils from DM-nuclei scattering, in particular direct detection experiments as described in \secref{sec:DirectDetection}, are very important for probing GeV - TeV scale DM such as WIMPs. 

The absence of a concrete WIMP signal across direct detection, indirect detection, and collider searches has shifted significant theoretical and experimental attention toward DM candidates with masses below
the GeV scale, coupled to the SM through light mediators
rather than through electroweak strength interactions at the weak scale~\cite{Battaglieri:2017aum}. In this case, DM of sub-GeV mass would evade current nuclear recoil direct detection experiments. Among the simplest and most widely studied scenarios, is the one where the dark photon, $A'$, (introduced in the following section) plays the role of the mediator between DM and the SM.
For DM masses in the
$\mathrm{MeV}$--$\mathrm{GeV}$ range, thermal freeze-out through this portal fixes a calculable relation between the mediator mass $m_{A'}$, the mixing
parameter $\epsilon$, the dark coupling $\alpha_D$, and the DM mass $m_\chi$ that reproduces the observed relic abundance. This relation defines the
\textit{thermal target} benchmark curves, conventionally expressed in terms of the parameter $y \equiv \alpha_D \, \epsilon^2 (m_\chi/m_{A'})^4$ (introduced below), against which the sensitivity of accelerator-based searches is typically
quoted~\cite{Berlin:2018bsc}. As mentioned above, such light masses do not deposit enough energy via nuclear recoils to be detected.

DM-electron scattering however, is a suitable mechanism in which sub-GeV DM particles can be directly discovered. Complementary to DM-electron scattering at direct detection experiments, accelerator-based searches give strong constraints on sub-GeV DM for many theoretical scenarios \cite{Battaglieri:2017aum}. The core subject of this thesis is sub-GeV (or light) DM, and using one of the leading types of experiments for these candidates, fixed target experiments, as a way to measure a potential DM signal.

\section{Dark Photons}
\label{subsection:DarkPhotons}

If thermal DM has a mass below a GeV, there must exist a new light mediator, providing the interactions between DM and SM particles. One possibility is a dark photon mediator, $\Ap$, arising from a new $U(1)'$ symmetry group under which all SM fields are singlets \cite{Fabbrichesi:2020wbt,Holdom:1985ag,Babu:1997st}. The mechanism responsible for the generation of the mass of this new gauge boson could come from the Stueckelberg mechanism \cite{Stueckelberg:1938hvi} or from an extended Higgs sector.
The presence of a new $U(1)'$ symmetry describing a new Abelian\footnote{An Abelian theory is one where the gauge group generators commute; $U(1)$ is the simplest example, having only a single generator, unlike non-Abelian groups such as $SU(2)$ or $SU(3)$.} gauge boson gives rise to \gls{KM} between the SM photon and the $\Ap$, a mixing of their kinetic terms that, once diagonalized, endows the $\Ap$ with an effective coupling to the SM electromagnetic current suppressed by $\epsilon$.
The KM with the SM hyper-charge \cite{Holdom:1985ag,Babu:1997st} is,
\begin{equation}
    \mathscr{L}_\text{KM} = -\frac{\epsilon}{2 \cos \theta_W}F'_{\mu\nu}B^{\mu \nu},
    \label{eq:Lkinetic_mixing}
\end{equation}
with the field strength tensor $X^{\mu\nu} \equiv \partial^\mu X^\nu - \partial^\nu X^\mu$ for a generic field $X_\mu$. $F'^{\mu\nu}$ and $B^{\mu\nu}$ are the  field strength tensors of the $U(1)'$ gauge field $\tilde{A}'_\mu$ and the SM hyper-charge gauge field $B_\mu$, respectively (both before field redefinitions). $\theta_W$ is the weak mixing angle, and $\epsilon$ is the \gls{KM} strength.

Diagonalizing the kinetic terms in \eqnref{eq:Lkinetic_mixing} results in the Lagrangian for the interaction between SM fermions and the $\Ap$\footnote{Assuming $m_{A'} \ll m_Z$},
\begin{equation}
    \mathscr{L}_{A'} = -\frac{1}{2} m_{\Ap}^2 {\Ap}^{\mu} \Ap_\mu - \frac{1}{4}{\Ap}^{\mu \nu} \Ap_{\mu \nu} - \epsilon e A'^\mu \sum_f q_f \bar{f} \gamma_\mu f,
    \label{eq:L_DP}
\end{equation}
 where $A'^\mu$ is the physical dark photon vector field after diagonalizing the kinetic terms and canonically normalizing, $A'^{\mu\nu}$ is its field strength tensor, $e$ is the elementary charge, $f$ is the SM fermion spinor, and $q_f$ is their electric charges. I have specifically left out the interaction terms between the $\Ap$ and the DM, since they are dependent on the nature of the DM particle. Throughout this thesis, we assume various DM models, thus different interaction Lagrangians between the $\Ap$ and DM. In \partref{part:StandardParadigm}, we consider minimal DM models where the DM candidate is a complex scalar or a fermion, described in the following subsection. We go beyond these minimal models in \chapref{chap:Models}, exploring vector DM frameworks with extended scalar sectors and new non-Abelian gauge symmetries.

 \subsection{Minimal Dark Matter Models}
 \label{sec:minimal_DM_models}
 
 The minimal DM models presented here, and used throughout a large part of this thesis, are simplified models where DM interacts with SM through the dark photon mediator \cite{Berlin:2018bsc}. These models are predictive, in the sense that their early universe dynamics are directly related to laboratory observables. DM annihilation into SM fermion pairs, through a virtual dark photon, is the dominant process setting the relic abundance, given $m_{A'}>2m_\text{DM}$. The DM annihilation cross section (times velocity), if we take $m_{A'} \gg m_\text{DM}$, scales as,
 \begin{equation}
     \sigma v(\chi \chi \to f\bar{f}) \propto \frac{\epsilon^2 \alpha_D m_\text{DM}^2}{m_{A'}^4} \equiv \frac{y}{m_\text{DM}^2}, \quad y\equiv \epsilon^2 \alpha_D \left( \frac{m_\text{DM}}{m_{A'}} \right)^4,
 \end{equation}
 where $\alpha_D\equiv g_D^2/4\pi$, $g_D$ is defined below, and we have introduced a new convenient parameter $y$, in which experimental searches are also sensitive to. This allows for easy comparison of the thermal targets, the curves in the $(y,m_\text{DM})$ plane consistent with the observed relic abundance measured by Planck \cite{Planck:2018vyg}, with the experimental sensitivity in the form of exclusion bounds. In this thesis, we consider complex scalar DM and Dirac fermion DM. Paper II and \chapref{chap:GlobalFits} studies and compares both scenarios in the context of global fits.

\subsubsection*{Complex Scalar Dark Matter}
The Lagrangian associated with complex scalar DM is,
 \begin{equation}
    \mathscr{L}_\text{DM} = |\partial_\mu \chi|^2 - m_{\text{DM}}^2|\chi|^2 + ig_{\text{D}}A'^\mu \big[ \chi^* \left(\partial_\mu \chi \right) - \left(\partial_\mu \chi^* \right) \chi \big] - g_D^2 A'^\mu A'_\mu |\chi|^2,
\end{equation}
where $\chi$ is a complex scalar.
We take this model as a default benchmark scenario throughout the thesis, in particular for Paper II, III, IV, V, and VII. Notably, for complex scalar DM the non-relativistic DM annihilation cross section into SM fermions, multiplied by velocity $v$, is proportional to $v^2$, where $v$ is velocity, and is therefore $p-$wave suppressed at late times (small velocities) \cite{Boehm:2003hm,Boehm:2020wbt}. Strong bounds from energy injection during recombination \cite{Slatyer:2009yq} are thus evaded in this scenario \cite{Balan:2024cmq}.

\subsubsection*{Dirac Fermion Dark Matter}
Next, we consider DM that is a Dirac fermion, with the Lagrangian given by,
\begin{equation}
    \mathscr{L}_\text{DM} =\bar{\chi}(i\slashed{\partial}-m_\text{DM})\chi +g_D A'^\mu \bar{\chi}\gamma^\mu \chi,
\end{equation}
where $\chi$ is a Dirac fermion. We only consider vector-like couplings\footnote{Axial-vector couplings can only arise in dark Higgs mechanisms \cite{Bell:2016uhg}.}. Contrary to the complex scalar DM case, DM annihilates via $s-$wave and therefore is not velocity suppressed, leading to strong bounds from indirect detection \cite{Balan:2024cmq}.

\part[The Standard Paradigm: Minimal Dark Matter Models at Fixed-Target Experiments]
{The Standard Paradigm: \\[0.4em]
{\LARGE\bfseries Minimal Dark Matter Models at Fixed-Target Experiments}}
\label{part:StandardParadigm}

 \chapter{Fixed-Target Experiments}
\label{chap:AcceleratorExperiments}

\section{Accelerator-Based Dark Matter Searches}

\begin{figure}[h]
    \centering
\includegraphics[width=0.8\linewidth]{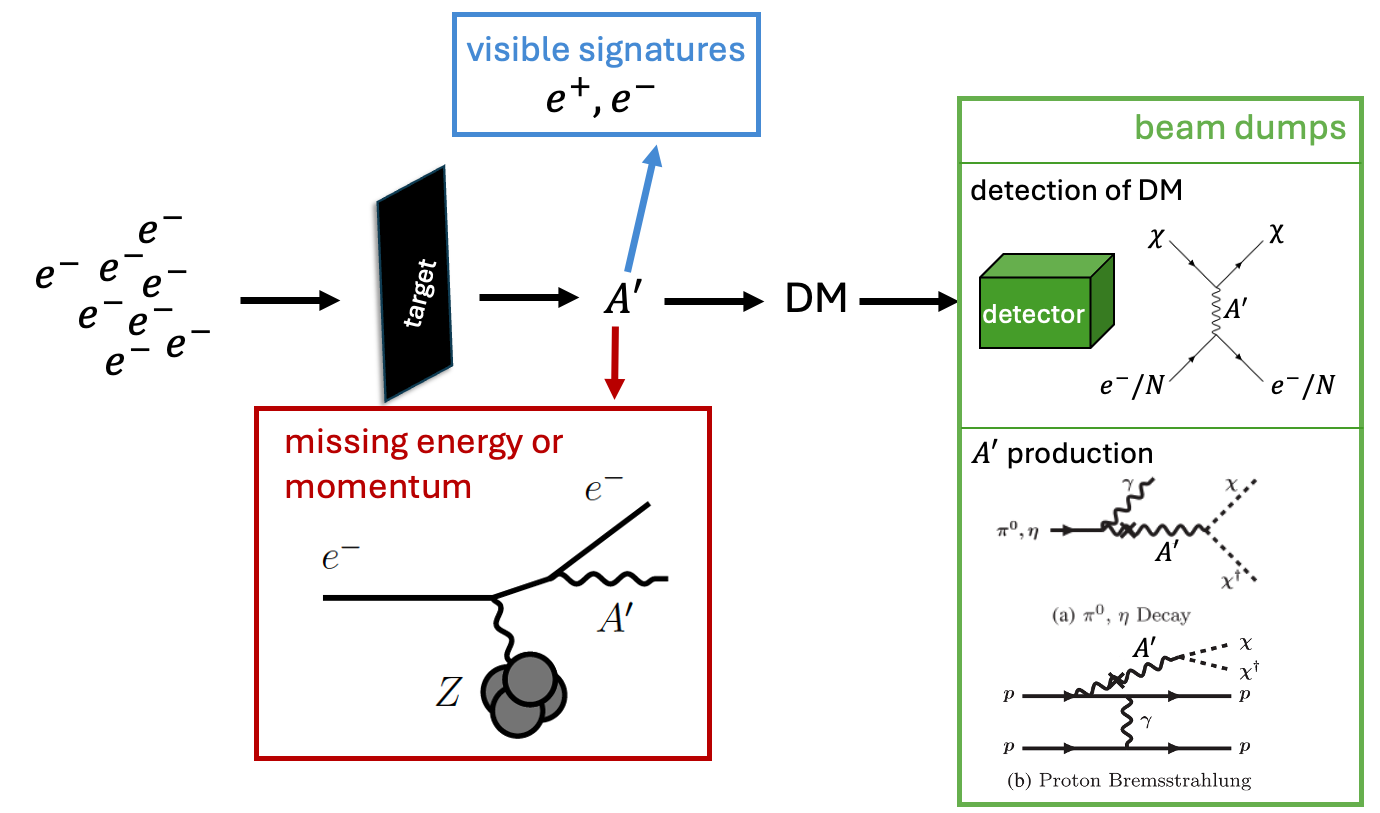}
    \caption{Illustration of a fixed target experiment, where an electron (or proton) beam impinges on a fixed target, producing a shower of particles, possibly including dark photons. There are three main types of searches with this set up, missing energy, visible signatures, and proton beam dumps.}
    \label{fig:fixed_target}
\end{figure}

DM with mass below a GeV deposits insufficient energy in nuclear recoils to exceed current detector thresholds at direct detection experiments, making accelerator-based searches an 
essential complementary probe of the sub-GeV mass range
\cite{Berlin:2020uwy,Bjorken:2009mm,deNiverville:2016rqh,Batell:2009di,Battaglieri:2017aum}. These experiments test for DM signals produced from high energy collisions of SM particles.  In fixed-target experiments, a relativistic beam of protons or electrons is incident on a fixed target, a dense material at rest in the lab frame, causing interactions with target nuclei as illustrated in \figref{fig:fixed_target} \cite{Batell:2022dpx,Gori:2022vri}. Through these interactions, a relativistic flux of dark sector particles (the DM itself or the mediator) can be produced and later detected through their missing energy/momentum, directly in a downstream detector, or through their visible decays into SM final states.

\begin{figure}[h]
    \centering
    \includegraphics[width=0.49\linewidth]{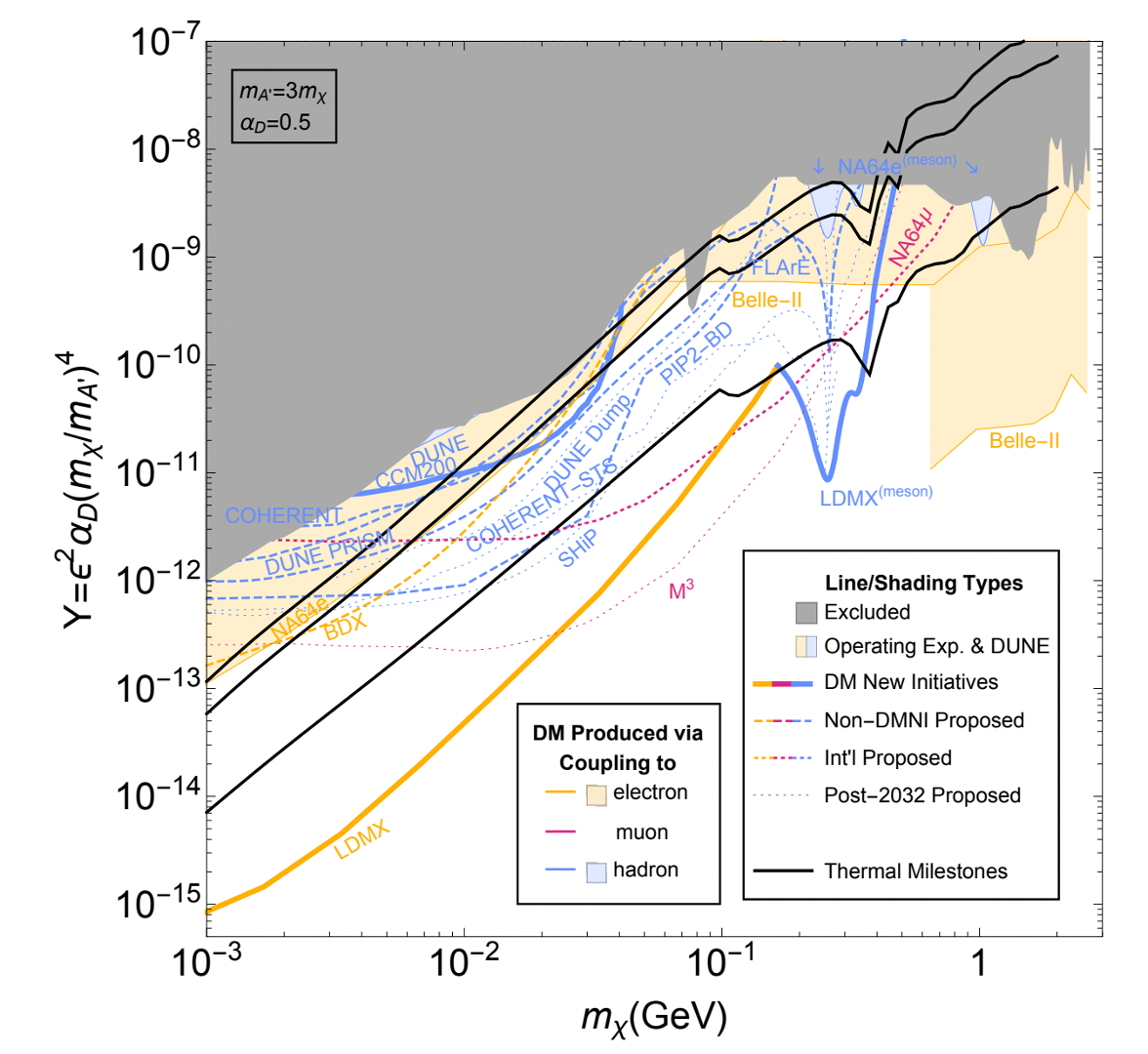}
    \includegraphics[width=0.49\linewidth]{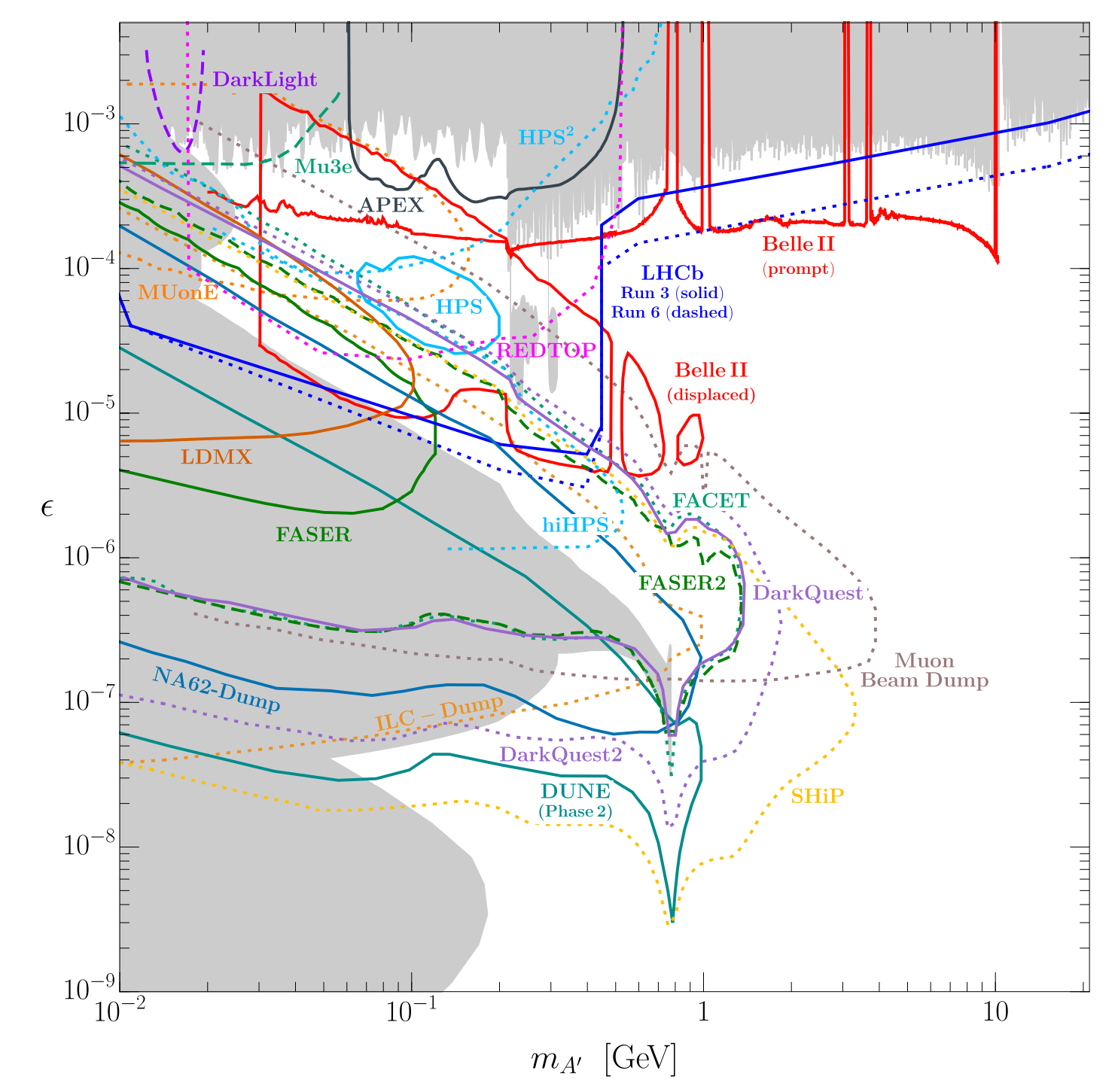}
    \caption{Current landscape of current and future accelerator experiment constraints on sub-GeV DM, in particular the elastic scalar DM with a dark photon mediator \cite{Krnjaic:2022ozp} for an invisibly decaying dark  (left) and a visibly decay dark photon (right) \cite{Batell:2022dpx}. Sensitivity estimates on the parameter $y \equiv \epsilon^2 \alpha_D \left( \frac{m_{\chi}}{m_{A'}} \right)^4$, where $\alpha_D = 0.5$ and $3 m_{\text{DM}} = m_\Ap$, are plotted as a function of the DM mass along with the relic target in black, on the left figure. The right panel plots existing constraints in grey and future experimental reach in colour on the \gls{KM} strength $\epsilon$ -- dark photon mass plane, with no relic target shown since this projection does not fix the dark sector parameters ($m_\chi$, $\alpha_D$) that the thermal target depends on.}
    \label{fig:accelerator_landscape}
\end{figure}

The current and future experimental constraints on the sub-GeV DM parameter space are summarized in \figref{fig:accelerator_landscape}. For an invisibly decaying dark photon, in the case of elastic scalar DM, sensitivities are plotted in the left figure. Visibly decaying dark photon limits are presented on the right. 
To put these constraints into perspective, the contour which reproduces the observed relic abundance assuming the freeze-out mechanism\footnote{See \secref{section:FreezeOut} for the calculation of the freeze-out relic abundance.} measured by \textit{Planck} \cite{Planck:2018vyg} is drawn for a benchmark set of parameters. These so-called \textit{thermal targets} serve as valuable benchmarks for experiments, giving theoretical context to the exclusion bounds and projected sensitivity. These thermal target curves conventionally accompany invisible dark photon searches but not visible ones. An invisible search assumes the mediator mass, in this case the dark photon, to be $ > 2 m_\chi$, so that a
resonantly produced dark photon decays promptly and dominantly to
$\chi\bar{\chi}$. The annihilation process, $\chi\bar{\chi}
\rightarrow A' \rightarrow f\bar{f}$, setting the relic
abundance, is controlled by the same combination of couplings as the DM production rate at invisible searches.
It is then possible to collapse the four-dimensional parameter
space $(m_{A'}, m_\chi, \epsilon, \alpha_D)$ onto a single
combination, 
\begin{equation}
    y \equiv \epsilon^2 \alpha_D (m_{\text{DM}}/m_{A'})^4,
\end{equation}
by fixing
the mass ratio $m_{A'}/m_\chi$. Plotting $y$ against $m_\chi$ then makes the thermal target well defined.
A visible search, by contrast, probes the regime $m_{A'} < 2m_\chi$
(or simply does not assume the presence of DM in the model at all), so that $A'$ can only decay to SM fermions. The
natural observable is the two-dimensional plane $(m_{A'}, \epsilon)$, which by construction contains no information about
$m_\chi$ or $\alpha_D$. For this reason, visible search exclusion
plots are conventionally left without a relic target.

Collider experiments provide additional sensitivity to the sub-GeV and broader dark sector parameter space. At $e^+e^-$ colliders, DM can be searched for via mono-photon signatures, in which a monochromatic single photon is produced along with missing energy and momentum, via the process $e^+e^- \to \gamma A' \to \gamma \chi\bar{\chi}$. The production of dark photons would result in a peak in the  missing mass spectrum, $m_X^2 = s - 2E^*_\gamma \sqrt{s}$, where $s$ is the centre of mass energy squared and $E^*_\gamma$ is the energy of the photon. BaBar~\cite{BaBar:2017tiz} 
and Belle~II~\cite{Belle-II:2018jsg,Ferber:2022ewf} have performed such searches. These colliders also set leading constraints on \textit{visible} dark photon decays~\cite{BaBar:2014zli,Belle-II:2018jsg,Ferber:2022ewf}, 
and KLOE~\cite{KLOE-2:2018kqf} provides complementary constraints through radiative 
$\phi \to \eta A'$ decays, where $\phi$ meson decays to an $\eta$ meson and a dark photon. At hadron colliders, mono-jet and mono-photon plus missing transverse energy searches at the \gls{lhc} are sensitive to DM 
production in association with hard QCD radiation, however, it is less competitive than fixed-target or $e^+e^-$ experiments in the sub-GeV mediator 
mass range. The dedicated long-lived particle experiment \gls{faser}~\cite{FASER:2023tle} 
exploits the highly boosted forward flux of light mesons in $pp$ collisions to search for dark photons decaying visibly to $e^+e^-$ pairs, setting new constraints at 
$m_{A'} \sim 10$--$100$\,MeV; an upgrade, FASER2, is planned for the High-Luminosity LHC era with significantly extended reach. Taken together, collider searches are most 
complementary to fixed-target experiments at higher mediator masses for invisible 
searches, where the reach of beam experiments is kinematically limited, while 
visible decay searches at colliders constrain overlapping but distinct regions of the dark photon parameter space to those probed by the fixed-target experiments.

Fixed-target experiments can be broadly divided into two categories according to the nature 
of the beam: \textit{proton beam dump} experiments and \textit{electron beam} experiments. In proton beam 
experiments, the high centre-of-mass energy available in proton--nucleus collisions drives 
DM production primarily through the decays of secondary mesons and via proton bremsstrahlung. 
However, the same hadronic collisions produce an irreducible flux of neutrinos, which 
constitute the dominant background for DM scattering searches. Furthermore, the large 
amount of hadronic activity and secondary particle production makes it impractical and rather messy to 
fully reconstruct the initial state on an event-by-event basis, precluding missing energy 
or momentum searches. Instead, proton beam dump experiments rely on detecting DM 
after it has traversed a shielding region that absorbs SM backgrounds, through 
its scattering signatures in a downstream detector. In electron beam experiments, by contrast, the comparatively clean initial state, a single electron on a nucleus,
allows precise accounting of the visible final-state energy and momentum on an 
event-by-event basis. The absence of hadronic activity also means the neutrino background 
is essentially absent. Together, these features enable missing energy (where the recoil electron energy is measured and the DM escapes unseen) or missing momentum 
(where the full recoil electron 4-momentum is reconstructed) as the primary search strategy, 
providing sensitivity to DM production without requiring a downstream scattering signal. 
The two approaches are therefore complementary: proton experiments benefit from higher 
production rates, while electron experiments achieve 
superior signal-to-background and are particularly powerful at lower mediator masses. 

The experiments considered in this thesis are presented below, starting with proton beam experiments and followed by electron beams.

\section{Proton Beams}
Proton beam dump experiments offer a complementary approach to electron fixed-target experiments for light DM searches \cite{Buonocore:2019esg,Batell:2009yf,Beacham:2019nyx,deNiverville:2016rqh}. In these experiments, a high-intensity proton beam impinges on a thick target, producing DM predominantly through the decays of secondary mesons and via proton bremsstrahlung. The resulting DM flux is then detected through scattering signatures in a downstream near detector. Compared 
to electron beam experiments, proton 
collisions grant access to a richer set of production channels, at the cost of a larger and less controllable neutrino background. Interestingly, these experiments serve a dual purpose: probing neutrino physics in addition to dark sectors \cite{deNiverville:2011it}.
Here I provide a brief non-exhaustive list of relevant proton beam experiments.
\smallskip

\noindent
{\bf MiniBooNE}
This beam dump experiment searches for electron and nuclear recoils from DM produced in interactions between an 8.9 GeV proton beam and a steel target \cite{MiniBooNE:2017nqe,MiniBooNEDM:2018cxm}. A DM flux can be produced from meson decay, dark bremsstrahlung, and less dominantly through direct parton-level processes. This DM flux can be detected from nuclear and electron recoils in the downstream detector made of mineral oil. No statistically significant signal excess was observed.
\smallskip

\noindent
{\bf LSND}
 The second beam dump experiment we consider uses an 800 MeV proton beam incident on a water or high atomic number metal target, producing DM mainly through meson decays (the energy is not sufficient for sizeable production through bremsstrahlung) \cite{LSND:2001akn,deNiverville:2011it,LSND:1996jxj}. Downstream, a detector filled with mineral oil is sensitive to DM-electron scattering signals; 
DM-nucleon scattering, by contrast, is not detectable, as the resulting nuclear recoil energy at this beam energy falls below LSND's detection threshold.
\gls{lsnd} reported 242 scattering events which include elastic scattering by neutrinos and potentially DM, with an expected SM background of 229 ± 28.
\smallskip

\noindent
{\bf COHERENT} 
Located in ``Neutrino Alley'' at the 
Spallation Neutron Source at Oak Ridge National Laboratory,  COHERENT~\cite{COHERENT:2017ipa} uses an approximately 1\,GeV proton beam which impinges on a mercury target producing an intense flux of neutrinos from pion and muon decays. COHERENT operates multiple detector subsystems, including a CsI scintillating crystal array and a liquid argon detector, and has made the first two measurements of coherent elastic neutrino--nucleus 
scattering (CE$\nu$NS) on CsI~\cite{COHERENT:2017ipa} and argon~\cite{COHERENT:2020iec}. 
A flux of light DM via $\pi^0$ decays and dark bremsstrahlung is produced, which COHERENT can search for through nuclear recoil signatures in its detectors~\cite{COHERENT:2021pvd}.
COHERENT's sensitivity to DM is complementary to higher-energy proton beam experiments such as DUNE.
\smallskip

\noindent
{\bf SHiP}
\gls{ship} is a beam dump experiment approved in 2024 for operation at the CERN \gls{sps} Beam Dump Facility, collecting $4\times10^{19}$ protons on target per year at 400\,GeV for a projected total of $6\times10^{20}$ protons over its lifetime \cite{SHiP:2020noy,SHiP:2025ows}. The experiment consists of two complementary detector systems: the \gls{snd}, which searches for light DM through recoil signatures off electrons and nucleons; and the Decay Spectrometer, which targets visible decays of feebly interacting long-lived particles. For invisible DM searches, light DM particles are produced in the tungsten target via several mechanisms: dominantly meson decays ($\pi^0,\,\eta,\,\omega$) and proton bremsstrahlung and subsequently scatter in the SND. In \secref{section:IVM} we explore the possibility of DM produced at SHiP from the decay of secondary vector mesons from exclusive photoproduction \cite{Schuster:2021mlr}.

\smallskip

\noindent
{\bf DUNE}
The Deep Underground Neutrino Experiment (DUNE) is a long-baseline neutrino oscillation experiment at Fermilab using a 120\,GeV proton beam impinging on a graphite target, with a projected total exposure of $1.1\times10^{21}$ protons on target per year for seven years \cite{DUNE:2024wvj,DUNE:2021tad}. For light DM searches, the relevant detector is the Near Detector complex, a liquid argon time projection chamber. Light DM particles are produced predominantly via $\pi^0$ and $\eta$ meson decays in the target and proton bremsstrahlung. DM is then detected via elastic scattering off electrons or nucleons in the LAr target \cite{Brdar:2022vum,DeRomeri:2019kic}
\begin{figure}[h!]
    \centering
    \includegraphics[width=0.7\linewidth]{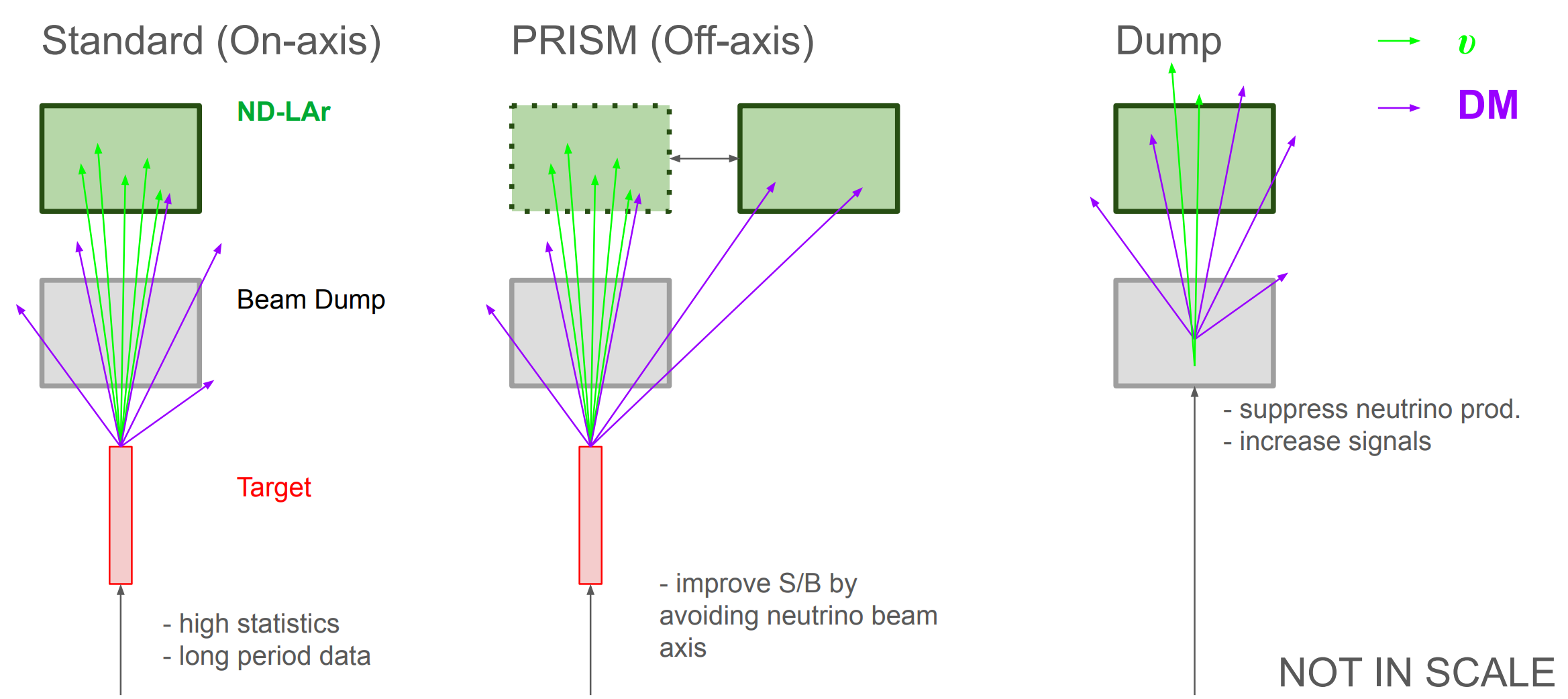}
    \caption{Three different operation modes of DUNE: standard on-axis (left), PRISM off-axis (middle), and beam dump (right) \cite{jang2025light}. The presence of the beam dump is present in all modes, however the beam dump mode does not have the target in front.}
    \label{fig:DUNE}
\end{figure}
DUNE can operate in three complementary 
modes. The standard on-axis configuration provides high statistics, however more background and systematic uncertainties. The DUNE-PRISM off-axis configuration \cite{DeRomeri:2019kic}, on the other hand, in which the detector is displaced transversely up to 30\,m from the beam axis \cite{Dawson:2025bpy}, exploits the different angular spectra of the neutrino background and the 
light DM signal to improve the signal-to-background ratio. Finally, a dedicated beam dump mode in which the hadronic target is removed, with a substantially reduced neutrino 
flux and a lower background for a light DM search.
The background for the on and off-axis modes are taken from \cite{Betancur:2021ect} and the beam dump mode is assumed to have a factor of $10^{-3}$ reduction relative to the standard on-axis mode \cite{Brdar:2022vum}.
DUNE is also sensitive to visibly decaying dark photons, as
studied in \cite{Berryman:2019dme}.

\smallskip

\noindent
{\bf NO$\bm{\nu}$A}
The long-baseline neutrino oscillation experiment, \gls{nova} ~\cite{NOvA:2018gge,NOvA:2025ykl}, at Fermilab, uses the same 120\,GeV proton beam on a graphite target as DUNE delivered via the NuMI beamline. The NO$\nu$A near detector is a liquid scintillator tracking calorimeter located downstream of the NuMI target and 100\,m underground at Fermilab, positioned off-axis. Using a total exposure of $2.55\times10^{21}$\,POT, a search for light DM has been performed using this detector~\cite{NOvA:2025ykl}, with light DM produced dominantly via $\pi^0$ and $\eta$ meson decays and proton bremsstrahlung and detected through DM-electron elastic scattering.
Background from neutrino-electron scattering is 
further reduced by kinematic cuts which we adopt from \cite{Buonocore:2019esg}. 

\smallskip

T2K \cite{Araki:2023xgb,deNiverville:2016rqh}, MicroBooNE \cite{MicroBooNE:2023gmv}, and 
SBND \cite{SBND:2025lha,SBND} are also worth briefly mentioning: all three search for 
invisible light DM signatures via DM-electron and DM-nucleon scattering in their respective 
near detectors, analogous to the searches discussed above.

\begin{figure}
    \centering
    \includegraphics[width=0.6\linewidth]{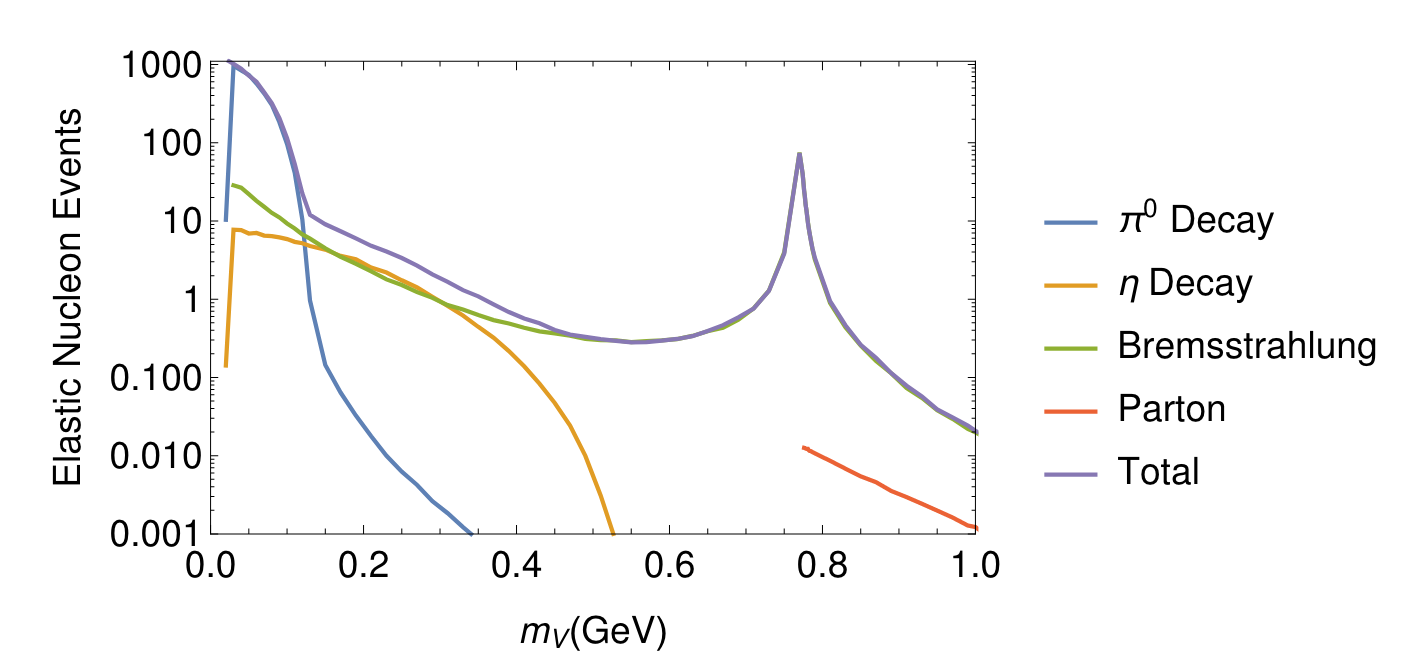}
    \caption{Dark photon production channels at proton beam dump experiments \cite{deNiverville:2016rqh}. Number of elastic nucleon scattering events as a function of dark photon mass.}
    \label{fig:fixedtarget_channels}
\end{figure}

Proton beam dump experiments provide a powerful setting to search for DM produced in proton-nucleus collisions. Dark photons can be produced in several ways; the main production channels are meson decay, dark proton bremsstrahlung, and direct parton level processes. Figure~\ref{fig:fixedtarget_channels} summarizes the rate of each process as a function of the dark photon mass. 
Dark bremsstrahlung dominates in the higher mass region $m_\Ap \gtrsim 500$ MeV, while $\pi^0$ decay dominates at lower masses. Direct parton level processes also give rise to dark photon production but are only relevant at masses above $\sim$ GeV.
In \secref{subsection:MesonDecay} the meson decay channel is described, and dark bremsstrahlung in \secref{subsection:DarkBrem}.

\begin{figure}[h]
    \centering
\includegraphics[width=0.7\linewidth]{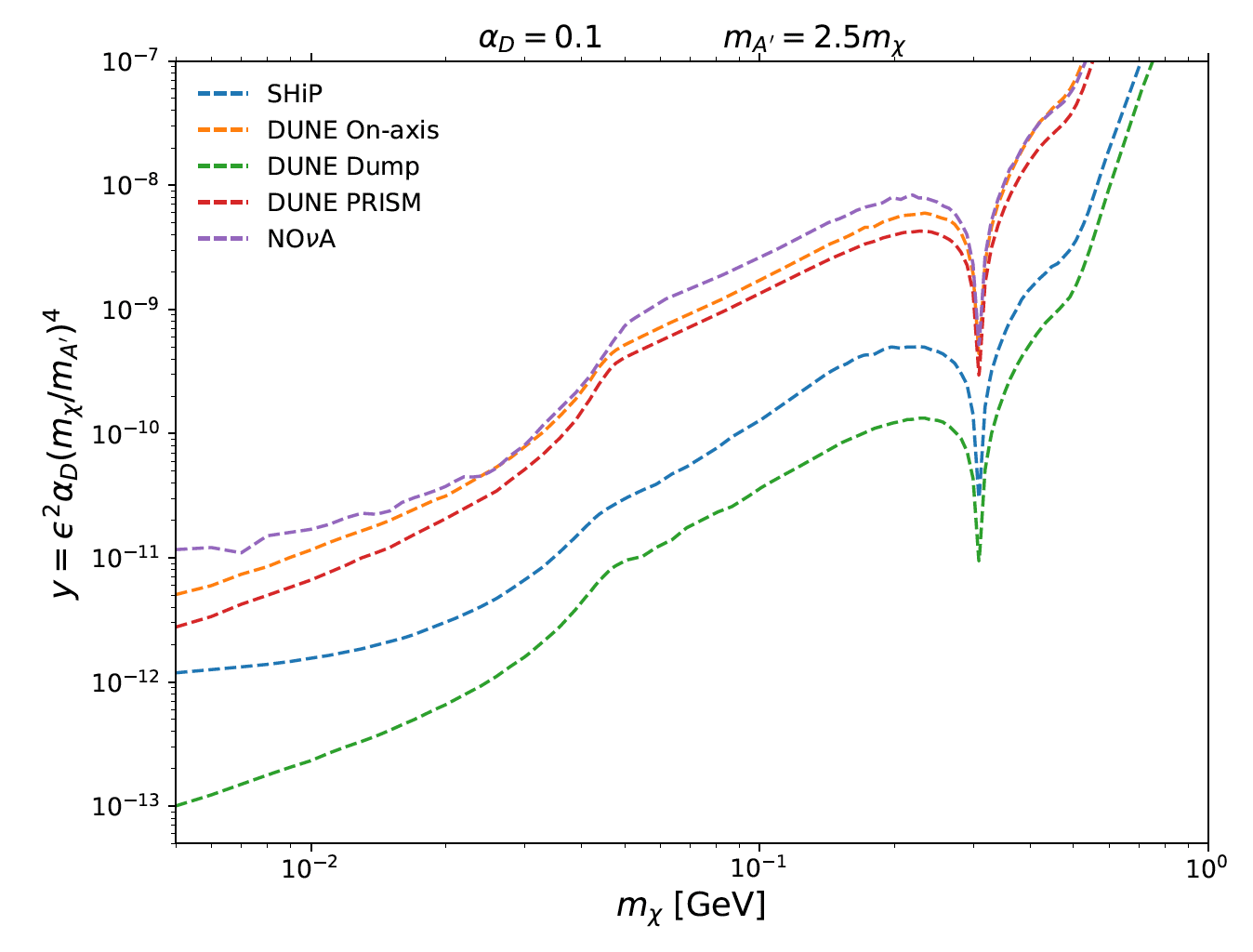}
    \caption{The projected sensitivity of next-generation high energy proton beam dump experiments SHiP \cite{SHiP:2025ows} (blue) and DUNE \cite{DUNE:2021tad} (orange for on-axis mode, green for beam dump mode, and red for PRISM mode with up to 24m off-axis), and of the current experiment NO$\nu$A (purple) (we follow \cite{deNiverville:2018dbu} and report on $2.97 \times 10^{20}$ POT), reported as 90\% \gls{cl} bounds \cite{Ledell2026}. The DM detection channel included is electron scattering, and the DM production channels considered are proton bremsstrahlung and meson decay. Sensitivities are reported for a benchmark set of parameters, $m_{A'}=2.5 m_{DM}$ and $\alpha_D=0.1$.}
    \label{fig:proton_beam_sensitivity}
\end{figure}

The projected sensitivities, reported as 90\% \gls{cl} bounds, for complex scalar DM mediated by an invisibly decaying dark photon, for the next-generation high energy proton beam dump experiments SHiP and DUNE and current experiment \gls{nova} are compared in \figref{fig:proton_beam_sensitivity}\footnote{See \cite{DeRomeri:2019kic} for a dedicated likelihood analysis of DUNE's off-axis search for DM. There, the flux systematic uncertainty is modelled as a nuisance parameter correlated across all off-axis positions and marginalized jointly with the data, allowing partial self-calibration of this uncertainty. This differs from the independent channel combination we employ, and largely explains why they find a stronger PRISM enhancement than reported here; treating the correlation explicitly would be expected to increase our combined sensitivity further.}
The results are reported in the $(m_\chi, y)$ plane. The curves are evaluated for a fixed coupling $\alpha_D = 0.1$ and mass ratio $m_{A'} = 2.5\,m_\chi$.
The mass ratio $m_{A'}/m_\chi = 2.5$ has been chosen since it is not yet in tension with the null result of current direct detection experiments \cite{Krnjaic:2025noj}.
All five curves share a common qualitative shape: $y$ grows steadily with $m_\chi$ up to the $\rho$ resonance, reflecting the suppression of the dark photon production cross section as the mediator mass increases. Due to the mixing of the dark photon with the $\rho$ and $\omega$ vector mesons, the production cross section is resonantly enhanced when the dark photon mass coincides with the $\rho/\omega$-meson mass, producing the sharp dip seen in all five sensitivity curves.

For each DUNE mode, we adopt a benchmark exposure of 7 years for illustrative purposes, allowing for a comparison of the modes' physics reach rather than a projection tied to any specific approved run plan. The on-axis and PRISM configurations correspond to operation in DUNE's standard neutrino-beam mode and are expected to accumulate exposure concurrently with the long-baseline oscillation program. The DUNE Dump configuration, by contrast, requires diverting the proton beam away from the neutrino production target and directly onto the beam dump, a configuration first proposed in Ref.~\cite{Brdar:2022vum} as a means of suppressing the neutrino-induced background that otherwise limits new-physics searches at near detectors. Because this mode necessarily competes with DUNE's primary neutrino oscillation program for beam time, it has not been adopted as part of the collaboration's approved running plan, and any future dump-mode exposure would likely be limited to a much shorter dedicated run rather than the full multi-year exposure assumed here. We nonetheless include it in Fig.~\ref{fig:proton_beam_sensitivity} to illustrate the best-case sensitivity such a configuration could offer, while emphasizing that its physical realization remains uncertain and would need to be weighed against its cost to the neutrino program.

We note that only DM elastic scattering with electrons have been included since its background is most well-studied. Including elastic and inelastic scattering with nuclei and deep inelastic scattering would further extend the sensitivity of these experiments. We leave a dedicated study of the expected background of these channels to future work.

\subsection{Simulation Tools for Proton Beam Dumps}

In this thesis, we employ a modified version of \BdNMC\footnote{Note that within the proton bremsstrahlung channel of \BdNMC, the splitting function contains a 
spurious factor of 4 in the denominator and substitutes $z$ for $1$ in 
one of the cross terms. Secondly, the dark photon yield integration is evaluated over 
$[p_{T,\mathrm{min}}, p_{T,\mathrm{max}}]$ rather than 
$[p_{T,\mathrm{min}}^2, p_{T,\mathrm{max}}^2]$ \cite{Svensson2025}. Once corrected, our results differ from 
those of the published SHiP sensitivity study~\cite{SHiP:2020noy}.}\cite{deNiverville:2016rqh}, a Monte Carlo simulation software for modelling production, propagation, and scattering in beam dump experiments to compute the sensitivity of proton beam experiments. We have adapted \BdNMC to incorporate additional DM models.
\BdNMC first simulates the production of dark photons from bremsstrahlung and meson decays, considering both on and off-shell contributions. Next it models the scattering of the produced DM inside the detector with nucleons and electrons, taking into account the detector material and geometry. In this thesis, additional DM models, such as Dirac fermion DM and various vector DM models described in \chapref{chap:Models}, were added to the software since only complex scalar DM is implemented natively. This involves computing and adding the relativistic DM-electron and DM-nucleon cross section (as described in Paper I and II) and branching ratios for the DM models of interest to the software. 
The kinematic distribution of pseudoscalar mesons produced at proton beam dumps are modelled by the Sanford-Wang distribution from \cite{MiniBooNE:2008hfu} for lower beam energy experiments, and by the so-called BMPT distribution \cite{Bonesini:2001iz} for higher energies relevant for experiments like SHiP. The BMPT distribution is a parametrization of data from proton beam experiments with beryllium targets, scaled by the mass number of the target nucleus to approximate other types of targets. Proton bremsstrahlung is implemented based on \cite{Blumlein:2013cua,Gorbunov:2014wqa}, described in \secref{subsection:DarkBrem}.

We also make use of another software on the market, \maddump \cite{Buonocore:2018xjk}.
\maddump generates DM production processes directly within the~\textsc{MadGraph5\_aMC@NLO} \cite{Alwall:2011uj} (which from now on we denote as \madgraph) framework, allowing arbitrary dark sector models defined via \feynrules-generated UFO files to be simulated without requiring dedicated implementation of new production cross sections, as is necessary in \BdNMC. This makes \maddump very flexible for implementation of different dark sector models. It however, requires meson distributions as user input which can be generated by Monte Carlo event generators such as \pythia \cite{Bierlich:2022pfr}. Proton bremsstrahlung is implemented similarly within \maddump.

We compare both simulation tools \BdNMC and \maddump for the simulation of DM signatures at SHiP \cite{Svensson2025}.
\begin{figure}[h]
    \centering
    \begin{minipage}[b]{0.49\textwidth}
        \centering
        \begin{subfigure}{\textwidth}
            \centering
            \includegraphics[width=\textwidth]{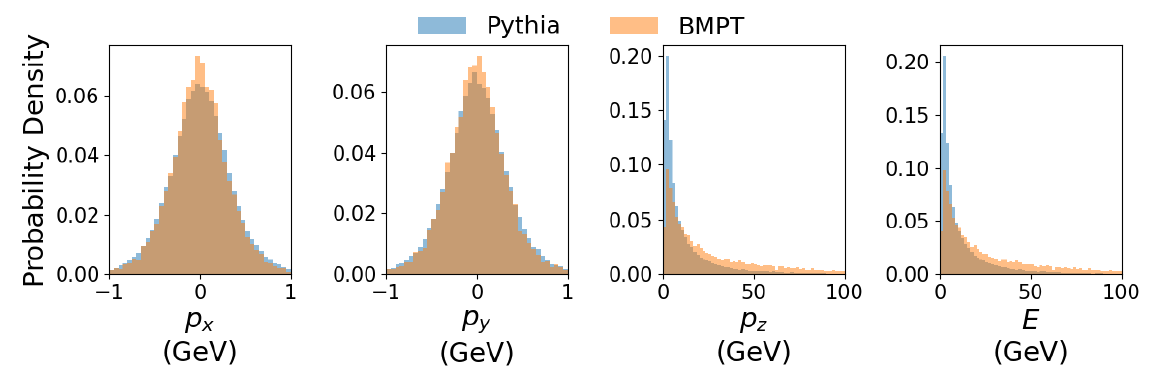}
            \caption{Kinematic distributions of pions produced at SHiP from \pythia simulations (blue) and BMPT (orange). }
            \label{fig:top_left}
        \end{subfigure}
        
        \vspace{0.5cm}
        
        \begin{subfigure}{\textwidth}
            \centering
            \includegraphics[width=\textwidth]{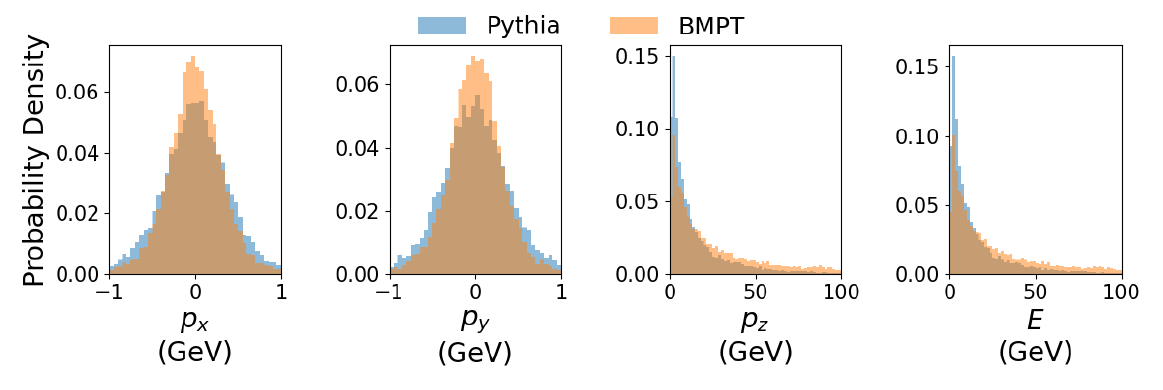}
            \caption{Kinematic distributions of $\eta$ mesons produced at SHiP from \pythia simulations (blue) and BMPT (orange).}
            \label{fig:bottom_left}
        \end{subfigure}
    \end{minipage}
    \hfill
    \begin{minipage}[b]{0.49\textwidth}
        \centering
        \begin{subfigure}{\textwidth}
            \centering
        \includegraphics[width=\textwidth, height=6.2cm]{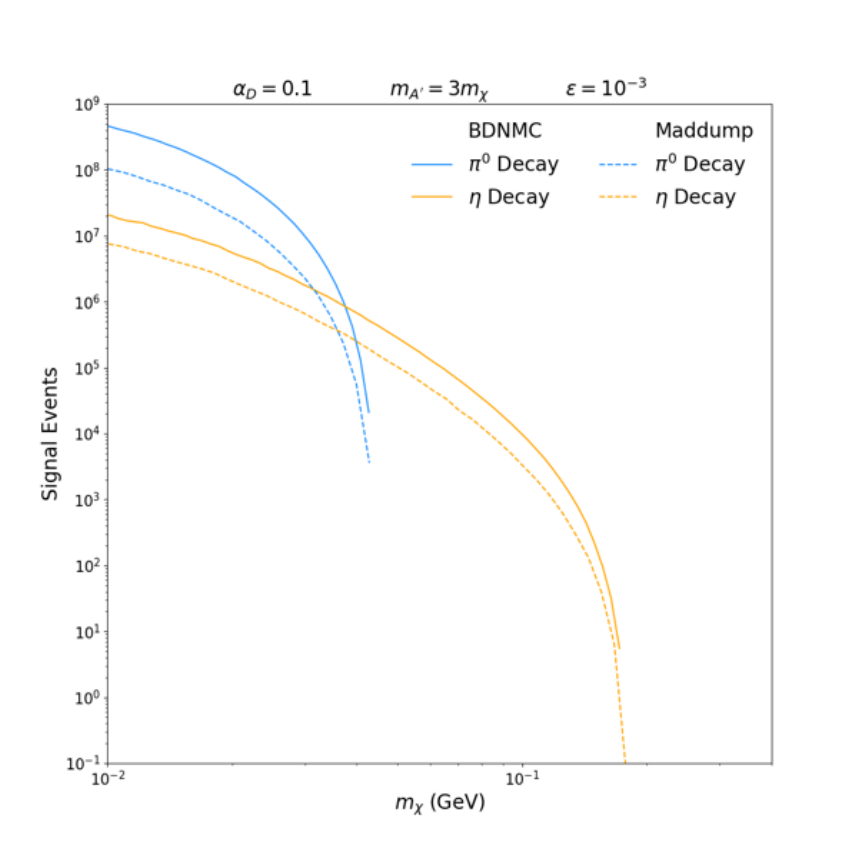}
            \caption{The number of predicted signal events at SHiP from $\pi^0$ decay (blue) and $\eta$ decay (orange) using \BdNMC (solid) and \pythia followed by \maddump (dashed).}
            \label{fig:right_fig}
        \end{subfigure}
    \end{minipage}
    \caption{Comparison of the modelling of pseudoscalar meson production at SHiP between BMPT \cite{Bonesini:2001iz} and \pythia \cite{Bierlich:2022pfr}, and their propagation into \BdNMC \cite{deNiverville:2016rqh} and \maddump \cite{Buonocore:2018xjk}, respectively \cite{Svensson2025}.}
    \label{fig:compare_mesons}
\end{figure}
The results for the comparison of the meson decay channel are summarized in \figref{fig:compare_mesons}. The transverse momentum distributions $p_x$ and $p_y$ agree well between the two approaches for $\pi^0$ mesons, however with the \pythia-generated spectrum slightly broader; this difference becomes more visible for $\eta$ mesons. The energy and $p_z$ (along the beam direction) distributions show more pronounced discrepancies: for both meson species, Pythia consistently produces softer spectra than the BMPT parametrization. These kinematic differences propagate into the predicted signal yields, shown on the right side of \figref{fig:compare_mesons}. \BdNMC systematically predicts a higher number of signal events across both decay channels.

\section{Electron Beams}
In the following, we briefly review a selection of electron beam experiments relevant to sub-GeV DM searches; this list is not intended to be complete. We leave details of the missing momentum experiment LDMX, a central topic of this thesis, to its dedicated \chapref{chap:LDMX}.
\smallskip

\noindent
{\bf NA64}
The fixed-target missing energy experiment NA64~\cite{NA64:2023wbi} is at the CERN \gls{sps} using a 100\,GeV electron beam incident on an active target. Interactions between the beam electrons and target nuclei can produce a dark photon via dark bremsstrahlung, which subsequently decays either 
invisibly to DM or visibly to SM final states, depending on the dark photon mass and the 
DM model parameters. For invisible decays, NA64 operates as a missing energy experiment: the recoil electron carries away less energy than the beam, with the deficit measured in the 
downstream calorimeters and attributed to the undetected dark photon and its decay 
products. For visible decays, NA64 instead searches for the dark photon decaying promptly 
to $e^+e^-$ or $\mu^+\mu^-$ pairs reconstructed in the calorimeters and tracking system \cite{NA64:2019auh}.
\smallskip

\noindent
{\bf E137}
An electron beam-dump experiment was conducted at SLAC in 1980--1982, in which a 
20\,GeV electron beam was dumped onto a set of aluminium plates~\cite{Batell:2014mga}. 
Dark photons produced via dark 
bremsstrahlung in the dump could decay invisibly to DM, which would then travel through 
shielding and a decay region before reaching an electromagnetic shower 
calorimeter, where DM-electron elastic scattering produces a shower. No candidate events were observed in either of two experimental runs. While the original experiment is no longer operating, constraints were 
derived from a reanalysis of archival data~\cite{Batell:2014mga}.

\smallskip

\noindent
{\bf Lohengrin}
Proposed as a missing momentum experiment at the Electron Stretcher Accelerator in Bonn, Lohengrin~\cite{Bechtle:2024atq} plans to use a high-intensity electron beam
incident on a thin target, producing dark photons via dark bremsstrahlung. Unlike NA64's missing 
energy approach, Lohengrin aims to reconstruct the full momentum of the recoiling electron using a tracking spectrometer, providing additional background rejection.
\smallskip

Also worth mentioning is BDX (Beam Dump eXperiment)\cite{Achenbach:2025ynn}, a proposed electron beam dump experiment at Jefferson Lab, which would search 
for invisible light DM via electron and nuclear recoils downstream, providing a modern, higher-intensity complement to the historical E137 search.

Electron beam experiments are not capable of producing large fluxes of mesons like for proton beams, however they allow for precise determination of the kinematics of the initial and final state beam electrons thus utilizing missing energy or momentum signatures of DM. Light DM is mainly produced through dark bremsstrahlung, as described below in \secref{subsection:DarkBrem}. At higher dark photon masses, secondary vector mesons decaying invisibly extend the sensitivity of these experiments notably \cite{Schuster:2021mlr}.

\section{Invisible Searches}

\subsection{Meson Decay}
\label{subsection:MesonDecay}
At proton beam fixed-target experiments, a large flux of pions and $\eta$ particles are produced \cite{Batell:2009di,deNiverville:2011it}.
The following chain of interactions occurs: $pp \to X \pi^0; \pi^0 \to \gamma A’$, where the pseudoscalar meson can also be an $\eta$ meson, $X$ denotes an unspecified/unmeasured set of particles and $A'$ can either be on-shell of off-shell.
\begin{figure}
    \centering
    \includegraphics[width=0.4\linewidth]{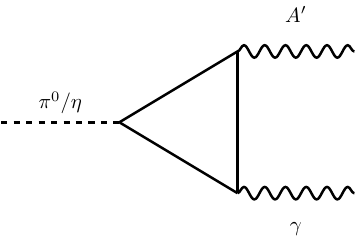}
    \caption{Feynman diagram of pion decay to a dark photon and ordinary photon. The diagram is analogous for $\eta $ and other scalar meson decays.}
    \label{fig:feynman_pion_decay}
\end{figure}
The  flux of $\pi^0$ (and $\eta$ if the beam energy is high enough) mesons produced from proton beam target interactions is determined from simulations, and will have a certain distribution in energy \cite{deNiverville:2016rqh}. These mesons could then decay to dark photons, which then decay to DM. In the case where $2m_{\text{DM}} < m_\Ap < m_{\pi^0}$, the dark photon decays on shell\footnote{We will restrict ourselves to the on-shell case in this thesis, but in the off-shell case the branching ratio involves a phase space integral that must be solved numerically, since the narrow width approximation no longer holds \cite{Kahn:2014sra}.} with the branching ratio \cite{deNiverville:2011it},
\begin{equation}
    BR(\pi^0 \to \gamma \chi \bar{\chi}) = Br(\pi^0 \to \gamma \gamma) 2 \epsilon^2 \left( 1 - \frac{m_\Ap^2}{m_{\pi^0}^2} \right)^3 BR(\Ap \to \chi \bar{\chi}),
\end{equation}
where DM and its antiparticle are denoted by $\chi$ and $\bar{\chi}$.
The total number of dark photons produced is then given by \cite{deNiverville:2016rqh},
\begin{equation}
    N_\Ap =  BR(\pi^0 \to \gamma \chi \bar{\chi}) \times \text{POT} \times N_{\pi^0 \text{per POT}},
\end{equation}
where POT is the number of protons on target, and $N_{\pi^0 \text{per POT}}$ is the number of pions produced per proton on target, which we approximate here to 1 \cite{deNiverville:2011it}.

\subsection{Dark Bremsstrahlung}
\label{subsection:DarkBrem}
\begin{figure}
    \centering
    \includegraphics[width=0.8\textwidth]{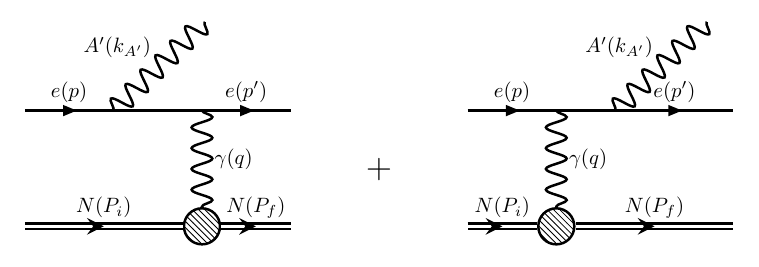}
    \caption{Dark photon bremsstrahlung feynman diagrams, where a dark photon is produced through a high energy electron, $e$, incident on a nuclear target, $N$. The analogous process with an incident proton is also represented by these diagrams.}
    \label{fig:feynman}
\end{figure}

An accelerating (negatively or positively) charge particle produces photons, a phenomenon called \textit{bremsstrahlung} (braking radiation). Bremsstrahlung occurs, for example, when high energy charged particles experience the electric field of an atom/nucleus/nucleon \cite{Bethe:1934za}.
Analogous to ordinary photons produced via bremsstrahlung, dark photons can be produced through dark bremsstrahlung. The dark photon, $\Ap$, bremsstrahlung process that we are interested in is,
\begin{equation}
    e^-(p) + N(P_i) \to e^-(p') + N(P_f) + \Ap(k_{A'}),
\end{equation}
 where $N$ is the nucleus in the target, with either electrons or protons depending on the type of beam. The Feynman diagrams for this process are drawn in fig. \ref{fig:feynman}.

\subsubsection{Electron Bremsstrahlung}
\label{section:electron_brem}

The process considered here, $e^-N \to e^-NA'$, describes a dark photon radiated from the
incoming or outgoing electron as it scatters elastically (or quasi-elastically) off a nucleus $N$. Since the full $2\to3$
phase-space integral is computationally demanding, the cross section is commonly evaluated
instead using the \gls{WW} approximation \cite{ImprovedWW_Tsai,Validity_Liu,Fermi:1924tc,Williams:1935dka,vonWeizsacker:1934nji}, explained in \appref{Chapter:Appendix}. In this approach,
the nucleus is treated as a source of a flux of quasi-real photons, factorizing the
full process into this photon flux multiplied by the cross section of the $2\to2$ subprocess
$e^-\gamma \to e^- A'$, evaluated at the minimal momentum transfer $t_{\rm min}$. The validity of
this factorization rests on the radiated photon being mostly collinear with the beam and on
neglecting recoil effects on the nucleus.

This is, however, an approximation, and not strictly justified across all of phase space. In
Paper~IV, we compare the WW cross section directly against the full tree-level calculation of
$e^-N \to e^-NA'$ and find that the assumption of a non-recoiling nucleus is not generally a good
one: the two calculations show substantial differences once one moves away from the kinematic
region for which the WW approximation was originally derived. This has also been investigated more thoroughly in \cite{Voronchikhin:2024vfu}. For this reason, rather than relying on the WW approximation, the signal predictions in this thesis are instead obtained from the numerically exact tree-level matrix element, evaluated via Monte Carlo phase space integration in \madgraph \cite{Alwall:2011uj}, without invoking the WW factorization.
The software is used with \gls{ufo} \cite{Degrande:2011ua} files as input, which include the new
vertices associated with the dark photon and DM, along with an approximation of the nucleus and its interaction with an ordinary photon. In this thesis, we consider a dark
photon with vector couplings to electrons/positrons. The nucleus is crudely approximated
as a heavy elementary particle with an electric charge corresponding to $Z e$, and vector coupling to the photon. To approximately account for nuclear structure, we dress the photon-nucleus vertex with the following
form factor \cite{Bjorken:2009mm,ImprovedWW_Tsai}\footnote{The inelastic form factor given in Eq. A19 of \cite{Bjorken:2009mm} has a typo: the expression should not be squared.},
\begin{equation}
    G(t) = \left( \frac{a^2 t}{1+a^2t}\right)^2 \left( \frac{1}{1+t/d} \right)^2 Z^2
    + \left( \frac{a'^2t}{1+a'^2 t} \right)^2
    \left( \frac{1+\frac{t}{4m_p^2}(\mu_p^2 -1)}{\left( 1 + \frac{t}{0.71~\text{GeV}^2} \right)^4}\right) Z,
    \label{eq:Bjorken_formfactor}
\end{equation}
where $t\equiv -q^2$, $q$ is the momentum transfer, $a = 111 Z^{-1/3}/m_e$ where $Z$ is the
atomic number of the nucleus, $d=0.164~\text{GeV}^2 A^{-2/3}$ where $A$ is the atomic mass of the
nucleus, $a' = 773 Z^{-2/3}/m_e$, $m_p$ is the proton mass, and $\mu_p=2.79$. The first term is the
elastic component, while the second is the inelastic component. In \chapref{chap:NuclearPhysics} and
Paper~VIII, we discuss the use of form factors computed from state-of-the-art first principles many-body methods,
going beyond this simple analytic parametrization and thus quantifying the impact of nuclear structure.

The treatment described above — coherent, elastic scattering of the electron off the nucleus as
a whole, dressed with the form factor of \eqnref{eq:Bjorken_formfactor} — corresponds to the
leading-order Bethe-Heitler process. Beyond this, higher-order contributions in $\alpha$ and the \gls{KM} parameter, considering a dark photon that couples not only to the beam electron but also to the nuclear target, have recently been computed for LDMX and
LOHENGRIN-type experiments \cite{Schurmann:2026qmt}. This introduces additional diagram topologies beyond the simple Bethe-Heitler picture: virtual
Compton scattering in which the dark photon is radiated directly from the nucleus rather
than from the electron line, and higher-order loop diagrams.
Since the dark photon couples to quarks and therefore, by extension, to nuclei, this nucleus-side emission channel is in principle present whenever a nuclear target is used, and is not captured by the electron-line-only treatment adopted in this thesis. A natural extension of the
many-body nuclear structure methods discussed in \chapref{chap:NuclearPhysics} and Paper~VIII would
be to compute the nuclear matrix elements entering this virtual Compton scattering dark photon emission directly,
rather than relying on the simple analytic form factor of \eqnref{eq:Bjorken_formfactor}. We leave
this as a direction for future work.

\subsubsection{Proton Bremsstrahlung}
Proton bremsstrahlung, $pN \to pNA'$ , is an important process to consider at proton beam dump experiments, becoming relevant for dark photon masses above $\sim500$ MeV above the mass of the mesons, as evident by \figref{fig:fixedtarget_channels}.

The initial proton and emitted dark photon have four-momentum $p=(E_p,0,0,|\vec{p}|)$ and $k_{A'}=(E_{A'},p_T\cos\phi,p_T\sin\phi,z|\vec{p}|)$, respectively, where $\phi$ is the azimuthal angle of the dark photon momentum about the beam axis, $p_T$ is the momentum of the $A'$ perpendicular to the beam axis, $z$ is the fraction of the beam momentum carried away by the dark photon.
The differential dark photon production rate for proton bremsstrahlung relies on the WW approximation \cite{Fermi:1924tc,Williams:1935dka,vonWeizsacker:1934nji} and can be written as \cite{Blumlein:2013cua,Gorbunov:2014wqa,deNiverville:2016rqh},
\begin{equation}
    \frac{d^2N_{A'}}{dzd^2p_{T}} = \frac{\sigma_{pA}(s')}{\sigma_{pA}(s)}w_{ba}(z,p_T^2)F_{1,N}^2(q^2),
\end{equation}
where $s'=2m_p(E_p-E_{A'})$, $s=2m_p E_p$, $m_p$ is the proton mass, $\sigma_{pA}(s)=f_A \sigma_{pp}(s)$ is the hadronic scattering cross section in which $\sigma_{pp}$ can be taken from data \cite{ParticleDataGroup:2014cgo} (where $F_A$ is assumed to cancel in the ratio), $F_{1,N}^2$ is the time-like form factor incorporating mixing with vector mesons from 
\cite{Faessler:2009tn}, and $w_{ba}(z,p_T^2)$ is the $p \to A' p$ splitting function \cite{Blumlein:2013cua},
\begin{equation}
\begin{aligned}  
    w_{ba}(z, p_T^2) &= \frac{\epsilon^2 \alpha}{2\pi H}
\left[
\frac{1 + (1-z)^2}{z}
- 2z(1-z)\left( \frac{2m_p^2 + m_{A'}^2}{H} - z^2 \frac{2m_p^4}{H^2} \right)
\right.
\\ &
\left.
+\, 2z(1-z)\left(1 + (1-z)^2\right) \frac{m_p^2 m_{A'}^2}{H^2}
+ 2z(1-z)^2 \frac{m_{A'}^4}{H^2}
\right],
\end{aligned}
\end{equation}
where $H = p_T^2 + (1-z) m_{A'}^2 + z^2 m_p^2$.

The formalism outlined above describes a dark photon produced off of the initial state proton. This is in contrast to the electron bremsstrahlung process, where the dark photon
is radiated from a lepton line. Since the electron interacts only electromagnetically and does not fragment, it remains a single well defined particle throughout the
collision regardless of whether the target nucleus is probed coherently, quasi-elastically off an
individual nucleon, or deep-inelastically off a parton; consequently, the initial state radiation and final state radiation diagrams attached to the electron line are both
well defined and can be summed coherently in every such regime. The proton bremsstrahlung case is
qualitatively different because the radiating particle, the proton, is not a fundamental particle. At beam dump energies this scattering is overwhelmingly
inelastic, and the radiating proton typically does not survive the collision as a single
identifiable particle, fragmenting instead into a complex multi-particle hadronic final state. As a
result, the standard formalism for proton bremsstrahlung restricts the calculation to initial state radiation: emission from the incoming beam proton, whose momentum and on-shell
condition are unambiguous prior to the collision. A well defined final state radiation diagram, by
contrast, only exists in the elastic or quasi-elastic limit, where the radiating proton survives intact. It is left as an interesting avenue of future research to understand the potential increase in sensitivity from final state radiation.

\subsection{Secondary Vector Meson Production}
\label{section:IVM}

\begin{figure}[h]
    \centering
\includegraphics[width=0.5\linewidth]{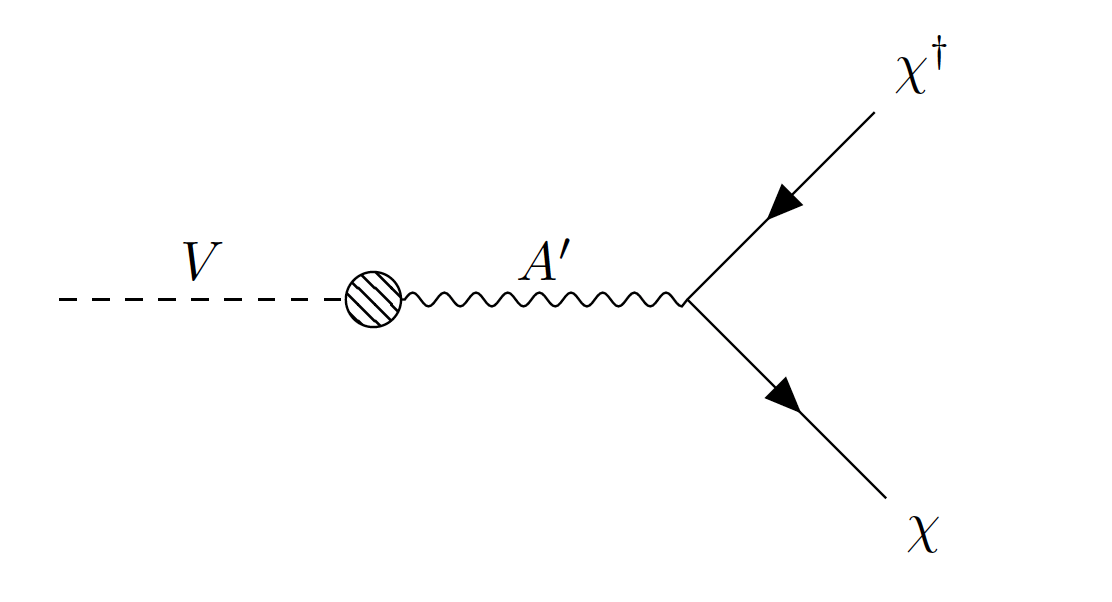}
    \caption{Feynman diagram for the vector meson decay to DM process through its mixing with the dark photon.}
    \label{fig:VMdecay_feynman}
\end{figure}

Beyond the dark photons produced through the channels discussed above, a large flux of ordinary photons are also expected to be produced through beam-target collisions. Photons are produced from bremsstrahlung and in large by the decay of mesons via $\pi \to \gamma \gamma$ (the latter only occurs for proton beams). These photons, if carrying sufficient energy, can re-scatter in the target (or detector) and produce vector mesons through \textit{exclusive photoproduction} \cite{Bauer:1977iq,Tang:2025qqe}. Since vector mesons mix with the dark photon, an additional source of dark photons are therefore generated. The process is depicted in \figref{fig:VMdecay_feynman}, where DM is produced and can be detected directly in a downstream detector or by missing energy/momentum. 

It has been shown in 
\cite{Schuster:2021mlr} that including this channel significantly increases the sensitivity in the $\sim 0.1 - 1$ GeV dark photon mass range for electron beam experiments such as NA64 and LDMX. One would expect that the larger number of photons produced at proton beam experiments from the large pion flux and their subsequent decays, would render this process even more relevant at proton beams and even increase the sensitivity to DM at experiments like SHiP. 

We investigate the secondary invisible vector meson decay process at the upcoming experiment SHiP by computing the expected signal rate of this process and the detection of these DM events. As a benchmark, we take the older ECN4 configuration of SHiP \cite{SHiP:2020noy} which has a thick molybdenum target and a lead detector. We consider specifically the secondary production of $\rho$, $\omega$, and $\phi$ mesons. 

This invisible vector meson production event rate is computed in the following steps: first the photon flux and its corresponding kinematic distributions are generated using \pythia, second the number of vector mesons produced through exclusive photoproduction are computed analytically, and thirdly the vector meson decay rate to DM is calculated.

\begin{figure}
    \centering
\includegraphics[width=\linewidth]{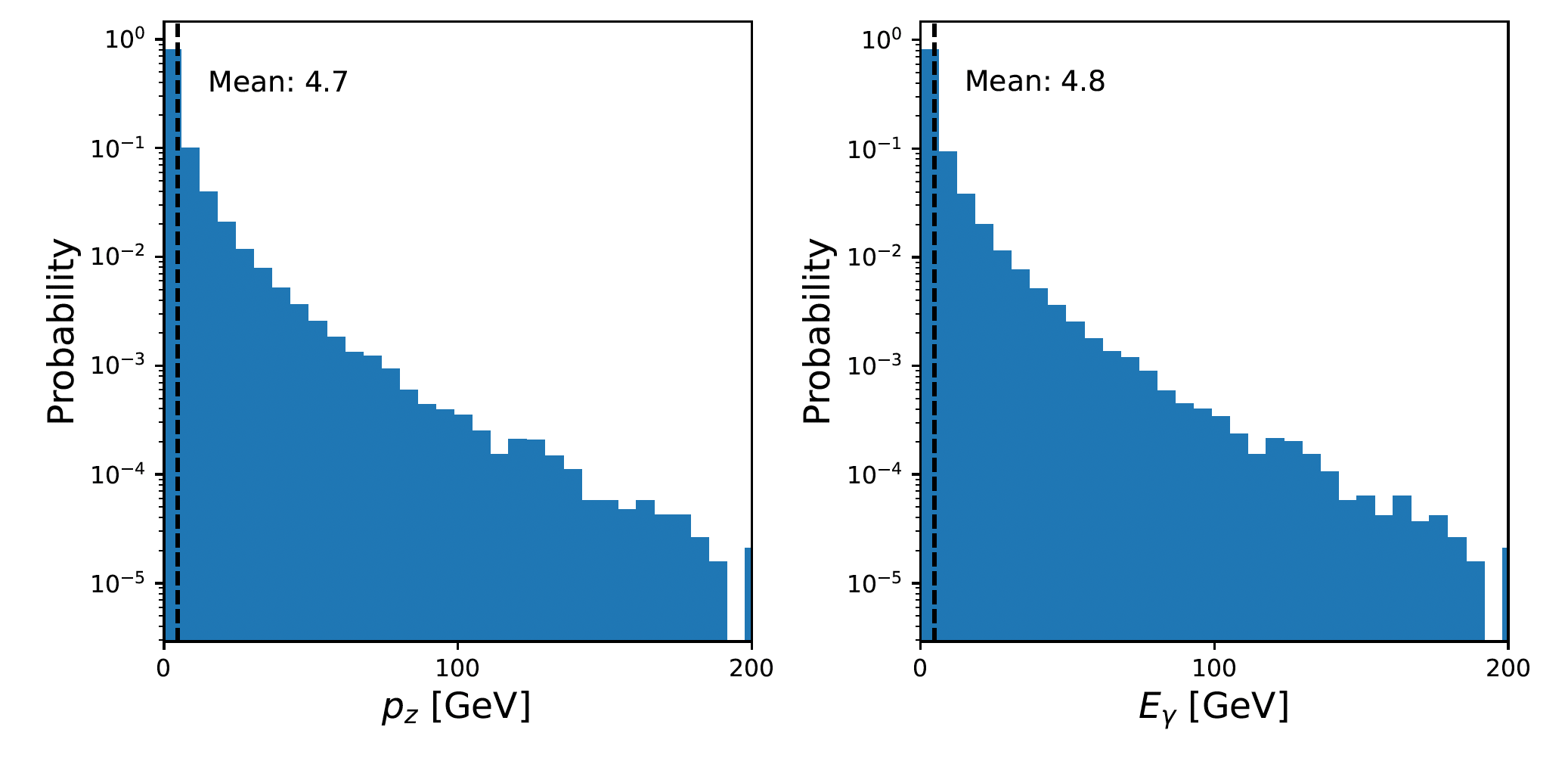}
    \caption{Kinematic distribution of photons produced at SHiP through neutral meson decays ($\pi^0$ and $\eta$), simulated using \pythia \cite{Ledell2026}. The longitudinal momentum component, $p_z$, is plotted on the left, and the energy on the right.}
\label{fig:photon_dists_IVM}
\end{figure}

The dominant process for photon production in proton-nucleus collisions, in this energy regime, is the decay of neutral mesons. The momentum along the beam-axis and the energy of these events are plotted in \figref{fig:photon_dists_IVM}.
The momentum components perpendicular to the beam-axis are not shown here but follow a Gaussian distribution centred at zero.
These proton-nucleus collision events are simulated using \pythia including only \textit{soft inelastic QCD processes}.
Additional sources of photons including direct production involving quarks and gluons, or bremsstrahlung also occur, however are expected to be sub-leading compared to meson decay.

These simulated events are next propagated into the analytic calculation of vector meson production, where these high energy photons scatter with a nucleus and produce a vector meson. We base the formalism of these calculations from \cite{Bauer:1977iq}.
Exclusive photoproduction of a vector meson $V$ off a nuclear target proceeds via two
distinct channels. In the \textit{coherent} channel, the nucleus remains in its ground state
and recoils as a whole, while in the \textit{incoherent} channel the meson is produced off
individual nucleons and the nucleus breaks up. Both processes can be described by a
Glauber optical model~\cite{Schuster:2021mlr, Bauer:1977iq}, in which the photon
fluctuates into the vector meson state before the nucleus, which then propagates through
the nuclear medium suffering multiple elastic scatters. The coherent differential cross section for vector meson photoproduction is,
\begin{equation}
\begin{aligned}
    \frac{d\sigma_c}{dt} &= \left.\frac{d\sigma_0}{dt}\right|_{\theta=0}
    \left|
    \int d^2b\, dz\, e^{i(\mathbf{q}_T \cdot \mathbf{b} + q_{\parallel} z)}\, n(b,z) \right.  \\ 
    &\left. \exp\!\left(
        -\frac{\sigma_V}{2}(1 - i\alpha_V)
        \int_z^{\infty} dz'\, n(b,z')
    \right)
    \right|^2,
    \label{eq:coherent_xsec}
\end{aligned}
\end{equation}
where $(d\sigma_0/dt)|_{\theta=0}$ is the forward differential cross section for photoproduction of $V$ off a single nucleon, $n(b,z)$ is the nucleon number density,
$\mathbf{q}_T$ and $q_{\parallel}$ are the transverse, longitudinal momentum transfers (bolded quantities are 3-momenta, unbolded are magnitudes), $b$ is the transverse coordinate vector and $z$ is the coordinate along the beam axis,
$\sigma_V$ is the total $V$-nucleon scattering cross section, and $\alpha_V$ is the ratio
of the real to imaginary parts of the forward $V$-nucleon scattering amplitude. The
exponential factor accounts for the absorption of the meson as it exits the nucleus, while
the phase $e^{i(\mathbf{q}_T\cdot\mathbf{b}+q_\parallel z)}$ encodes the difference in
phases arising from production at different longitudinal positions $z$ within the nucleus.
The incoherent differential cross section receives an additional shadowing correction
accounting for destructive interference between photoproduction at different points in the
nuclear volume,
\begin{align}
    \frac{d\sigma_i}{dt} &= \frac{d\sigma_0}{dt}
    \int d^2b\, dz\, n(b,z)
    \exp\!\left(
        -\sigma_V \int_z^{\infty} dz'\, n(b,z')
    \right) \nonumber \\
    &\quad \times
    \left|
    1 - \int_{-\infty}^{z} dz''\, n(b,z'')\,
    \frac{\sigma_V}{2}(1-i\alpha_V)\,
    e^{iq_{\parallel}(z''-z)} \right.
    \\ & \left.\exp\!\left(
        -\frac{\sigma_V}{2}(1-i\alpha_V)
        \int_{z''}^{z} dz'''\, n(b,z''')
    \right)
    \right|^2.
    \label{eq:incoherent_xsec}
\end{align}
The nuclear density $n(r)$ is modelled by a Woods-Saxon distribution. The single-nucleon
forward cross section $(d\sigma_0/dt)|_{\theta=0}$ is fixed to data, and the full
differential cross section at non-zero momentum transfer is obtained by multiplying by an
exponential form factor $e^{-B|t|}$, with slope parameter $B$ measured experimentally
for each meson species.

Finally, the vector meson mixes with the dark photon and then decays into DM through the diagram depicted in \figref{fig:VMdecay_feynman}.
The partial decay width of a vector meson $V$ into a dark matter pair mediated by a dark photon $A'$ is given by,
\begin{equation}
\Gamma(V \to A'^* \to \chi \bar{\chi}) = \begin{cases} 
\frac{\alpha_D (\epsilon e)^2 f_V^2}{12} \frac{(m_V^2 - 4m_\chi^2)^{3/2}}{(m_{A'}^2 - m_V^2)^2 + m_{A'}^2 \Gamma_{A'}^2} & \text{Complex Scalar DM} \\[10pt] 
\frac{\alpha_D (\epsilon e)^2 f_{V}^2}{3} \frac{(m_V^2 + 2m_\chi^2)\sqrt{m_V^2 - 4m_\chi^2}}{(m_{A'}^2 - m_V^2)^2 + m_{A'}^2 \Gamma_{A'}^2} & \text{Dirac Fermion DM} 
\end{cases},
\end{equation}
where $\alpha_D \equiv g_D^2 / 4\pi$ is the dark fine-structure constant, $e$ is the elementary electromagnetic charge, and $\epsilon$ denotes the \gls{KM} parameter. $m_V$ and $m_\chi$ represent the masses of the vector meson and the dark matter candidate, respectively, while $m_{A'}$ and $\Gamma_{A'}$ are the mass and total decay width of the dark photon mediator. The hadronic physics governing the meson vertex is parameterized by the vector current form factor $f_V$; here, we adopt the numerical values from Table III of Ref.~\cite{Schuster:2021mlr}, which are ultimately derived from the light-cone sum rules framework established in Ref.~\cite{Bharucha:2015bzk}.
A review of both the complex scalar and Dirac fermion dark matter models is presented in \chapref{chap:Models}. This process can be easily extended to other DM models. In Paper VI, we compute this vector meson decay process into vector DM, where the decay rate is written in the appendix (see \chapref{chap:Models} for more on spin-1 DM).

The total number of DM particles produced are,
\begin{equation}
    N_\text{DM} = N_V \text{BR}(V\to\chi\bar{\chi}),
\end{equation}
with the number of vector mesons produced \cite{Schuster:2021mlr},
\begin{equation}
    N_V = N_\gamma p_V,
\end{equation}
from $N_\gamma$ high energy photons and the probability of exclusive photoproduction $p_V$ given by \cite{Schuster:2021mlr},
\begin{equation}
    p_V \approx \int_0^L \exp \left( -\int_0^x \frac{7 \rho(x')}{9X_0(x') dx'} \right) \frac{\rho(x)\sigma_0^V}{m_p} f_V^{\text{nuc}}(x)dx,
\end{equation}
where $\rho(x)$ is the material density, $L$ is the total length, $X_0$ is the radiation length, and $\sigma_0^V$ is the exclusive photoproduction cross section with a nucleon. The cross section of photoproduction on the whole nucleus, given by \eqnref{eq:coherent_xsec} and \eqnref{eq:incoherent_xsec}, is $\sigma_{\gamma N \to V N}=f_\text{nuc}^V A \sigma_0^V$, with $f_\text{nuc}=f_\text{inc}+f_\text{coh}$, the sum of incoherent and coherent components.

\begin{figure}
    \centering
    \includegraphics[width=\linewidth]{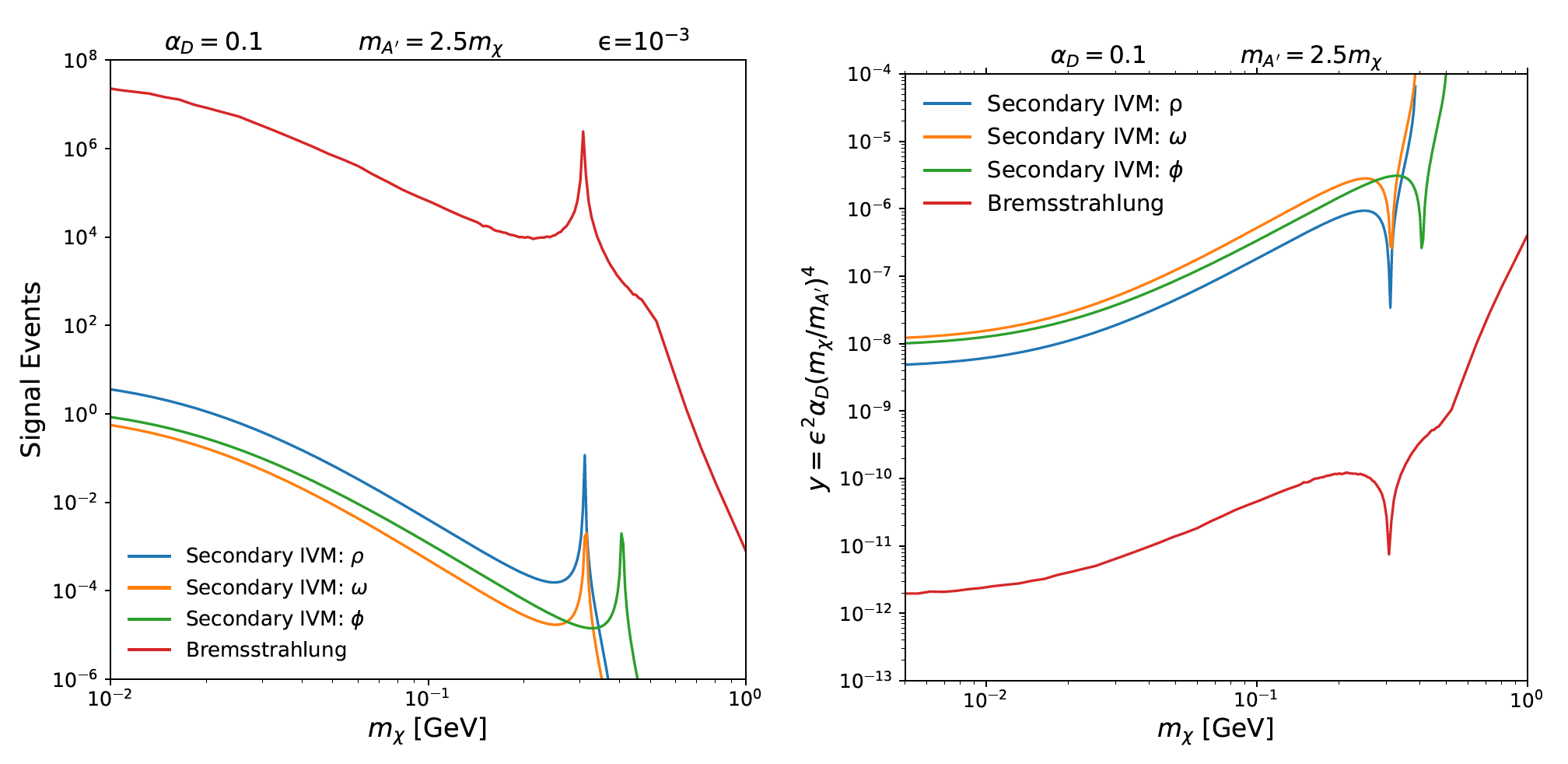}
    \caption{The number of complex scalar DM signal events (left) and the projected sensitivity at 90\% \gls{cl} (right) as a function of DM mass from invisible vector meson decay at SHiP \cite{Ledell2026}. The vector meson channels considered are $\rho$ (blue), $\omega$ (orange), and $\phi$ (green), with dark photon bremsstrahlung (red) for reference. The DM detection channel included is electron scattering.}
    \label{fig:IVMSensitivity}
\end{figure}
Figure~\ref{fig:IVMSensitivity} shows the projected number of DM signal events (left) and the 90\% \gls{cl}\ sensitivity in the parameter $y = \epsilon^2 \alpha_D
(m_\chi/m_{A'})^4$ (right) at SHiP as a function of the DM mass $m_\chi$, for the
invisible vector meson decay channels $\rho^0$, $\omega$, and $\phi$, alongside
the dark photon bremsstrahlung channel for reference (in red). We take complex scalar DM, and the benchmark parameters are fixed
to $\alpha_D = 0.1$ and $m_{A'} = 2.5\, m_\chi$, with $\epsilon = 10^{-3}$. The three meson channels exhibit a broadly similar shape, falling
with increasing $m_\chi$ as the available phase space for the decay $V \to A' \to
\bar{\chi}\chi$ closes, with sharp resonant peaks appearing near the respective meson
masses, $m_\chi \sim m_V / 2.5$, where the dark photon goes on-shell. Beyond these
peaks the signal drops sharply, as the kinematic condition $m_V > 2m_{\chi}$ can no longer be satisfied. The $\rho^0$ channel (blue) yields the largest event
rate among the meson channels, owing to its relatively large photoproduction cross
section and branching fraction, while the $\phi$ (green) and $\omega$ (orange) channels are weaker.  Bremsstrahlung (red) dominates over the meson channels across most of the
mass range. It is perhaps surprising that dark photon bremsstrahlung so strongly dominates the secondary vector meson channels at a proton beam dump such as SHiP. The origin of this large hierarchy is not fully understood and warrants further investigation.

\subsection{$e^+$ $e^-$ Annihilation}

A complementary production mechanism to the others discussed above arises from the annihilation of secondary positrons,
generated within the electromagnetic component of the shower, with electrons bound in the target atoms. As the beam showers in the target, $e^+e^-$ pairs from photon conversion populate a spectrum of secondary positrons which can annihilate either resonantly, $e^+e^-\to A^\prime$, or non-resonantly, $e^+e^-\to\gamma A^\prime$, at any positron energy above threshold. This channel is relevant for electron beam dumps \cite{Marsicano:2018krp}, proton beam dumps \cite{Celentano:2020vtu} via $\pi^0\to\gamma\gamma$ decays, and has been exploited by NA64 \cite{Andreev:2021fzd,NA64:2023wbi,NA64:2023ehh} via resonant
$e^+ e^- \to A' \to \chi \bar{\chi}$.

The strong enhancement of the production rate in the $200$--$300$~MeV mass window at NA64 follows directly from the resonant behaviour of the $e^+e^-\to A^\prime$ vertex. Unlike dark bremsstrahlung or the non-resonant $e^+e^-\to\gamma A^\prime$ process, where the dark photon is produced off an
internal off-shell propagator, resonant annihilation proceeds through an $s$-channel pole: the $A^\prime$ is produced on-shell whenever the centre of mass energy of the $e^+e^-$ system matches its mass, $\sqrt{s}=m_{A^\prime}$. For a positron of laboratory energy $E_{e^+}$ annihilating on an atomic electron approximately at rest, the resonance condition fixes a single positron energy,
$E_{e^+}^{\rm res} \approx m_{A^\prime}^2/2m_e$
at which the cross section exhibits a sharp Breit--Wigner peak whose width is set by the total decay width of the dark photon.

\section{Visible Searches}
\label{sec:VisibleSearches}
If the dark photon is lighter than $2m_\chi$ (or below threshold for any other dark sector final state), the only kinematically accessible decay is to SM particles via \gls{KM} \cite{Fabbrichesi:2020wbt,Berlin:2018bsc,Andreas:2013lya,LDMX:2025bog,Batell:2022dpx,Ilten:2022lfq}. Even when the invisible channel $A' \to \chi\bar\chi$ is open, the visible
channel can still dominate the total width in the regime $\epsilon \gtrsim g_D$.  Below $2m_\mu$ the only open
visible channel is $A' \to e^+e^-$; above $2m_\mu$ the dimuon channel opens, and above $2m_\pi$ hadronic final states become accessible. Depending on whether the proper decay length is short or long compared to the detector geometry, a visibly decaying $A'$ then appears either as a localized (possibly displaced) energy deposition inside the detector, or as missing energy/momentum if it escapes the detector volume before decaying. Here I briefly review the experimental signatures of the visible regime, first at lepton/hadron colliders and then at fixed-target experiments.

At lepton colliders, the dominant production mechanism for sub-GeV dark photons is $e^+e^- \to \gamma A'$, with $A' \to \ell^+\ell^-$ reconstructed as a narrow peak in the dilepton invariant mass spectrum (BaBar, Belle~II) \cite{BaBar:2014zli,Belle-II:2018jsg}; at hadron colliders, prompt or displaced di-muon resonance searches provide complementary sensitivity at higher $m_{A'}$. Since collider luminosities are typically far below those
achievable in fixed-target setups, and the relevant phase space favours larger $m_{A'}$, these searches are subdominant to fixed-target experiments in the sub-GeV
mass range that is the focus of this thesis.

Fixed-target searches for visibly decaying dark photons fall into two broad classes,
distinguished by target thickness and the resulting decay topology \cite{Battaglieri:2017aum,Krnjaic:2022ozp}. Thin-target experiments pass a beam through a target thin enough that the $A'$ is produced via dark bremsstrahlung but the beam itself is not absorbed; decay products are reconstructed promptly, or with a small displacement. Thick-target, or beam-dump,
experiments absorb the primary beam entirely in a dense dump, suppressing SM backgrounds by many orders of magnitude, and a far detector downstream searches for displaced visible decays or missing energy from the long-lived $A'$ that escape the dump before decaying.

For a beam dump type search, dark photons are produced through dark bremsstrahlung off the target nuclei, and the number of $A'$ that decay visibly within the active volume of a downstream detector is \cite{Berlin:2018bsc},
\begin{equation}
    N_{\text{vis}} = N_{A'} \big( e^{-z_{\text{min}}/\gamma c \tau} - e^{-z_{\text{max}}/\gamma c \tau}  \big),
\end{equation}
where $N_{A'}$ is the number of dark photons produced, $z_{\text{min}}$ and $z_{\text{max}}$ are the distances from the production point to
the beginning and end of the decay volume, respectively, $\gamma$ is the boost of the $A'$, and $c\tau$ is its proper decay length.

In Paper VI, we explore a light DM scenario that features a light $Z'$ mediator such that it dominantly decays visibly. We consider constraints from NA64 \cite{NA64:2019auh}, electron beam dump experiments E774, E141, Orsay, KEK, E137 \cite{Andreas:2012mt}, proton beam dumps CHARM and NuCal \cite{Blumlein:2011mv,Tsai:2019buq}, and LDMX \cite{LDMX:2025bog}.

Beyond the freeze-out paradigm, we demonstrated that freeze-in \cite{Giudice:2000ex,Bernal:2017kxu} is being tested by visible fixed-target experiments \cite{Cosme:2021baj}.

\chapter{Light Dark Matter eXperiment}
\label{chap:LDMX}

This chapter focuses on \gls{ldmx}, the experiment underpinning a large part of the work presented in this thesis \cite{LDMX:2025bog,LDMX:2018cma}. The concept for LDMX was first proposed by Izaguirre et al. in 2014~\cite{Izaguirre:2014bca} as a tracker-based fixed-target search targeting the invisible decay regime of the dark photon discussed in \secref{subsection:DarkPhotons}, with projected sensitivity extending well below the leading existing constraints from NA64~\cite{NA64:2023wbi} and BaBar~\cite{BaBar:2017tiz}, while remaining
deployable parasitically on existing electron beam infrastructure, such as the beam lines available at SLAC. Among the missing momentum and missing energy strategies introduced in Chapter~\ref{chap:AcceleratorExperiments}, NA64 represents the most mature implementation, using a thick, active electromagnetic calorimeter as both target and energy sink to identify candidate events through an energy deficit relative to the incident beam energy~\cite{NA64:2023wbi}. 
The missing momentum approach on which LDMX is based, instead uses a thin target together with high precision tracking upstream and downstream of it, reconstructing the momentum of the recoiling electron directly rather than solely inferring an energy deficit calorimetrically. By measuring both the momentum and scattering angle of the recoiling beam electron, the experiment retains kinematic information such as the transverse momentum from the production process, that an energy only measurement does not have. Because each beam electron typically undergoes at most one hard interaction in the thin target, this strategy allows the kinematics of the production vertex to be reconstructed almost exactly.
This, combined with a system of electromagnetic and hadronic calorimeter vetoes surrounding the target region, gives LDMX strong rejection against SM backgrounds that can otherwise mimic a new physics missing momentum signature.

LDMX is designed to operate parasitically using an 8 GeV electron beam at SLAC, with an initial pilot run of $4 \times 10^{14}$ \gls{eot}, and a subsequent run of $10^{16}$ EOT. Together with NA64, BaBar's mono-photon search, and the other fixed-target experiments considered in \chapref{chap:AcceleratorExperiments}, LDMX forms part of a broader experimental landscape probing the dark photon and other light DM mediator scenarios across complementary regimes of coupling and mass.
The theoretical motivation for LDMX spans both minimal DM theory frameworks and a broader class of hidden sector physics accessible through the
same missing momentum signature. At its core, LDMX is designed to test thermal freeze-out as the origin of sub-GeV DM: for scalar, pseudo-Dirac, and Majorana DM candidates (see \chapref{chap:Models} for vector DM) coupled to the SM through a kinetically mixed dark photon (discussed in \secref{sec:minimal_DM_models}), the requirement that the dark and visible sectors reach thermal equilibrium in the early Universe fixes a calculable relic target coupling strength as a function of the DM mass~\cite{Berlin:2018bsc}. These thermal targets fall in a mass and coupling range that is largely inaccessible to nuclear recoil direct detection experiments, which lose sensitivity below roughly a GeV because the kinematics of scattering off a much heavier target nucleus become unfavourable for light DM. Accelerator production does not suffer from this, and the missing momentum technique employed by LDMX is sensitive to couplings at and below the thermal targets across the sub-GeV mass range. The same experimental signature extends naturally to a wider landscape of light new physics beyond minimal DM models: millicharged particles, axion-like particles \cite{LDMX:2026xum} and other scalar states, and additional light gauge bosons \cite{Banerjee:2025ejz} can all be produced in the same electron-nuclear fixed-target collisions and identified either through missing momentum if they escape the detector, or through prompt or displaced visible decays if they do not~\cite{Berlin:2018bsc}. LDMX is therefore sensitive not to a single benchmark scenario but to a broad and theoretically well motivated class of sub-GeV dark sector phenomena.

The remainder of this chapter is organized as follows. Section
\secref{sec:ldmx-setup} gives an overview of the LDMX experimental setup relevant to
Papers I-VIII. \secref{sec:ldmx-signal} details the new physics signatures searched for at LDMX and the corresponding modelling of these signal production with the resulting recoil electron kinematics. \secref{sec:ldmx-background} discusses the dominant SM background sources and their modelling.

\section{Experimental Setup}
\label{sec:ldmx-setup}
\begin{figure}[h]
    \centering
    \includegraphics[width=0.9\linewidth]{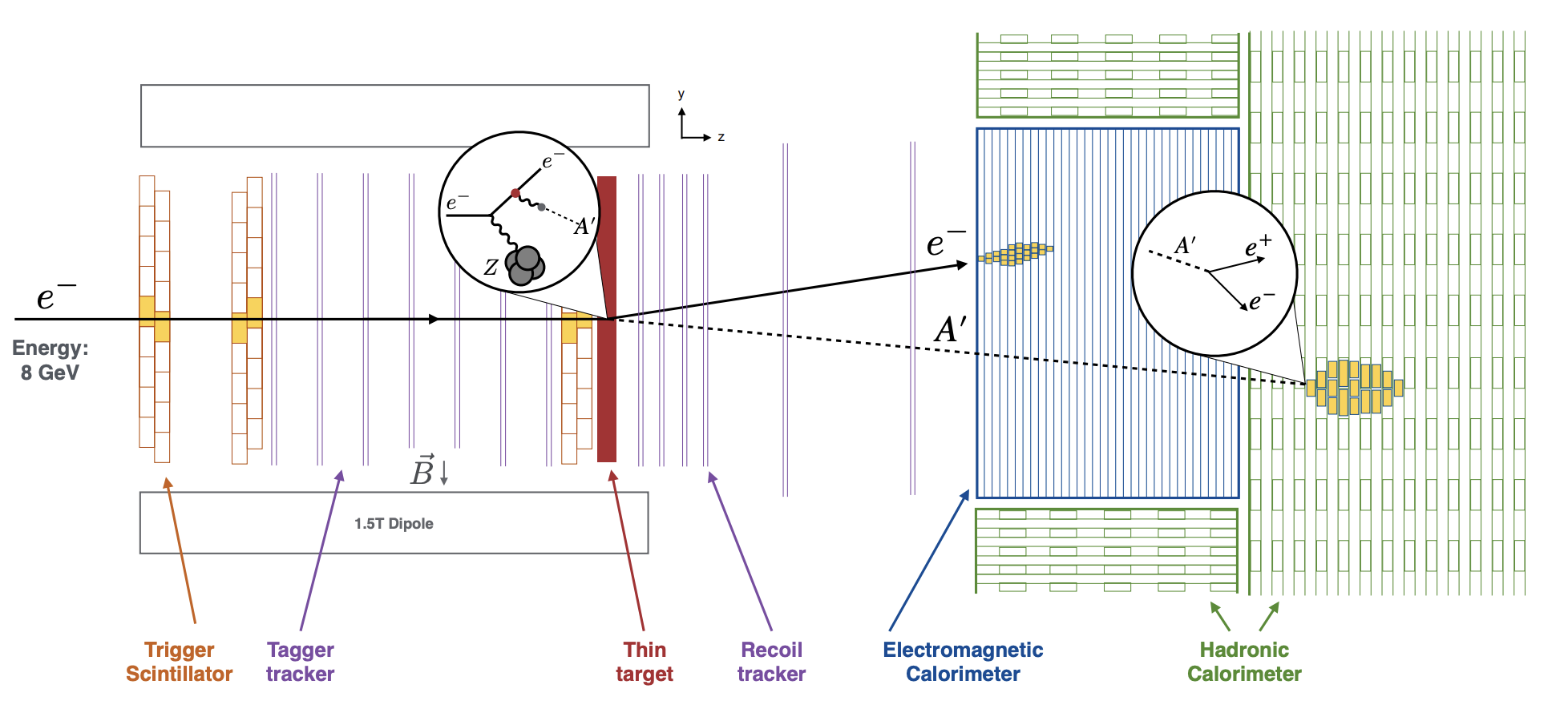}
    \caption{The experimental setup design of LDMX with the electron beam, trigger scintillator, tagger and recoil trackers, thin target where the dark photons are produced, followed by the ECal and HCal \cite{LDMX:2026xum,LDMX:2025bog}.}
    \label{fig:LDMX_diagram_detail}
\end{figure}
LDMX is proposed to operate at SLAC in End Station A, using an 8~GeV electron beam delivered via the Linac to End Station A beamline, which diverts bunches from the LCLS-II drive beam that are unused by the
photon science program. The beam is structured into trains at a 1~MHz repetition rate, each containing up to 20 bunches spaced by 26.9~ns, with an average occupancy of 1--2 electrons per bunch; a sparse, high-rate structure specifically chosen such that individual beam electrons can be properly tagged and reconstructed. The full anticipated dataset corresponds to $10^{16}$ \gls{eot}, with an initial pilot run of $4\times10^{14}$ EOT serving as a commissioning and proof of concept phase~\cite{LDMX:2025bog}.

The setup of the experiment is depicted in \figref{fig:LDMX_diagram_detail}.
Beam electrons are incident on a thin tungsten target, $\sim 10$\% radiation length thick, chosen so that the large majority of electrons undergo at most one hard interaction in the target. Immediately upstream of the target, a \gls{ts} counts the number of incoming electrons in each bunch, providing the normalization per bunch needed to set an appropriate missing energy threshold downstream. The beam can have contamination, hence it is important to measure the beam electrons both before and after the interaction with the target. The target sits between two silicon tracking systems: a \gls{stt} upstream, which measures the incoming beam electron's momentum, and a \gls{srt} downstream, which measures the momenta of outgoing charged particles, in particular the recoiling beam electron. Both trackers are housed within a single refurbished dipole magnet of 1.5~T, with the STT positioned in the high, uniform central field needed for precise beam-energy tagging, and the SRT positioned in the magnet's fringe field.

Following the SRT, the \gls{ecal} is a high granularity, high density sampling calorimeter based on the design of the CMS High Granularity Calorimeter developed for the CMS Phase 2 upgrade. The ECal is made of tungsten absorber planes interleaved with silicon planes, with both full containment of electromagnetic showers and the spatial resolution needed to resolve individual shower products. The ECal measures the energy of the recoil electron and radiated photons. 
Surrounding and following the ECal is the \gls{hcal},
a steel scintillator sampling calorimeter adapted from the Cosmic Ray Veto design developed for Mu2e. The HCal identifies neutral hadron backgrounds such as neutrons and kaons in addition to muons.
Although the missing momentum analysis relies on the HCal primarily as a veto, the same calorimeter system also allows LDMX to operate as a fully instrumented beam dump sensitive to visibly decaying long-lived particles~\cite{LDMX:2025bog}, a complementary search strategy that we explore in Paper VI.

\section{Signal Modelling}
\label{sec:ldmx-signal}
The high energy electron beam impinging on the tungsten nuclear target creates an environment to probe rare new physics processes that could occur in these beam-nucleus collisions. Any new particle in the MeV-GeV mass range that couples to electrons (or to quarks/nuclei if virtual Compton scattering is accounted for \cite{Schurmann:2026qmt}) can be probed at LDMX through the missing momentum technique it employs.
Dark photons can be produced through dark bremsstrahlung, described in \secref{section:electron_brem}, and in the case of dominantly invisible decays where $BR_{A'\to \text{DM\:DM}} \gg BR_{A' \to \text{SM\:SM}}$, inducing a measurable transverse momentum kick and missing energy to the recoil electron combined with an invisible final state.
\begin{figure}[h]
    \centering
\includegraphics[width=0.9 \linewidth]{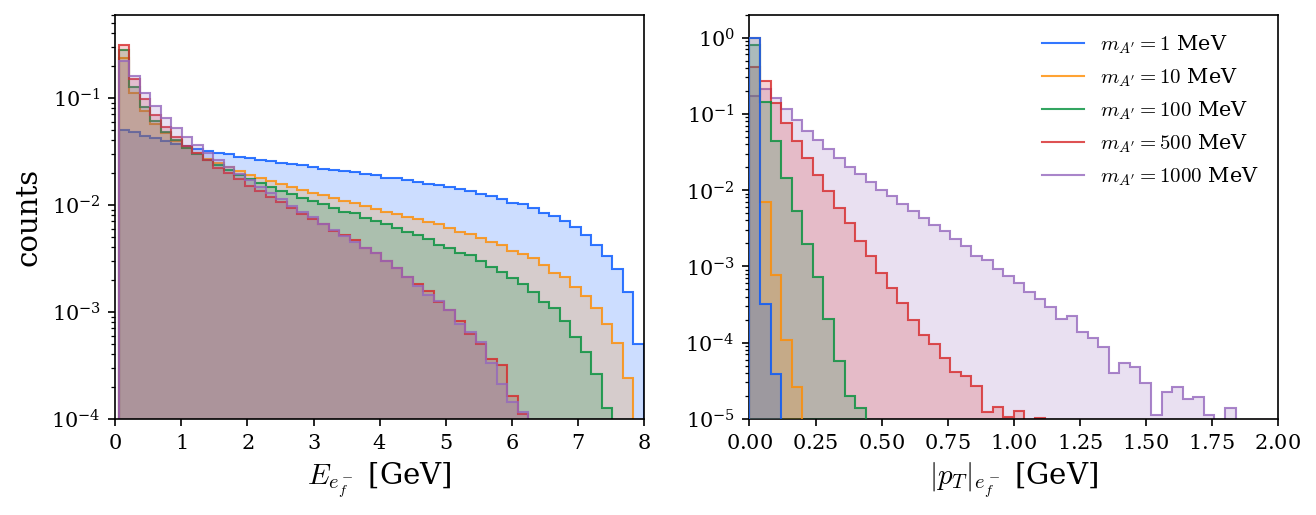}
    \caption{Normalized energy (left) and transverse momentum (right) distributions of the recoil electron from dark bremsstrahlung at LDMX at varying dark photon masses, simulated using \madgraph for an 8 GeV beam.}
\label{fig:LDMXdistributions}
\end{figure}
The energy, $E_{e^-_f}$, and transverse momentum, $|p_T|_{e^-_f}$, distributions of the recoil electron are plotted in \figref{fig:LDMXdistributions}. These distributions are computed using \madgraph with the methods described in \secref{section:electron_brem}, where the tungsten nucleus is implemented as an elementary particle dressed with the analytic form factor of \eqnref{eq:Bjorken_formfactor} within the UFO framework. The larger the dark photon mass, the more missing energy is carried away, and the lower the recoil electron energy. The transverse momentum distributions have the opposite effect, with larger dark photon masses leading to higher $p_T$. These distributions can be combined into a two-dimensional distribution, shown in \figref{fig:BGandSignal_LDMX}, which is used within a two-dimensional likelihood framework discussed in \chapref{chap:Likelihood}. The total number of dark photon events used to properly normalize these distributions, is,
\begin{equation}
    N_{A'}(\epsilon,m_{A'}) =  \sigma_{A'}(\epsilon,m_{A'}) \left(\frac{T X_0 N_e N_0}{A} \right),
\end{equation}
where $\sigma_{A'}$ is the cross section of dark photon production through bremsstrahlung in units of $cm^2$, $T$ is the number of radiation lengths of the target material, $X_0$ is the radiation length of the target in $g/cm^2$, $A$ is the atomic mass of the target nucleus in $g/mol$, $N_e$ is the number of \gls{eot}, and $N_0$ is Avogadro's number \cite{Izaguirre:2014bca}. For a given set of model parameters, $(\epsilon,m_{A'})$, $N_{A'}$ is computed and used to normalize the kinematic distributions.

The kinematic observables used to characterize the dark photon signal, the recoil electron's energy $E_{e^-_f}$ and the magnitude of its transverse momentum $|p_T|_{e^-_f}$, are predicted directly from the matrix elements describing dark photon production. In practice, however, neither quantity is measured exactly: each reconstructed value carries an associated uncertainty set by the intrinsic resolution of the detector, so that an electron with a given true energy or momentum is not always reconstructed with that same value. The net effect on the predicted distributions is twofold: sharp features present at truth level are smoothed out, and events migrate between bins of the observable in a way that depends on the local detector resolution. Any comparison between predicted and observed event yields must therefore account for this migration explicitly, by smearing the truth-level prediction with the appropriate detector response function before it is compared to data or used in the statistical analysis described in \chapref{chap:Likelihood}.
For LDMX, the relevant resolutions have been characterized in~\cite{LDMX:2025bog}. The $|p_T|_{e^-_f}$ resolution, $\sigma_{p_T}$, is obtained from the combined performance of the STT and SRT.
The ECal energy resolution, $\sigma_E$, is characterized
analogously, by injecting electrons and photons of known energy into a simulation of the calorimeter and fitting the resulting distribution of reconstructed versus true energy.

The simulated dark photon signal sample is restricted to the kinematic region in which the recoiling electron is measurable.
Three cuts are applied to the recoil electron at event generator
level: (1) a lower energy cut corresponding to the minimum momentum at which a recoil track can be reliably reconstructed $E_{e^-_f} > 50\,\mathrm{MeV}$, (2) an angular cut $|\theta_{e^-_f}| < 40^\circ$, covering the angular acceptance of the recoil tracker and ECal, and (3) $E_{e^-_f} < 3.16$ GeV due to the experimental trigger threshold and ensuring the recoil electron has undergone substantial energy loss consistent with dark sector particle production.

\section{Background Modelling}
\label{sec:ldmx-background}

\begin{figure}
    \centering
    \includegraphics[width=0.95\linewidth]{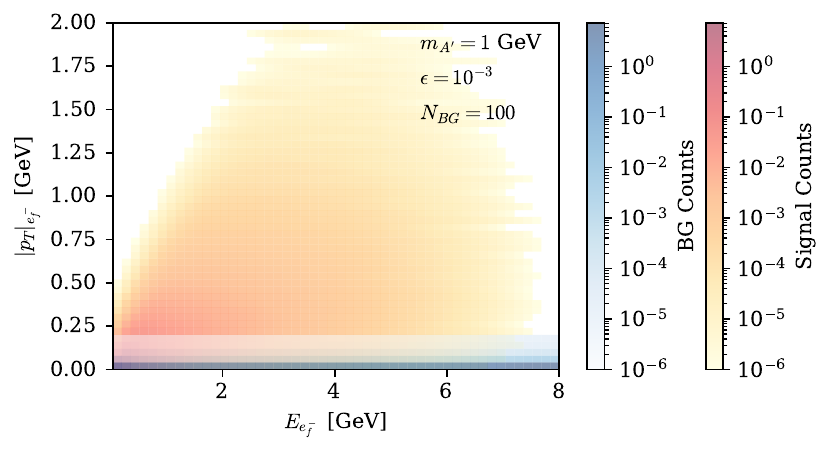}
    \caption{The two-dimensional signal (yellow/red) and background (blue) distributions in the recoil electron $|p_T|_{e^-_f}$ vs $E_{e^-_f}$ plane for LDMX. We take the dark photon model parameters to be $m_{A'}=1$ GeV and $\epsilon=10^{-3}$, and the total background normalization to be 100, for illustration.}
    \label{fig:BGandSignal_LDMX}
\end{figure}

Backgrounds from neutrino production lie far below the sensitivity floor at the EOT levels relevant to LDMX, therefore there are no irreducible backgrounds and all surviving backgrounds are instrumental, arising from a failure to detect the visible particles that accompany SM processes~\cite{LDMX:2025bog}. The dominant surviving class consists of hard
bremsstrahlung photons that fail to deposit their energy as an ordinary electromagnetic shower in
the ECal, for example through photo-nuclear interactions~\cite{LDMX:2025bog,LDMX:2018cma}.

For the statistical framework developed in Paper VII, this background is modelled in the two
observables entering the likelihood, the recoil electron energy $E_{e^-_f}$ and transverse
momentum $|p_T|_{e^-_f}$, by combining results from two existing LDMX simulation studies. The
$|p_T|_{e^-_f}$ shape is taken from the distribution of ECal photo-nuclear events passing the missing-energy trigger, shown in Fig.~6 of~\cite{LDMX:2023zbn}. The corresponding $E_{e^-_f}$ shape is taken from the inclusive single-electron background distribution plotted alongside the signal energy spectra in Fig.~10 of~\cite{LDMX:2018cma}. The two
one-dimensional distributions are combined via an outer product into a two-dimensional
$(|p_T|_{e^-_f}, E_{e^-_f})$ background template, under the simplifying assumption that the two
variables are uncorrelated for background events. This approximation is sufficient to illustrate
how the overall background level affects projected sensitivity, but a dedicated two-dimensional
characterization, including a full estimate of the absolute normalization at $10^{16}$~EOT,
would be needed for a complete analysis.
The background is plotted in blue in \figref{fig:BGandSignal_LDMX}, along with the signal distributions. The signal overwhelmingly dominates in the higher $p_T$ regime, and similarly for lower energies (more missing energy).

 \chapter{Statistical Framework for LDMX}
\label{chap:Likelihood}

Statistical inference is central to particle physics; where a model's parameters must be inferred, or a new physics hypothesis confirmed or excluded, from a finite sample of data, often hidden within background (see \cite{Cowan:2010js,Lista:2016chp,Segura:2024srj,Cranmer:2014lly} for statistics reviews). The field has historically relied predominantly on frequentist statistics, in part because its hypothesis testing conventions allow a discovery or exclusion claim to be made without specifying a prior probability distribution over the parameter of interest~\cite{Cousins:1994yw}. The resulting significance scale, expressed in terms of p-values and the $5\sigma$ discovery threshold~\cite{Lyons:2013yja}, has become an established field-wide convention. Nuisance parameters are, however, still constrained using a probability distribution function of the same form as a Bayesian prior.
Bayesian statistics, while less commonly used for discovery and exclusion claims in particle physics, is nonetheless an established complementary approach, particularly suited to parameter inference and model comparison.
The two frameworks differ fundamentally in how probability itself is interpreted, and this distinction underlies the two parallel analyses presented in this chapter.
In the context of comparing statistical methodologies, the dichotomy is famously summarized by Louis Lyons~\cite{Lyons:2013yja}:
\begin{quotation}
    \small \itshape
    ``Bayesians address the question everyone is interested in by using assumptions no-one believes, while Frequentists use impeccable logic to deal with an issue of no interest to anyone.''
    \vspace{0.2cm}
    
    \raggedleft --- Louis Lyons
\end{quotation}
In the frequentist interpretation, probability is understood as a frequency of an event (the number of occurrences over the number of trials) in the limit of infinite trials: model parameters are treated as fixed, unknown quantities, and statements of uncertainty are made about the behaviour of an estimator or test statistic under hypothetical repetition of the experiment. The Bayesian interpretation instead treats probability as a degree of belief, allowing the model parameters themselves to be assigned a probability distribution. A prior distribution, encoding what is known about the parameters before the data are considered, is updated by the likelihood via Bayes' theorem into a posterior distribution, from which credible regions are derived directly. 
The likelihood function remains common ground between the two paradigms regardless, and presenting it directly allows the reader to construct either a frequentist or Bayesian analysis; a practice this chapter and Paper VII follow by developing a single likelihood as the shared starting point for both analyses presented below.
We adopt the frequentist framework where the established conventions for formal discovery and exclusion claims apply most naturally, and the Bayesian framework is used for posterior-based parameter inference and model comparison.

Counting experiments in particle physics are governed by Poisson statistics: for a rare process observed over a fixed exposure, the number of events $d$ recorded follows,
\begin{equation}
    f_P(d,\nu) = \frac{\nu^d}{d!}e^{-\nu},
\end{equation}
where $\nu$ is the expected number of events under a given model. This single bin description is adequate whenever signal and background differ only in their overall rate, so that the total event count alone carries the discriminating information.

If, instead, the signal and background follow a distribution over some observable(s) whose shape depends on the model parameters, this shape information can be exploited within the statistical analysis. In this case, the data are sorted into bins of the observable, and a Poisson count is associated with each bin individually, $f_P(d_i|\nu_i) = \nu_i^{d_i} e^{-\nu_i}/d_i!$, with $\nu_i = S_i + B_i$ the predicted number of signal plus background events of bin $i$\footnote{An unbinned analysis of LDMX is left as a potential future research direction.}. 
At LDMX, the \gls{KM} strength $\epsilon$ and mediator mass $m_{A'}$
leave their mark not only on the total number of events, but on the shape of the recoil electron kinematic distribution. Therefore, collapsing the
data to a single count would discard exactly the information needed to distinguish one hypothesis from another with a similar yield. As described in \chapref{chap:LDMX}, the observables used at LDMX are the transverse momentum, $p_T \equiv \sqrt{p_x^2 + p_y^2}$, where the beam axis is pointing along $\hat{z}$, and the energy, $E_{e^-_f}$, of the recoil electron. The distributions of these kinematic quantities vary with $m_{A'}$ and $\epsilon$. The signal plus background expectation, $\nu$, is thus binned into finite $E_{e^-_f}$ and $|p_T|_{e^-_f}$ bins.
The product of Poisson factors over bins is used to build the full likelihood, which treats the bin counts as statistically independent. We can then write the likelihood as,
\begin{equation}
    \mathcal{L}(\vec{d}|\epsilon,m_{A'}) = \prod_i \frac{\left(S_i(\epsilon,m_{A'}) + B_i\right)^{d_i}}{d_i!}e^{-\left(S_i(\epsilon,m_{A'}) + B_i\right)}.
\end{equation} 

Uncertainties on the signal and background predictions can be incorporated into the likelihood through nuisance parameters, which are subsequently profiled or marginalized depending on the statistical framework adopted. Including these uncertainties in the likelihood causes the inferred parameter regions to broaden accordingly.
One method is to marginalize over the uncertainty \cite{Conrad:2002kn,GAMBIT:2017qxg}, where a nuisance parameter $\xi$ is introduced as a multiplicative factor,
\begin{equation}
\begin{aligned}
    \mathcal{L}(\vec{d}|\epsilon,m_{A'}) &= \prod_i \int \frac{[S_i(\epsilon,m_{A'})\xi_S+B_i \xi_B]^{d_i}}{d_i!} \\ &e^{-\left(S_i(\epsilon,m_{A'})\xi_S+B_i \xi_B\right)} P(\xi_S|\sigma_{\xi_S}) P(\xi_B|\sigma_{\xi_B})  d\xi_S d\xi_B,
\end{aligned}
\end{equation}
where $P(\xi|\sigma_{\xi})$ is the probability distribution of $\xi$ with standard deviation $\sigma_\xi$ (which for example can be a log-normal distribution). This method is \textit{semi-Bayesian} since the nuisance parameter is being marginalized over \cite{Conrad:2002kn}.
Another method is to introduce nuisance parameters that characterize the uncertainties and instead profile over them. Let us now consider that we have the uncertainty associated with both the signal, $S$, and background, $B$, of a given bin, therefore we introduce two nuisance parameters: $\theta_S$ and $\theta_B$. The likelihood of a given bin is now \cite{Lista:2016chp},
\begin{equation}
\begin{aligned}
     \mathcal{L}(\vec{d}|\epsilon,m_{A'},\theta_S,\theta_B) &= \frac{[S(\epsilon,m_{A'})\theta_S + B\theta_B]^d}{d!} \\ & e^{-(S(\epsilon,m_{A'})\theta_S+B\theta_B)} \mathcal{P}(\theta_B| \sigma_B^2) \mathcal{P}(\theta_S | \sigma_S^2),
\end{aligned}
\end{equation}
where we take $\mathcal{P}(\theta | \sigma^2)$ to be log-normal distributions defined by,
\begin{equation}
    \mathcal{P}(\theta | \sigma^2) = \frac{1}{\sqrt{2 \pi} \sigma \theta } \exp \bigg[ - \frac{1}{2} \left( \frac{\ln \theta}{\sigma}\right)^2 \bigg],
\end{equation}
where the signal (background) estimates are recovered when $\theta_S$ ($\theta_B$) is one, and are allowed to shift according to the probability distribution.
In the case of the Bayesian analysis, the probability distribution functions over the nuisance parameters are instead the priors of $\theta_B$ and $\theta_S$, therefore they are not included at the likelihood level, but appear within the posterior.

In Paper VII, where this statistical analysis is performed for LDMX using \madgraph computed cross sections,
$\sigma_S$ is fixed to 0.1 since the statistical uncertainty of the cross section as reported by \madgraph is $\sim 10 \%$. For the frequentist analysis, we fix $\sigma_B$ to $0.1$, and for the Bayesian analysis we introduce a hyperprior on $\sigma_B$ allowing this degree of uncertainty itself to be informed by the data rather than fixed a priori.

Since we have two observables at our disposal, we combine both the $|p_T|_{e^-_f}$ and $E_{e^-_f}$ information into a single joint likelihood, where the bins are now defined as a two-dimensional grid. We can write the full likelihood as a function of the parameters,
\begin{equation}
\begin{aligned}
    \mathcal{L}(\vec{d}|\epsilon,m_{\Ap},\theta_S,\theta_B) &= \prod_{i}^{N_E} \prod_j^{N_{p_T}} \frac{\big[ S_{ij}(\epsilon,m_{\Ap})\theta_S + B_{ij}\theta_B\big]^{d_{ij}}}{d_{ij}!} e^{- \big[ S_{ij}(\epsilon,m_{\Ap})\theta_S + B_{ij}\theta_B\big] } \\ & \times \mathcal{P}(\theta_B| \sigma_B^2) \mathcal{P}(\theta_S | \sigma_S^2),
    \label{eq:likelihood}
\end{aligned}
\end{equation}
where $N_E$ and $N_{p_T}$ are the number of $E_{e^-_f}$ and $|p_T|_{e^-_f}$ bins, respectively.

\section{Frequentist Analysis}
When searching for new physics, whether for making an exclusion or discovery, we can define a statistical test which determines if there is enough evidence in the data to reject a given hypothesis $H_0$ in favour of an alternative $H_1$~\cite{Cowan:2010js,Cowan:2013pha,Cranmer:2014lly}. The identity of $H_0$ depends on which question is being asked. For a discovery claim,
$H_0$ is the background-only (SM) hypothesis, tested against the alternative $H_1$ that
a new physics signal is also present; rejecting $H_0$ constitutes a discovery. For an exclusion limit, the hypothesis under test is instead $H_1$, the signal-plus-background hypothesis at a specific value of the parameters of interest (in this case $m_{A'}$ and $\epsilon$), tested against the alternative that the data prefers less signal than this; rejecting $H_1$ in this case excludes that specific value of $(m_{A'},\epsilon)$.

The choice of test statistic is not unique, but the Neyman--Pearson lemma establishes that, for two simple hypotheses, the likelihood ratio $\mathcal{L}(H_1)/\mathcal{L}(H_0)$ is the most powerful choice, since it maximizes the probability of correctly rejecting $H_0$ when $H_1$ is true~\cite{Neyman:1933wgr}. Since we also generally have nuisance parameters, as introduced above, the natural generalization of the likelihood ratio is the profile likelihood ratio. The profile likelihood ratio is defined as \cite{Cowan:2010js},
\begin{equation}
    \lambda(\bm{\psi}) = \frac{\mathcal{L}(\bm{\psi},\hat{\hat{\bm{\theta}}})}{\mathcal{L}(\hat{\bm{\psi}},\hat{\bm{\theta}})},
\label{eq:likelihood_lambda}
\end{equation}
where $\bm{\psi} \equiv (\psi_1,\psi_2, \dots)$ is the set of parameters of interest, $\bm{\theta}\equiv (\theta_1,\theta_2,\dots)$ is the set of nuisance parameters, $\hat{\hat{\bm{\theta}}}$ is the maximum-likelihood value of $\bm{\theta}$ at the specific $\bm{\psi}$ value at which $\lambda(\bm{\psi})$ is being evaluated (the conditional maximum-likelihood estimator), while the denominator is the likelihood evaluated at the unconditional (global) maximum-likelihood estimates of $\bm{\psi}$ and $\bm{\theta}$, namely $\hat{\bm{\psi}}$ and $\hat{\bm{\theta}}$.

A test statistic is then constructed from the profile likelihood ratio, 
\begin{equation} 
t_{\bm\psi} \equiv -2\log\lambda(\bm\psi),
\label{eq:test_statistic}
\end{equation} 
a sign chosen so that $t_{\bm{\psi}}\geq0$, with larger values corresponding to data increasingly incompatible with the tested hypothesis (signal $S$ evaluated at $\bm{\psi}$, $S(\bm{\psi})$), and we take the $\log$ for better control of the likelihood which often varies over many orders of magnitude. The asymptotic distribution of $t_{\bm{\psi}}$ can be approximated analytically in certain cases, as discussed below~\cite{Cowan:2010js}. 

Having observed some value $t_{\bm{\psi}}^{\rm obs}$ of the test statistic, the compatibility of the data with the hypothesis $S(\bm{\psi})$ is then quantified by the $p-$value, defined as the probability, under the assumption that $\bm{\psi}$ is the true value of the parameters of interest, of obtaining a value of the test statistic at least as extreme as the one actually observed, 
\begin{equation} 
p_{\bm\psi} = \int_{t_{\bm\psi}^{\rm obs}}^{\infty} f(t_{\bm\psi}\,|\,\bm\psi)\,dt_{\bm\psi}, 
\end{equation} 
where $f(t_{\bm\psi}\,|\,\bm\psi)$ is the probability density function of the test statistic under the hypothesis $\bm\psi$. A small $p-$value indicates that the observed data are unlikely to have arisen under $\bm\psi$, and it is on this basis that $\bm\psi$ is either excluded or, when $\bm\psi$ corresponds to the background-only hypothesis, rejected in favour of a discovery claim.

In an analysis of already existing data, $t_{\bm\psi}^{\rm obs}$ is simply the value of the
test statistic computed from the observed data set. The present analysis, however, is a projection of the sensitivity expected from LDMX Phase~II rather than an analysis of data that has actually been collected, so no such observation exists. In its place, we
follow standard practice and substitute a value representative of the expected outcome of the experiment, taken to be the median of the
test statistic's distribution~\cite{Bhattiprolu:2020mwi},
with this median estimated from an ensemble of toy pseudo-experiments generated under $H$.
The identity of $H$ differs between the exclusion and discovery cases: for exclusion $H$ is the background only hypothesis, and for discovery the signal-plus-background hypothesis. The resulting $p$-value should accordingly be understood as the median expected outcome of the projected experiment, rather than the result of any single realization of it.

To compute the $p-$value, we must run pseudo-experiments to generate the distributions of the test statistic $f(t_{\bm{\psi}}|\bm\psi)$, which can be cumbersome computationally. It is therefore useful to have an analytic approximation for the test statistic distribution. It was demonstrated by Wilks and Wald that $f(t_{\bm\psi}|\bm\psi)$ approximately follows a chi-squared distribution with the difference in the number of free parameters between the alternative and the null hypotheses as the degrees of freedom~\cite{Wilks:1938dza,Wald1943}. This result only holds in the large sample limit, when the number of events in each bin is sufficient. In the LDMX scenario, this is unfortunately not always the case. Especially for smaller values of $\epsilon$ or background normalization ($N_\text{BG}=1$), there are bins that can have very few expected events.
We therefore forgo these asymptotic approximations and instead compute the test statistic distributions from full pseudo-experiments. While computationally more demanding, this comes at the benefit of a more statistically robust result, free from the systematic inaccuracies that the asymptotic approximation would otherwise introduce in this low-count regime.

\subsection{Exclusion}
We now specialize the above formalism for the setting of LDMX.
For determining whether a given set of model parameters $(m_{A'},\epsilon)$ is excluded at a given \gls{cl}, we define the exclusion test statistic based off of \eqnref{eq:test_statistic}. We consider a signal hypothesis excluded at a C.L. of $1-\alpha$ if the exclusion $p-$value is $\leq \alpha$, and we take the convention of $\alpha=0.1$ corresponding to a $90\%$ C.L..
Expressed in terms of the model parameters of our scenario the exclusion test statistic is \cite{Cowan:2010js},
\begin{equation}
    q_\epsilon =
    \begin{cases}
    -2 \log \frac{\mathcal{L}\big(\vec{d}(\epsilon')\: |\: \epsilon, \: \hat{\hat{\theta}}_S, \: \hat{\hat{\theta}}_B \: ; \: m_{\Ap} \big)}{\mathcal{L}\big(\vec{d}(\epsilon')\:|\:\hat{\epsilon},\: \hat{\theta}_S,\: \hat{\theta}_B \: ; \: m_{\Ap} \big)} & \hat{\epsilon} \leq \epsilon \\
    0 & \hat{\epsilon} > \epsilon, 
    \end{cases}
    \label{eq:q_eps}
\end{equation}
where $\epsilon'$ is the value of $\epsilon$ under which the pseudo-data $\vec{d}$ are generated, and the definitions of the hat and double hat parameters are given under \eqnref{eq:likelihood_lambda}. We enforce the one-sided nature of the test by setting $q_\epsilon$ to zero if the data prefers a signal strength larger than the tested parameter value, since this should not be evidence against the tested hypothesis, and we leave $m_{A'}$ fixed at the value which we want to evaluate. We construct the distribution of $q_\epsilon$ under the signal-plus-background hypothesis at the desired $(m_{A'},\epsilon)$, $f(q_\epsilon|\epsilon)$. Since we are making a projected exclusion bound, we also construct $f(q_\epsilon|0)$, the test statistic distribution under background-only, and take its median value as the expected $q^{\text{obs}}_\epsilon$. Both of these distributions are generated by performing full pseudo-experiments, via random sampling from a Poisson distribution. We compute the exclusion $p-$value,
\begin{equation}
    p_\epsilon = \int_{\text{med}[f(q_\epsilon|0)]}^{\infty} f(q_\epsilon|\epsilon)\, dq_\epsilon.
\end{equation}

\begin{figure}
    \centering
    \includegraphics[width=0.8\linewidth]{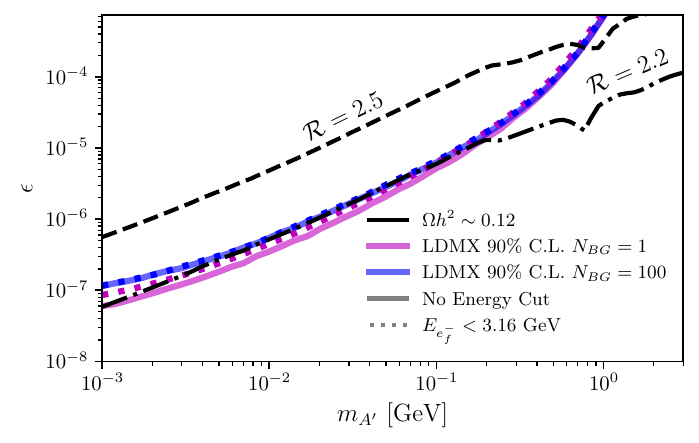}
    \caption{Projected $90\%$ C.L. exclusion bounds on the kinetic
mixing parameter $\epsilon$ as a function of the dark photon mass $m_{A'}$, obtained from the frequentist likelihood framework developed in this chapter and Paper VII. The region above a given curve is excluded at $90\%$ C.L., while the region below remains experimentally
viable. The magenta and blue curves assume a background of $N_{\rm BG}=1$ and $N_{\rm BG}=100$ events, respectively. Dotted curves additionally impose a cut on the recoil electron energy, $E_{e^-_f}<3.16$~GeV, while solid curves retain the full electron energy range. The black dashed and dash-dotted lines show the complex scalar DM thermal relic targets for $\mathcal{R}\equiv m_{A'}/m_{\rm DM}=2.5$ and $2.2$, respectively.}
    \label{fig:likelihood_exclusion}
\end{figure}

The contours for which $p_\epsilon=0.1$, corresponding to 90\% C.L., in the $\epsilon$ vs $m_{A'}$ plane, are drawn in \figref{fig:likelihood_exclusion}.
The projected exclusion contours are plotted for the two background scenarios, $N_\text{BG}=1$ in magenta and $N_\text{BG}=100$ in blue, and for the scenario with (dotted) and without (solid) the $E_{e^-_f}<3.16$ GeV energy cut. These exclusion contours are overlaid on top of the complex scalar relic target curves: the \gls{KM} required to reproduce the observed relic abundance,
$\Omega h^2\sim0.12$, via freeze-out (see \secref{sec:minimal_DM_models} for more about complex scalar DM and \secref{section:FreezeOut} for the freeze-out mechanism). These results indicate that LDMX is projected to exclude a large part of the $\mathcal{R}=2.2$ relic target. 

\subsection{Discovery}
We evaluate the discovery potential of LDMX by testing a set of four benchmark parameter values which sit on the relic target. The benchmarks are given in \tabref{tab:discovery}.
We use the test statistic for discovery, $q_0$, to determine whether or not each benchmark is \textit{discoverable} at LDMX \cite{Cowan:2010js},
\begin{equation}
    q_0 =
    \begin{cases}
    -2 \log \frac{\mathcal{L}\big(\vec{d}(\epsilon') \: | \: \epsilon=0, \: \hat{\hat{\theta}}_S, \: \hat{\hat{\theta}}_B \: ; \: m_{\Ap} \big)}{\mathcal{L}\big(\vec{d}(\epsilon') \:|\:\hat{\epsilon}, \:\hat{\theta}_S, \: \hat{\theta}_B \: ; \: m_{\Ap} \big)} & \hat{\epsilon} \geq 0 \\
    0 & \hat{\epsilon} < 0,
    \end{cases}
    \label{eq:q0}
\end{equation}
where the notation is the same as for \eqnref{eq:q_eps}.
$q_0$ is constructed such that it is set to zero whenever $\hat\epsilon<0$. Physically, $\hat\epsilon<0$ corresponds to a fit that prefers fewer events than the background-only prediction -- a deficit, rather than an excess, of events. Since a deficit of events cannot itself constitute evidence for a new physics signal (which can only ever add events on top of the background, never subtract from it), the test statistic is set to zero in this case, ensuring that downward fluctuations of the background are not mistakenly interpreted as evidence in favour of discovery.

The discovery $p-$value is given by,
\begin{equation}
    p_0 = \int_{\text{med}[f(q_0|\epsilon)]}^{\infty} f(q_0|0)\, dq_0.
\end{equation}

The $p$-value is conventionally translated into an equivalent significance $Z$, defined via the upper-tail probability of a standard Gaussian,
\begin{equation}
    Z = \Phi^{-1}(1-p_0),
\end{equation}
where $\Phi$ is the standard normal cumulative distribution function. This transformation is purely a convention for expressing the $p$-value on a more intuitive scale, in units of standard deviations, and carries no implication that
$f(q_0|0)$ is itself Gaussian distributed. A discovery is conventionally claimed once
$Z\geq5$, corresponding to $p\lesssim2.9\times10^{-7}$, a threshold adopted across particle physics \cite{Lyons:2013yja}.

\begin{figure}
    \centering
    \includegraphics[width=\linewidth]{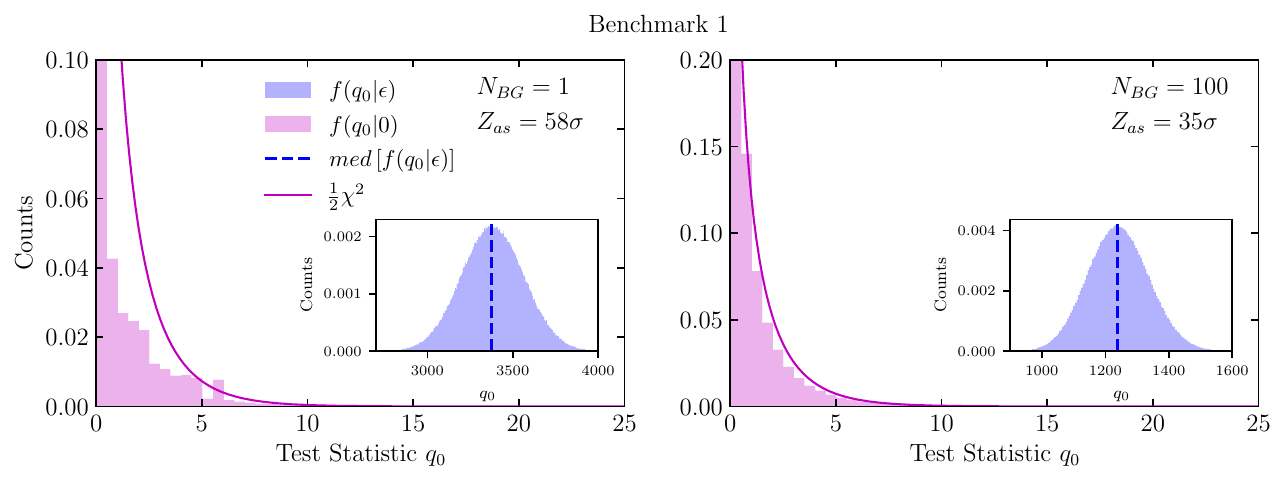}
        \includegraphics[width=\linewidth]{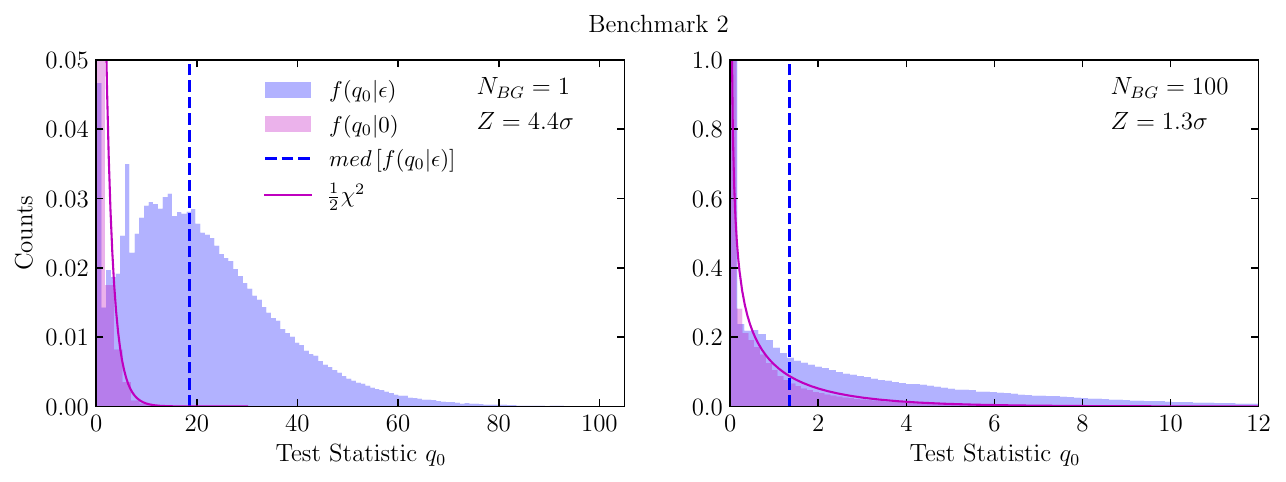}
    \caption{Distributions of the discovery test statistic $q_0$ for benchmark points 1 and 2, evaluated for background scenarios $N_{\rm BG}=1$ (left column) and $N_{\rm BG}=100$ (right column). The pink histogram shows the distribution of $q_0$ under the background-only hypothesis, $f(q_0|0)$, while the blue histogram shows the corresponding distribution under the signal-plus-background hypothesis, $f(q_0|\epsilon)$, both obtained from full pseudo-experiments.  Overlaid in magenta is the asymptotic half-$\chi^2$ approximation to $f(q_0|0)$~\cite{Cowan:2010js}.  The vertical dashed blue line marks the median of $f(q_0|\epsilon)$, from which the median expected discovery significance $Z$ (or its asymptotic counterpart $Z_{\rm as}$, substituted when $Z\gg5$) is obtained. See Paper VII for all four benchmarks.}
    \label{fig:likelihood_distributions}
\end{figure}

Analogously to the exclusion case, the distributions of the discovery test statistic, under the background-only and signal-plus-background hypothesis, are constructed from full pseudo-experiments from random Poisson draws. Figure~\ref{fig:likelihood_distributions} plots the discovery test statistic distribution under background-only, $f(q_0|0)$, and under signal-plus-background, $f(q_0|\epsilon)$, for benchmarks 1 and 2.
These two benchmarks are shown together to contrast a high yield, more readily discoverable case (benchmark 1) against a low yield, more challenging case (benchmark 2). As discussed earlier, in the large sample limit there is an asymptotic approximation of the test statistic. $f(q_0|0)$ can be approximated by a chi-squared and a delta function, each weighted by one half \cite{Cowan:2010js}. The computational effort required to accurately generate the distributions and thus compute $p_0$ increases with the discovery significance $Z$. We therefore report the asymptotic approximation of the significance, $Z_{\text{as}}$ for benchmarks with $Z_{\text{as}} \gg 5$. The asymptotic approximation is drawn as a magenta curve in \figref{fig:likelihood_distributions}. The agreement, or lack thereof, between the pseudo-experiment derived pink histogram and the magenta asymptotic approximation illustrates directly where this approximation breaks down, motivating the use of full pseudo-experiments throughout this chapter and Paper VII.

\renewcommand{\arraystretch}{1.2}
\begin{table}[h]
\centering
\caption{Expected discovery sensitivity of LDMX Phase~II at four benchmark points lying along the thermal relic targets of Figure~\ref{fig:likelihood_exclusion}, with mass ratio $\mathcal{R}\equiv m_{A'}/m_{\rm DM}=2.5$ (benchmarks 1 and 3) and $\mathcal{R}=2.2$ (benchmarks 2 and 4). For each benchmark, $m_{A'}$ and $\epsilon$ give the assumed dark photon mass and \gls{KM}, and $S$ the corresponding
expected signal yield; the expected background count $B$ is instead fixed externally to one of two scenarios, $B=1$ or $B=100$, indicated by the subscript on each $Z$
column rather than shown as a separate column. The median expected discovery significance $Z$ is computed from the discovery test statistic $q_0$ using full pseudo-experiments, except where $Z\gg5$, in which case the asymptotic approximation is used instead. A benchmark point is classified as \textit{discoverable} if $Z\geq5$. The superscript ``cut'' labels columns computed
with an additional requirement on the recoil electron energy, $E_{e^-_f}<3.16$~GeV.}
\label{tab:discovery}
\setlength{\tabcolsep}{2pt}
\begin{tabular}{lcccccccc}
\toprule \toprule
& $m_{\Ap}$ [GeV] & $\mathcal{R}\equiv \frac{m_{\Ap}}{m_\text{DM}}$ & $\epsilon$ & $S$ & $Z_{B=1}$ & $Z_{B=100}$ & $Z^\text{cut}_{B=1}$ & $Z^\text{cut}_{B=100}$ \\
\hline
\textbf{BM 1} & 0.01 & 2.5 & 4.8 $\times 10^{-6}$ & 310 & $\gg 5$ & $\gg 5$ & $\gg 5$ & $\gg 5$ \\
\textbf{BM 2} & 0.01 & 2.2 & 5.2 $\times 10^{-7}$ & 4   & 4.4  &  1.3 & 4.3 & 1.3 \\
\textbf{BM 3} & 0.1  & 2.5 & 6.3 $\times 10^{-5}$ & 250 & $\gg 5$ & $\gg 5$ & $\gg 5$ & $\gg 5$ \\
\textbf{BM 4} & 0.1  & 2.2 & 6.0 $\times 10^{-6}$ & 2   & 4.2  & 2.2 & 3.9 & 1.7 \\
\toprule \toprule
\end{tabular}
\end{table}

The degree of separation between the two distributions, under background-only and signal-plus-background, reflects how distinguishable a true signal at this benchmark would be from a
background-only fluctuation. Benchmarks with distributions more separated correspond to larger discovery significances, such as benchmarks 1 and 3, as reported in \tabref{tab:discovery}.
This is not surprising, since benchmarks 1 and 3 predict many more signal events than benchmarks 2 and 4 do. We find that LDMX is capable of discovering benchmarks 1 and 3 at $> 5\sigma$, with and without the energy cut.

\subsection{Maximum Likelihood Parameter Estimation}
\begin{figure}
    \centering
    \includegraphics[width=\linewidth]{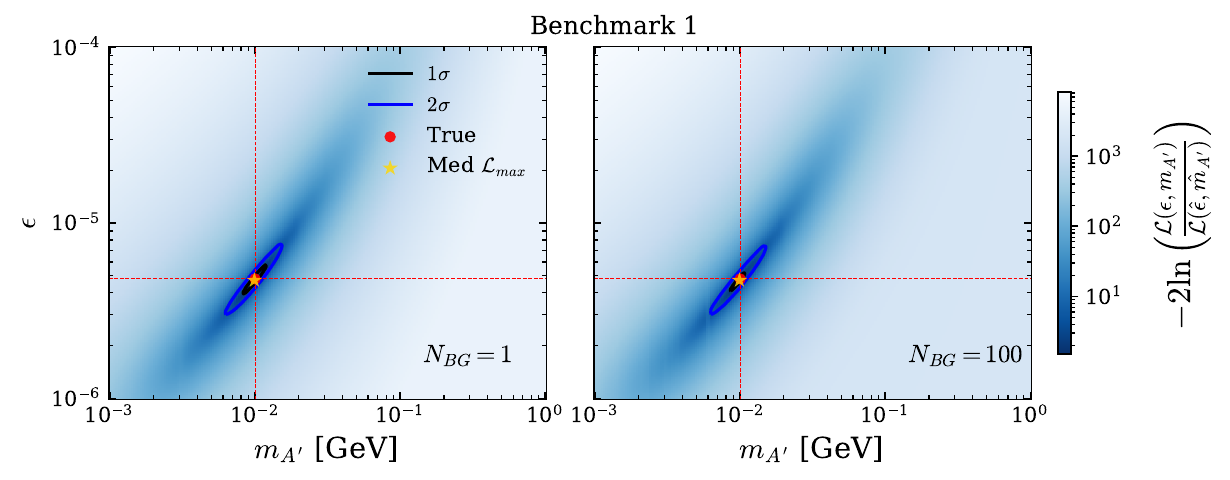}
    \caption{The profile likelihood ratio in the $(\epsilon, m_{A'})$ plane for benchmark 1, evaluated under two background scenarios: $N_{\rm BG}=1$ (left) and $N_{\rm BG}=100$ (right). At each point in this plane, the nuisance parameters $\theta_S,\theta_B$ are profiled out, so that the colour scale shows $\Delta(-2\log\mathcal{L})$ as a function of $\epsilon$ and $m_{A'}$ alone. The red point marks the true parameter values $(\epsilon, m_{A'}) = (4.8 \times 10^{-6}, 0.01 \text{ GeV})$. The yellow star is the median maximum-likelihood estimate across these pseudo-experiments, while the black and blue contours denote the $1\sigma$ and $2\sigma$ confidence regions, respectively, defined under the asymptotic $\chi^2$ approximation with $k=2$ degrees of freedom. See Paper VII for the other three benchmarks.}
    \label{fig:likelihood_profileL}
\end{figure}

Once (if) a discovery is achieved, parameter estimation can be performed on the data. We explore the performance of maximum likelihood parameter estimation of all four benchmarks, using the profile likelihood ratio,
\begin{equation} \Delta(-2\log \mathcal{L}) = -2\log  \frac{\mathcal{L}(\vec{d}(\epsilon',m_{A'}')\,|\,m_{A'},\epsilon, \hat{\hat{\theta}}_S,\hat{\hat{\theta}}_B)}{\mathcal{L}(\vec{d}(\epsilon',m_{A'}')\,|\,\hat{m}_{A'}, \: \hat{\epsilon}, \:\hat{\theta}_S, \: \hat{\theta}_B)},
    \label{eq:profile_llr}
\end{equation}
where $\epsilon'$ and $m_{A'}'$ are the parameter values used in generating the pseudo-data, taken to be the parameters of the benchmark.
The pseudo-data are generated randomly, and we scan over $\epsilon$ and $m_{A'}$ and compute the profile likelihood. The resulting scan is plotted in \figref{fig:likelihood_profileL} for benchmark 1 under two background scenarios.
To represent a typical, rather than a single anomalous, outcome of the experiment, this
profile likelihood ratio is averaged over 25 independent pseudo-data sets generated
at the benchmark's true parameter values, marked by the red point.
The $1\sigma$ and $2\sigma$ confidence regions are defined as the set of parameter
points satisfying $\Delta(-2\log\mathcal{L}) < 2.30$ and $6.18$, respectively. These
thresholds follow from Wilks' theorem \cite{Wilks:1938dza,Wald1943}, under which $\Delta(-2\log\mathcal{L})$ is asymptotically distributed as a $\chi^2$ random variable with $k=2$ degrees of freedom, one for each of the two parameters of interest, $\epsilon$ and $m_{A'}$. The
values 2.30 and 6.18 are the thresholds enclosing the same cumulative probability,
under this $\chi^2_2$ distribution, as a $1\sigma$ ($68.3\%$) and $2\sigma$
($95.5\%$) interval would enclose for a single Gaussian-distributed parameter. This asymptotic
construction is itself only approximate at the low expected bin counts relevant to LDMX. We use the median maximum-likelihood point as the expected recovered parameter values.

Benchmark 3 also recovers the true parameters for both background scenarios. As expected, benchmarks 2 and 4 perform markedly worse, since the small expected signal yield leaves the resulting confidence regions dominated by Poissonian fluctuations, varying substantially in size and shape across different pseudo-data realizations and precluding a reliable parameter estimate.

\section{Bayesian Analysis}
Bayes theorem relates the probability of the model parameters given the data $\vec{d}$, $P(\bm{\psi},\bm{\theta}|\vec{d})$, to the likelihood of the data given the parameters, $\mathcal{L}(\vec{d}|\bm{\psi},\bm{\theta})$, and the prior probabilities of the parameters, $\pi(\bm{\psi},\bm{\theta})$,
\begin{equation}
    P(\bm{\psi},\bm{\theta}|\vec{d}) = \frac{\mathcal{L}(\vec{d}|\bm{\psi},\bm{\theta}) \pi(\bm{\psi},\bm{\theta})}{P(\vec{d})}.
\end{equation}
We use the likelihood defined in \eqnref{eq:likelihood}, and we take log-normal priors for $m_{A'}$ and $\epsilon$ over the range $m_{A'} \in \big[ 10^{-3}, 1 \big]$ GeV and $\epsilon \in \big[ 10^{-20}, 1.0 \big]$. For the nuisance parameters, we adopt log-normal priors for $\theta_S$ and $\theta_B$, and we introduce a hierarchical half-normal hyperprior on $\sigma_B$.

Sampling from the posterior is carried out using nested sampling, implemented via the package \texttt{dynesty}~\cite{Speagle:2019ivv}. We use the dynamic variant of nested sampling, in which the number of live points is adaptively allocated during the run rather than fixed in advance, concentrating computational effort on the regions of parameter space that have more posterior volume.
\begin{figure}
    \centering
    \includegraphics[width=\linewidth]{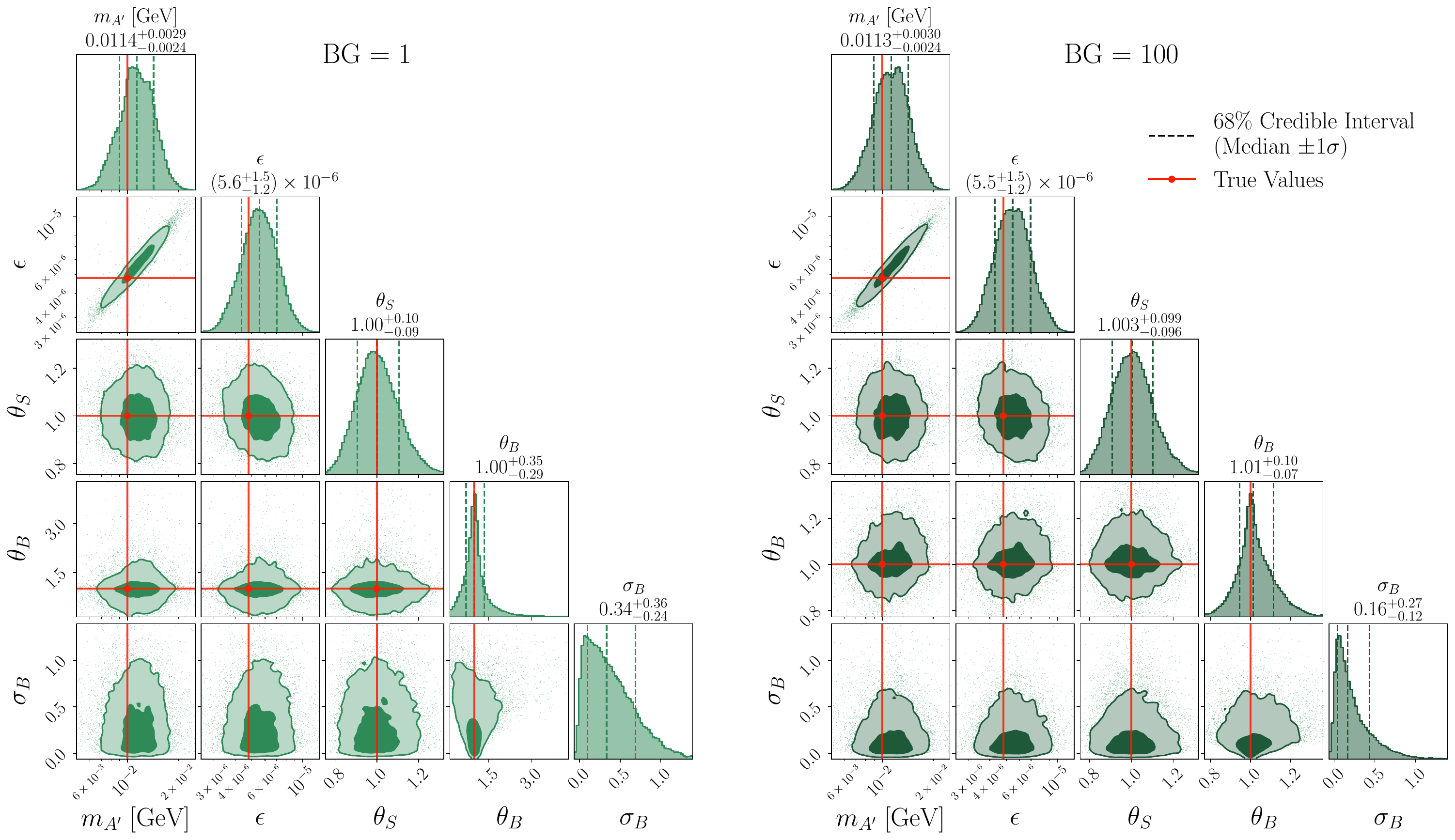}
    \caption{Corner plots with the marginalized one- and two-dimensional posterior distributions for benchmark 1 $(m_{A'},\epsilon) = (0.01 \:\text{GeV},4.8 \times 10^{-6})$, recovered via dynamic nested sampling, under background scenarios $N_{\rm BG}=1$ (left) and $N_{\rm BG}=100$ (right), each from a single representative pseudo-dataset \cite{Berger}. Here $g_f$ represents $\epsilon$. The diagonal panels show the one-dimensional marginal posterior for each of the five sampled parameters individually, obtained by integrating the full five-dimensional posterior over the remaining four; the off-diagonal panels show the corresponding two-dimensional marginal posteriors for each pair of parameters. The five sampled parameters are the two parameters of interest, the dark photon mass $m_{A'}$ and \gls{KM} $\epsilon$, together with the three nuisance parameters $\theta_S$, $\theta_B$, and $\sigma_B$, governing the signal normalization, background normalization, and background uncertainty width, respectively. In the diagonal panels, dashed vertical lines mark the posterior median and the bounds of the 68\% credible interval. In the off-diagonal panels, solid contours enclose the 68\% and 95\% credible regions. Red points and lines mark the true injected parameter values used to generate the pseudo-dataset. See Paper VII for the other three benchmarks.}
    \label{fig:likelihood_bayesian}
\end{figure}
The results of the Bayesian scan are summarized in \figref{fig:likelihood_bayesian} for benchmark 1, where corner plots of the posterior distributions are plotted for all five sampled parameters. One representative pseudo-data set is plotted for each background scenario.

Across all four benchmark points, the recovered posteriors divide naturally into two qualitatively distinct regimes, governed by how large the expected signal yield is relative to Poisson fluctuations in the background. At benchmarks 1 and 3, where the
signal yield is large, the posteriors on both $m_{A'}$ and $\epsilon$ are well-localized and approximately Gaussian, with credible intervals of order $10\%$ and posterior medians consistent with the injected values to within $1\sigma$ in both background scenarios. The two-dimensional posteriors additionally reveal a positive correlation between $m_{A'}$ and $\epsilon$, since a larger mediator mass suppresses the expected
signal rate in a way that a correspondingly larger \gls{KM} can compensate for.
At benchmarks 2 and 4, by contrast, the expected signal contribution is comparable to or smaller than the Poisson fluctuations expected in the background alone, so that the likelihood becomes effectively independent of $\epsilon$ over much of the prior range;
the resulting posteriors on $m_{A'}$ and $\epsilon$ are correspondingly uninformative,
spanning the full prior range with wide credible intervals. Taken together, these results indicate
that the Bayesian framework recovers precise and reliable parameter estimates along the $\mathcal{R}=2.5$ relic target, while along the $\mathcal{R}=2.2$ target, the small expected signal yield leaves the posterior dominated by the prior.

The Bayesian framework presented here also allows us to test the ability of LDMX in discriminating between different signal hypotheses, given they result in different recoil electron kinematics. In \secref{sec:models_bayesian_comparison} and Paper VII, we perform a model comparison analysis on various dark photon models \cite{Catena:2025fsl}. 
The Bayes factor is computed and serves as a quantitative measure of the relative support the data provide for one model over another,
\begin{equation}
   K_{12} = \frac{\mathcal{Z}_1}{\mathcal{Z}_2}
    = \frac{\int \mathcal{L}(\vec{d}\mid\vec{\nu}_{\mathcal{M}_1})\,\pi(\vec{\nu}_{\mathcal{M}_1})\,\mathrm{d}\vec{\nu}_{\mathcal{M}_1}}
    {\int \mathcal{L}(\vec{d}\mid\vec{\nu}_{\mathcal{M}_2})\,\pi(\vec{\nu}_{\mathcal{M}_2})\,\mathrm{d}\vec{\nu}_{\mathcal{M}_2}},
\end{equation}
where $\mathcal{Z}$ is the Bayesian evidence and $\vec{\nu}_{\mathcal{M}}$ is the set of model parameters for model $\mathcal{M}$.

\section{Validating a Potential Signal}

A signal observed at LDMX, in the form of an excess in the electron recoil energy and transverse momentum distributions over the expected background, would on its own only establish that an invisible particle carrying a significant fraction of the beam energy has been produced. It would not, by itself, establish that this particle plays the required role of the mediator for the DM in the cosmological history of the universe. Confirming the DM origin of such a signal therefore requires independent corroboration from outside LDMX itself. This section develops a strategy for obtaining that corroboration by combining a hypothetical LDMX signal with data from a next-generation DM direct detection experiment \cite{Catena:2024yed}. The goal is to answer the following question: what direct detection exposure is required to assert the compatibility of the hypothetical LDMX signal with a DM origin?

\begin{figure}
    \centering
    \includegraphics[width=0.75\linewidth]{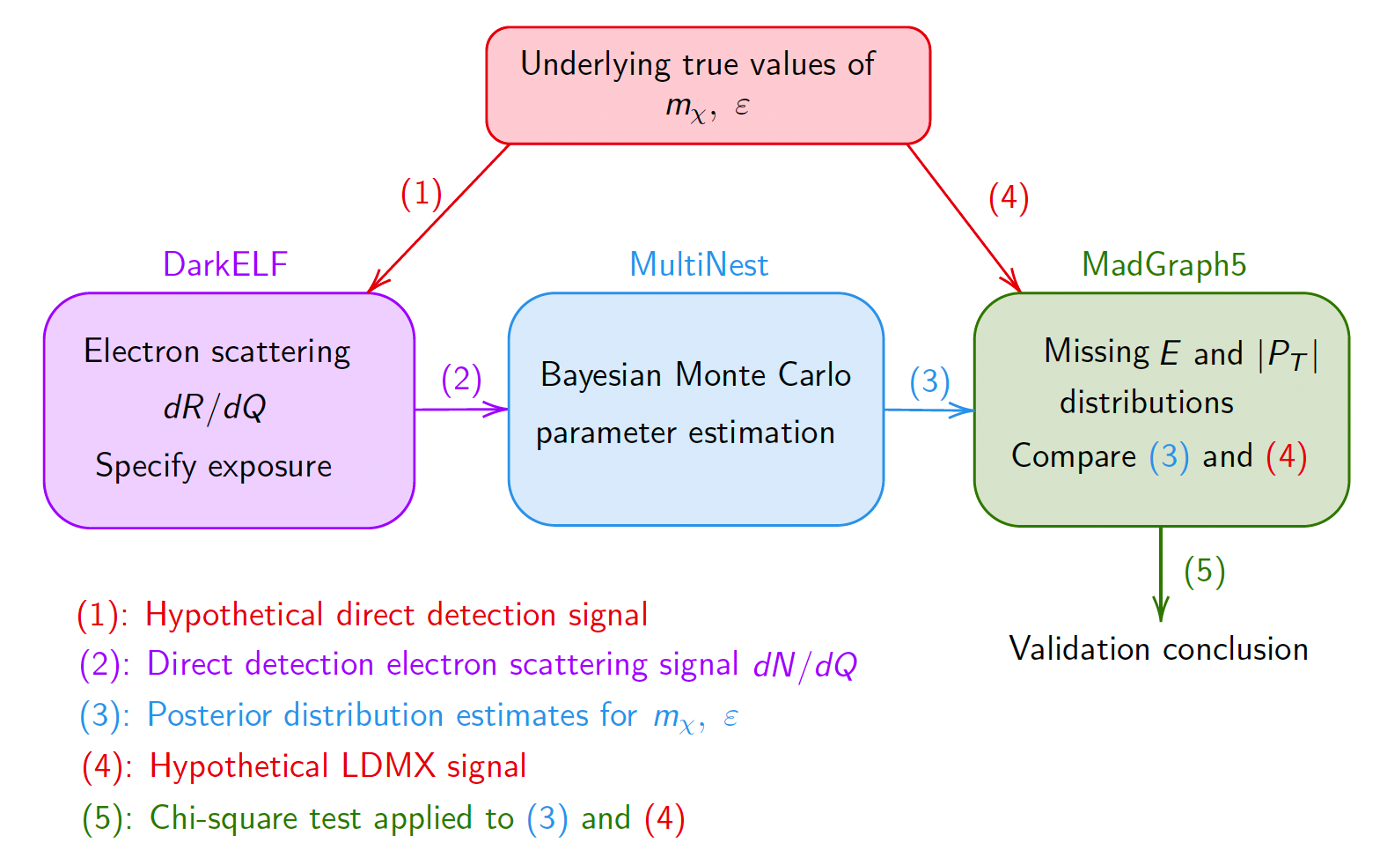}
    \caption{Schematic of the methodology for validation of a potential LDMX signal with direct detection data \cite{Lund2024}. \textsc{DarkELF} computes the direct detection signal, \textsc{MultiNest} samples the posterior \cite{Feroz:2008xx}, and \madgraph simulates the LDMX signal \cite{Alwall:2011uj}.}
    \label{fig:likelihood_validation_schematic}
\end{figure}

The strategy proceeds in four steps. First, a signal is recorded at LDMX, observed (or, as in this analysis, simulated) as an excess in the electron recoil energy and transverse momentum distributions. Second, a direct detection experiment searching for electron recoils is operated with a given exposure, generating its own independent dataset that is sensitive to the same underlying DM parameters. Third, a posterior probability distribution function for the DM mass and coupling is extracted from the direct detection data, and used to predict what the LDMX electron recoil energy and transverse momentum distributions should look like if the two signals share a common origin. Fourth, this prediction is compared against the distributions actually observed at LDMX using a chi-square test. The minimum direct detection exposure required to confirm a common origin, were one to exist, rather than erroneously concluding that the two signals are unrelated, is then extracted.
Figure~\ref{fig:likelihood_validation_schematic} outlines the methodology of Paper III.

The direct detection side of this comparison uses \textsc{DarkELF} \cite{Knapen:2021bwg}, which calculates the rate of DM induced electronic transitions in a target material, here taken to be silicon. Each transition produces a discrete number of electron-hole pairs, $Q$. For a given DM mass and DM-electron  cross section, related to $\epsilon$ once the mediator mass is fixed\footnote{$\mathcal{R}=3$ was chosen before the latest results from DAMIC-M \cite{DAMIC-M:2025luv} and PandaX-4T \cite{PandaX:2025rrz} were released, which exclude this mass ratio.} via $\mathcal{R}=m_{A'}/m_\text{DM} = 3$ and $\alpha_D$ is fixed to 0.5, \textsc{DarkELF} predicts the expected number of events in each $Q$ bin. A simulated direct detection dataset for a given experimental exposure is then obtained by randomly sampling, from a Poisson distribution with this expected mean, the number of events actually observed in each $Q$ bin.
Given this simulated dataset, a posterior probability for $(m_\text{DM},\sigma_e)$ is obtained via Bayesian inference.
The posterior pdf is then used to predict what the LDMX electron recoil energy and transverse momentum distributions should look like.

The predicted distributions, $N^{E_e}_{{\rm DD},i}$ and $N^{p_T}_{{\rm DD},i}$, are compared against the distributions actually recorded at LDMX, $N^{E_e}_{{\rm LDMX},i}$ and $N^{p_T}_{{\rm LDMX},i}$, using the test statistic,
\begin{align}
\chi^2_{\rm TS}= \sum_{i=1}^{n_b} \frac{\left(N^{E_e}_{\textrm{LDMX},i} - N^{E_e}_{\textrm{DD},i}  \right)^2}{N^{E_e}_{\textrm{LDMX},i} + N^{E_e}_{\textrm{DD},i}}
+  \sum_{i=1}^{n_b} \frac{\left(N^{P_T}_{\textrm{LDMX},i} - N^{P_T}_{\textrm{DD},i}  \right)^2}{N^{P_T}_{\textrm{LDMX},i} + N^{P_T}_{\textrm{DD},i}}\,,
\label{eq:chi2}
\end{align}
where $n_b$ is the number of $p_T$ and $E$ bins. 

The quantity of interest is the threshold exposure, defined as the exposure at which the observed value of $\chi^2_{\rm TS}$ equals the critical value $\chi^2_*$ corresponding to a 95\% confidence level rejection of the null hypothesis. Below the threshold exposure, the direct detection experiment has not yet accumulated enough exposure to distinguish a genuine common origin from statistical independence: the chi-square test would reject the DM origin of the LDMX signal even in a scenario where the two datasets do, in fact, share a common origin. Above the threshold exposure, this confusion no longer occurs, and the test is capable of correctly identifying the two datasets as compatible. It is worth being precise about what ``compatible" means in this context: failing to reject the null hypothesis establishes that the two distributions could share a common DM origin, not that they definitely do.

\begin{figure}[h]
    \centering
    \includegraphics[width=0.75\linewidth]{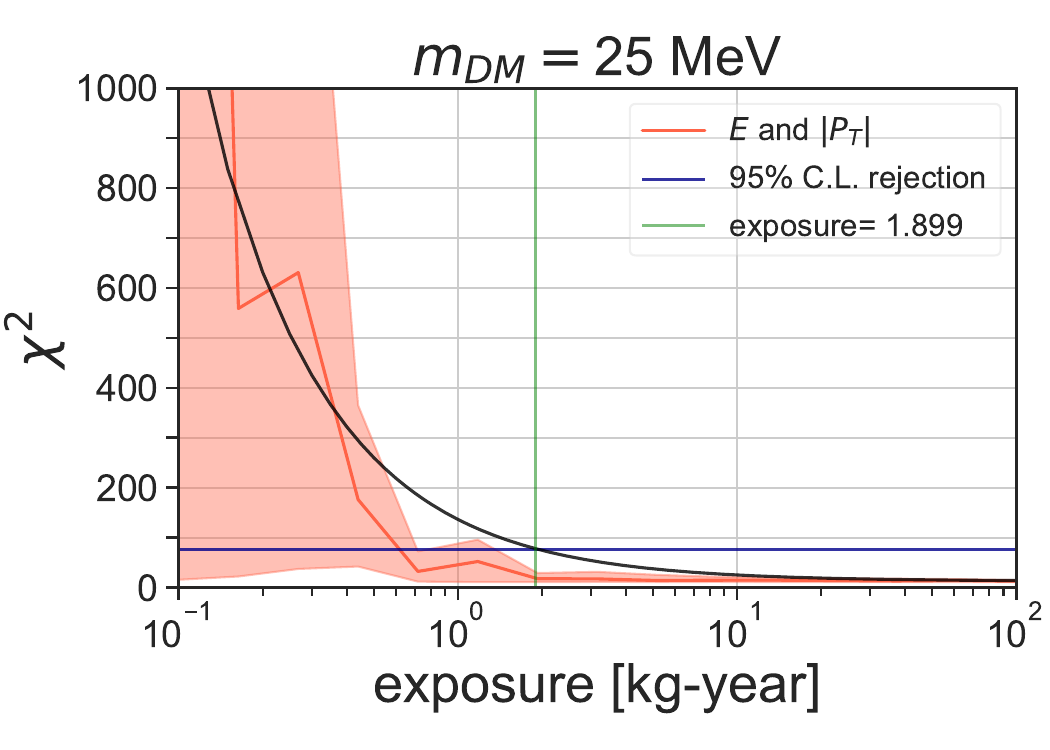}
    \caption{Chi-square test statistic from \eqnref{eq:chi2}, as a function of the direct detection exposure, for a complex scalar DM particle mass of $m_\chi=25$~MeV. The horizontal blue line, $\chi^2_*$, the critical value corresponding to a 95\% C.L. rejection of the common DM origin of the LDMX and direct detection distributions. The vertical green line marks the threshold exposure, read off from the intersection of this fitted curve with $\chi^2_*$. To the right of this line the chi-square test would infer the compatibility of the two distributions, while to the left, the test would erroneously reject this common origin due to the direct detection exposure being insufficient, irrespective of whether the LDMX signal is in fact of DM origin.}
    \label{fig:likelihood_validation}
\end{figure}
Figure~\ref{fig:likelihood_validation} shows $\chi^2_{\rm TS}$ as a function of the direct detection exposure, for a benchmark DM mass of $m_\chi=25$ MeV lying on the thermal relic target, with both the electron recoil energy and transverse momentum information included. At small exposure, $\chi^2_{\rm TS}$ lies well above the critical value $\chi^2_*$ (blue 95\% C.L. horizontal line), reflecting the fact that a direct detection experiment with little exposure has not yet collected enough events to meaningfully constrain $(m_\text{DM},\epsilon)$, so that the predicted LDMX distributions remain far from the true signal purely as a consequence of statistical noise. As the exposure increases, the direct detection posterior narrows around the true parameter values, the predicted LDMX distributions converge towards the true signal, and $\chi^2_{\rm TS}$ falls correspondingly. The threshold exposure, at which $\chi^2_{\rm TS}$ crosses $\chi^2_*$, is read off from the intersection of a fit to the mean $\chi^2_{\rm TS}$ (from ten realizations of the direct detection data) curve with the critical value, and for this benchmark is found to be $\simeq 1.9$~kg-year.
This value is the largest of the threshold exposures found across the DM mass range considered, which spans 0.0013~kg-year at $m_\chi=4$~MeV to 1.9~kg-year at $m_\chi=25$~MeV.  Even at the upper end of this range, the required exposure remains within reach of near future direct detection experiments: by comparison, the current exposure achieved by DAMIC-M is approximately $0.00356 \text{ kg-year}$\cite{DAMIC-M:2025luv}.

\part[Beyond The Standard Paradigm: Global Fits, Dark Sector Theories, and Nuclear Modelling]
     {Beyond The Standard Paradigm: \\[0.4em]
      {\LARGE\bfseries Global Fits, Dark Sector Theories, and Nuclear Modelling}}
 
  \chapter{Global Fits for Light Dark Matter}
\label{chap:GlobalFits}

The conventional way of determining excluded regions of the model parameter space for new physics scenarios, typically in the coupling-mass plane, is to draw separate exclusion limits at a given \gls{cl} for each dataset, creating a union of exclusion contours. The rate at which a parameter value on the exclusion contour is falsely excluded for one experiment is given by $1-\text{C.L.}$, where $\text{C.L.}$ is taken commonly in particle physics to be $90\%$ or $0.9$. However, when combining multiple experiments, this ``patching" of exclusion limits results in a false reporting of the C.L.. The actual false parameter exclusion rate, or error, is $1-\text{C.L.}^N$, where $N$ is the number of experiments \cite{AbdusSalam:2020rdj}. For a $\text{C.L.}=0.9$ and 3 experimental constraints, the error is then $1-0.9^3 \approx 0.27$, a larger error than the 0.1 that one could naively expect.

In addition to the desire for a more statistically robust method of handling multiple experimental constraints, the large parameter space that comes with many DM models can be non-trivial to navigate. In particular, making statements about the regions of parameter space still unexcluded by current data is a multidimensional problem. The approach explored in this thesis is to instead perform so called \textit{global fits} of the model parameter space, where the likelihoods of many constraints are combined in a statistically consistent framework.
This chapter is based on the work of paper II \cite{Balan:2024cmq}.

\section{Combining Likelihoods}

When performing global fits, we consider the total likelihood, or probability of the data given the model and nuisance parameters, which is simply the product of every relevant likelihood,
\begin{equation}
    \mathcal{L}_{\text{total}} = \mathcal{L}_{\text{DD}} \times \mathcal{L}_\text{collider} \times \mathcal{L}_\text{ID}\times \mathcal{L}_\text{cosmology}\times  \cdots
\end{equation}
such as the likelihood from direct detection, collider, indirect detection, and cosmology, which we take to be independent of each other. Depending on the statistical nature of the experiment or observation, the likelihood will take a certain form.
For example, the likelihood for a counting experiment such as direct detection and accelerator experiments takes the form of a Poisson distribution, which for $d$ observed events is given by,
\begin{equation}
    \mathcal{L}(d) = e^{-(s+b)}\frac{(s+b)^d}{d!},
\end{equation}
where $s$ is the number of signal events predicted from theory and $b$ is the number of expected events from known background processes.

\subsection{\GAMBIT}
\GAMBIT v2.5.0 is a publicly available software package which performs global fits on DM models, with the goal of being flexible and extendable to other particle physics models and likelihood analyses \cite{GAMBIT:2017yxo}. Within \GAMBIT there exists a growing number of likelihoods categorized into modules that can all be analyzed at once in the context of a particular DM model. Within these modules there are different analyses involving backends to simulation and calculation software with varying levels of complexity. Of particular interest to the accelerator-based probes in Paper II, the ColliderBit module calculates high energy collider physics observables \cite{GAMBIT:2017qxg}.

\section{Light Dark Matter Analysis}
With the many existing and future probes of sub-GeV DM, each dominating in their own region of parameter space in a complementary way, global fits are able to provide us with a status update on this class of DM candidates. We determine the viable regions of parameter space that are not already ruled out by experiments and observations, and compare them to projections of future experiments. In particular, we consider two theoretical scenarios: a complex scalar and a Dirac fermion DM candidate, each interacting with the SM fermions through a dark photon $\Ap$ mediator as described in \secref{sec:minimal_DM_models}.
We consider the case where DM has a mass from 1 MeV to 1 GeV and $m_\Ap > 2m_\text{DM}$, therefore dark decays of the $\Ap$ are allowed and dark bremsstrahlung occurs on-shell.

DM annihilations to SM species would inject energy into the photon-baryon plasma in the early Universe, leading to potential deviations in CMB measurements. Limits on this exotic energy injection are placed on the DM annihilation cross section. 
In addition, DM annihilating in our galaxy could produce secondary signals: the resulting electron-positron pairs would up-scatter low energy photons via inverse Compton scattering, producing X-rays detectable by telescopes \cite{Cirelli:2023tnx}\footnote{It was found recently in \cite{Balaji:2025afr} that these X-ray constraints from \cite{Cirelli:2023tnx} are much weaker due to a mischaracterization of the instrument's solid angle.}. CMB constraints remain, nevertheless, the dominant indirect detection constraints on sub-GeV DM, placing particularly stringent limits on certain sub-GeV DM models.
In particular, Dirac fermion DM, where the non-relativistic thermally averaged cross section, $\langle \sigma v \rangle_{\chi \bar{\chi} \to f\bar{f}}$\footnote{The thermally averaged annihilation cross section in the non-relativistic limit can be expanded in powers of $v^2$ in a partial-wave expansion $\langle \sigma v \rangle \approx a + bv^2 + \cdots$, where $a$ is the s-wave term, $b$ the $p-$wave, etc. The spin of the incoming and outgoing particles can forbid certain terms due to conservation of angular momentum. For scalar DM, $s-$wave annihilations through an s-channel $\Ap$ are forbidden, leading to p-wave or $v^2$ dependent annihilations.}, for DM annihilation to SM, is independent of the velocity. This leaves a large part of the parameter space excluded, including parts where the relic density constraint is satisfied. 
However, $\langle \sigma v \rangle_{\chi \bar{\chi} \to f\bar{f}} \sim v^2$ for complex scalar DM, leading to significantly weaker constraints from CMB and X-ray measurements.

We explore three ways in which these tight constraints on Dirac fermion DM can be relaxed. The first way is through resonant enhancement of the annihilation cross section during freeze-out, which occurs if $2 m_{\text{DM}} \sim m_\Ap$. If DM freezes-out at the time/temperature at which there is a resonance in $\langle \sigma v \rangle_{\chi \bar{\chi} \to f\bar{f}}$, more DM annihilations will occur before freezing-out, leading to an overall smaller abundance -- therefore smaller couplings are required to reproduce the observed abundance. We define the resonance parameter, 
\begin{equation}
    \epsilon_R \equiv \frac{m_\Ap^2 - 4m_{\text{DM}}^2}{4m_{\text{DM}}^2},
\end{equation}
to parameterize the closeness to resonance. 
The next way is through an asymmetry between the number of DM particles and anti-DM particles by letting,
\begin{equation}
    \eta_\text{DM} \equiv \frac{n_{\chi}-n_{\bar{\chi}}}{s},
\end{equation}
be different from 0, where $s$ is the entropy density and $n_\chi$ is the DM number density.
This leads to a suppression of the expected indirect detection signals since only the symmetric component can annihilate. The last way we consider to relax the strong constraints is if the DM particle only constitutes a portion of the total DM relic abundance, namely sub-component DM. The relative abundance is, 
\begin{equation}
    f_{\text{DM}} \equiv \frac{\Omega_{\text{DM}}h^2}{\Omega_{\text{DM,obs}}h^2} < 1.
\end{equation}

\begin{figure}[h]
    \centering
\includegraphics[width=0.7\linewidth]{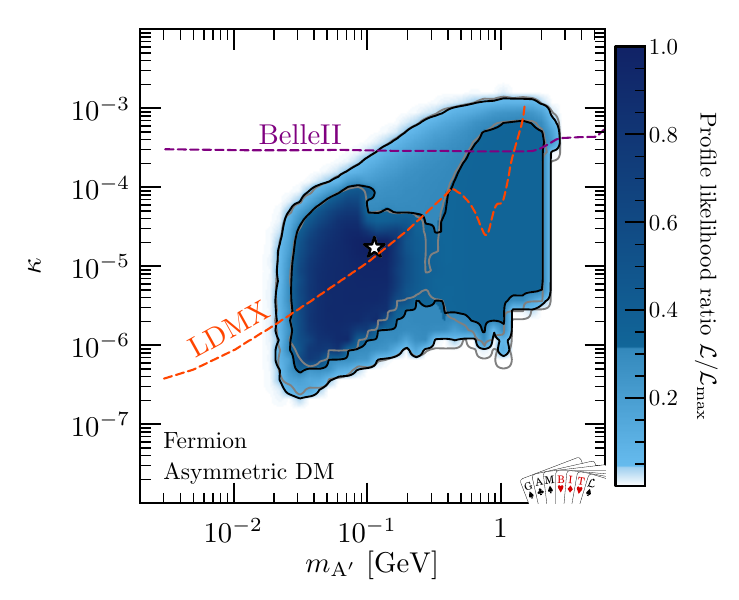}
    \caption{Profile likelihood ratio of Dirac fermion DM as a function of \gls{KM} parameter $\kappa$ and the $\Ap$ mass, with 1$\sigma$ and 2$\sigma$ confidence regions drawn in black and shaded in blue for $\Omega h^2 \approx 0.12$, and in grey for $\Omega h^2 < 0.12$ (subcomponent DM). Here we allow for an asymmetry in the DM to anti-DM abundance. The star indicates the best-fit. The expected exclusion bound from the future experiments LDMX and Belle-II are drawn in dashed red and purple, respectively. Taken from Paper II \cite{Balan:2024cmq}.}
    \label{fig:global_fit}
\end{figure}

For both DM models, we combine constraints from cosmology, astrophysics, accelerators, and direct detection experiments in both frequentist and Bayesian analyses. In \figref{fig:global_fit}, the profile likelihood from the frequentist analysis is plotted as a function of the \gls{KM} strength parameter $\kappa$\footnote{We use $\kappa$ (called $\epsilon$ everywhere else in this thesis) in this chapter to facilitate easy comparison with Paper II.} and the dark photon mass, $m_\Ap$, with the $1\sigma$ (68\% confidence level) and $2\sigma$ (95 \% confidence level) contours shaded in blue, and the best fit point represented by a star.
Small dark photon masses are not preferred due to the cosmological observations. In particular, observations of light element abundances constrain the amount by which \gls{bbn} can be altered by \gls{bsm} physics. When the $\Ap$ is too light, it decouples from the thermal bath too late, contributing too long to the energy density and hence the number of effective degrees of freedom.
Large couplings are excluded by fixed target experiments, most dominantly NA64. The positively sloped shape of the $1\sigma$ contour comes from the number of expected events at fixed-target experiments falling off as an inverse power of $m_\Ap$, with flattening at $\sim 1 \: \text{GeV}$ attributed to limits from the BaBar experiment. Below a certain $\kappa$, DM would be overproduced in the universe, inconsistent with Planck measurements of the relic abundance -- explaining the excluded region for small $\kappa$. The prior range on $m_\text{DM}$ extends only to 1 GeV, and since $m_{A'}=2m_\text{DM}\sqrt{\epsilon_R+1}$, the right side of the confidence regions is at $m_{A'}\sim 2$ GeV depending on the value of $\epsilon_R$ taken to maximize the profile likelihood.
Projections of the future experiments LDMX, described in \chapref{chap:LDMX}, and Belle-II are drawn as dashed curves, demonstrating the large sensitivity potential. LDMX is expected to probe 64\% of the posterior volume.

Bayesian analyses are also given in Paper II, where the main difference is that finely tuned parameters are penalized and instead posterior mean and median values strongly depend on the volume of the posterior support.

\chapter{Light Dark Matter Theories}
\label{chap:Models}

The realm of currently possible theoretical explanations of DM is overwhelming and arguably infinite. As discussed in \chapref{chap:ParticleDarkMatter}, the existing literature explores a wide range of models describing DM's particle nature. Within the literature for sub-GeV DM with a dark photon mediator, the standard paradigm (and the focus of \partref{part:StandardParadigm}) has centered on spin-0 and spin-1/2 DM candidates coupled to a dark photon through kinetic mixing. 

DM could equally well be a new spin-1 boson, $X$, which interacts with SM particles through a dark photon mediator, or more generally, through multiple mediators \cite{Catena:2023use,Catena:2022fnk,Dent:2015zpa}. Likewise, the dark photon itself need not couple to SM fermions solely via kinetic mixing: higher dimensional operators can endow it with electromagnetic-type moments analogous to magnetic dipole, electric dipole, anapole, and charge radius interactions familiar from within the SM, giving rise to qualitatively distinct phenomenology at fixed-target experiments.

This chapter moves beyond the standard paradigm explored in \partref{part:StandardParadigm} to examine these two extensions. We explore the phenomenology of light vector DM in \secref{sec:VectorDM}, and the signatures of dark photon higher-order electromagnetic moments at fixed-target experiments in \secref{section:DarkMoments}.

\section{Vector Dark Matter}
\label{sec:VectorDM}

Spin-1 (vector) DM constitutes a theoretically motivated and increasingly explored class of sub-GeV DM candidates \cite{Hambye:2008bq,Choi:2019zeb,Hu:2021pln,Nomura:2020zlm,Nomura:2021aep,Nomura:2021tmi,Ramos:2021txu,Abe:2020mph,Chu:2023zbo}. Such states arise naturally in models with an extended dark gauge sector. Vector DM candidates have been examined across a range of phenomenological contexts, including their direct detection signatures \cite{Krnjaic:2023nxe,Catena:2019hzw,Catena:2018uae}, potential gravitational wave signals from an associated dark sector phase transition \cite{Benincasa:2025tdr,Belyaev:2025zse}, and production via freeze-in rather than thermal freeze-out \cite{Krnjaic:2022wor}. At fixed-target experiments, previous work has largely focused on the on-shell production of vector DM, where the mediator is heavier than the DM and decays invisibly into a DM pair. Extending this framework to the complementary, less explored regime with an inverted mass hierarchy between the DM and mediator opens up qualitatively new phenomenology, including suppressed off-shell DM production alongside enhanced visible mediator decay channels.
 
We first study a set of \textit{simplified} spin-1 DM models \cite{Catena:2023use}, where simplified models extend the SM with only a DM candidate and a mediator with interactions parametrized by a set of operators consistent with the model building assumptions (those that can be generated by either Abelian or non-Abelian dark sectors), without specifying a \gls{uv} completion \cite{Dent:2015zpa,Catena:2017xqq,Catena:2019hzw}.
A key advantage of this simplified model framework is that it enables direct comparison with the simplified spin-0 and spin-1/2 benchmark models presented in \secref{sec:minimal_DM_models}. The relic abundance in all cases is set by pair annihilation into SM particles via the dark photon mediator, the resulting thermal targets occupy the same parameter space, and the fixed-target search sensitivity can be straightforwardly compared across DM spins.
These simplified models are, in some regions of parameter space, subject to strong unitarity violations at high energies, motivating the construction of UV-complete vector DM models that embed the DM candidate within an extended dark gauge and scalar sector, together with an explicit mass generation mechanism, rather than introducing DM and its mediator as isolated new degrees of freedom. We study the phenomenology of these models, determining the experimental sensitivity and regions of parameter space consistent with the observed relic density \cite{Planck:2018vyg}. This chapter is based on the work presented in \cite{Catena:2023use, Banerjee:2025ejz}, where we refer to Papers I and VI for the details.

\subsection{Simplified Models}
Simplified models are theories which minimally extend the SM, introducing only the new degrees of freedom needed to describe the physics accessible to experimental searches \cite{LHCNewPhysicsWorkingGroup:2011mji}. Unlike effective field theories, simplified models include a particle mediator between the visible and dark sectors. However, they still neglect UV-completeness and the mass giving mechanism that a complete theory would include.

We extend the SM with one complex vector field $X^\mu$ and one real vector field $A'^\mu$: the first is the DM candidate and the second is the mediator \cite{Catena:2017xqq}, remaining agnostic about the underlying gauge structure. Since neither field is charged under the SM gauge group, a possible way to generate mediator-SM couplings is through kinetic mixing between $A'^\mu$ and the SM hypercharge field \cite{Fabbrichesi:2020wbt}.
The simplicity of this extension allows for direct comparison between scalar and fermionic DM models which have already been investigated in the literature \cite{Berlin:2018bsc}. Writing down the most general, Lorentz invariant, and of dimension 4, Lagrangian describing the interactions between $X^\mu$ and $\Ap^\mu$ gives,
\begin{align}
    \mathscr{L} \supset &-\left[i b_5 X_\nu^\dagger \partial_\mu X^\nu A'^\mu + b_6 X_\mu^\dagger \partial^\mu X_\nu A'^\nu + \text{h.c.} \right]
    \nonumber \\
    &-\left[b_7 \epsilon_{\mu \nu \rho \sigma} \left( X^{\dagger \mu} \partial^\nu X^\rho \right) \Ap^\sigma + \text{h.c.} \right] \nonumber \\ 
    &- h_3 \Ap_\mu \bar{f}\gamma^\mu f , 
    \label{eq:Lsimplified}
\end{align}
where $b_i$ are the dark sector couplings, $h_3$ is the SM-mediator coupling, and $f$ is the SM fermion field which includes electrons, muons, taus, and quarks, and does not include neutrinos. The dark sector couplings, without loss of generality, are such that $b_5$ can be taken to be real, while $b_6$ and $b_7$ are complex.
The first and second lines of \eqnref{eq:Lsimplified} can arise in models where $X^\mu$ is part of a non-abelian and abelian field theory, respectively. The third line describes the interaction between the $\Ap$ and the SM fermions arising from the \gls{KM}, as described in \secref{subsection:DarkPhotons}. We consider the cases where only one of the $b_i$ couplings is non-zero, and $\epsilon$ is always a non-zero free parameter of the model. To facilitate easy comparison between other DM models considered in previous literature, we label the dark coupling as $g_D$, where $g_D$ is one of $b_5$, $\Re[b_6]$, $\Im[b_6]$, $\Re[b_7]$, or $\Im[b_7]$, and $\alpha_D \equiv \frac{g_D^2}{4\pi}$. $\alpha_D = 0.5$ throughout this chapter and Paper I.

\subsubsection*{Violations of Perturbative Unitarity}

By imposing conservation of probability in quantum mechanics, the S matrix must be unitary, $S^\dagger S = 1$. From this, it follows that a bound arises on the partial wave amplitudes, $\mathcal{M}_{if}^J$, which are scattering amplitudes with fixed total angular momentum $J$. For example, we consider the process $i_1 i_2 \to f_1 f_2$, where the following bounds come from the unitarity condition \cite{Chanowitz:1978mv},
\begin{equation}
    \begin{aligned}
        & 0 \leq \Im(\mathcal{M}_{ii}^J) \leq 1 \\
        & |\Re (\mathcal{M}_{ii}^J)| \leq \frac{1}{2},
    \end{aligned}
\end{equation}
where $\mathcal{M}_{if}^J$ is the diagonalized matrix of matrix elements of all possible two particle states, and $\mathcal{M}_{ii}^J$ are the eigenvalues \cite{Schuessler:2007av,Kahlhoefer:2015bea}.
For a given theory, if this \textit{unitarity bound} is violated in any process at tree level (leading order in perturbation theory), this signifies the necessity of either including higher-order terms (perturbativity has broken down) or that the theory requires additional fields -- each of which could restore unitarity. The former case arises for incomplete theories such as effective field theories and simplified models at high scattering energies, couplings, or for particular masses.
Due to our simplified spin-1 DM models having momentum dependence introduced from the vertices and from the longitudinal component of the polarization vectors, our simplified models violate unitarity at certain scattering energies. We consider violations of unitarity from DM self scattering \cite{Chang:2023cki} and DM-$e^-$ scattering.

Figs. 4 and 5 of Paper I \cite{Catena:2023use} show the resulting calculated unitarity ``bound", or region of theoretical validity of the simplified model. This bound is strong at relevant energies, and even stronger at higher energies, highlighting the UV incompleteness of the theory.

\subsection{Non-Abelian SIMPs}
With motivation to avoid unwanted unitarity violations as described in the previous section, we consider a second, more complete model for spin-1 DM where the SM is extended by a local $SU(2)_X\times U(1)_\Zp$ symmetry group, with gauge couplings $g_X$ and $g_{Z'}$, under which the SM species are not charged \cite{Choi:2019zeb}. In this model, DM is a \gls{SIMP}, characterized by $3 \to 2$ annihilations of the DM in the early universe setting the relic density \cite{Choi:2017zww}. It was found in the work \cite{Hochberg:2014dra} that SIMPs can achieve the correct relic density with masses $\sim 0.1$ GeV.

This model includes an extended Higgs sector with a scalar singlet, $S$, and a scalar $H_X$ which can take several representations under $SU(2)_X$. The representation $I$ of $SU(2)_X$ in which $H_X$ transforms can be e.g. a doublet ($I=1/2$), a triplet ($I=1$), and a quadruplet ($I=3/2$). The gauge bosons associated with the new $SU(2)_X$ symmetry are $X_{\mu,i}$ ($i=1,2,3)$, or equivalently $X_\mu \equiv (X_{\mu,1} + i X_{\mu,2})/\sqrt{2}$ and $X_\mu^\dagger \equiv  (X_{\mu,1} - i X_{\mu,2})/\sqrt{2}$, and the gauge boson of $U(1)_{Z'}$ is $Z'_\mu$. The \glspl{vev}, $v_S$ and $v_X$, of the dark scalars, $S$ and $H_X$ respectively, spontaneously break the $SU(2)_X\times U(1)_\Zp$ symmetry, giving the dark gauge bosons mass and combining $X_{\mu,1}$ and $X_{\mu,2}$ to give the complex gauge boson $X_\mu$, the DM candidate, and generating two linear combinations of $Z'_\mu$ and $X_{\mu,3}$ which we denote by $\tilde{Z}'_\mu$ and $\tilde{X}_{\mu,3}$. 

A gauge \gls{KM} is introduced, leading to interactions between $\tilde{Z}'$, $\tilde{X}_{3}$, and SM fermions. Due to $v_X$, $\tilde{X}_{\mu,3}$ has a different mass from $X_{\mu}$, therefore there are two mediators between the dark and visible sectors, the $\tilde{Z}'$ (analogous to the dark photon) and the $\tilde{X}_{3}$.

By construction this model is not subject to the unitarity violations described in the previous section.
The interaction Lagrangian is given by,
\begin{equation}
  \begin{aligned}
  \mathscr{L} \supset &-i g_X\cos\theta'_X \left[
\left(\partial^\mu X^\nu -\partial^\nu X^\mu\right)
 X^\dagger_\mu {\tilde X}_{3,\nu} -
\left(\partial^\mu X^{\nu\dagger} -\partial^\nu X^{\mu\dagger}\right)
 X_\mu {\tilde X}_{3,\nu} \nonumber \right. \\
& \left. + X_\mu X^\dagger_\nu\left(\partial^\mu {\tilde X}^\nu_3 -\partial^\nu {\tilde X}^\mu_3\right)
\right] \nonumber \\
&-i g_X\sin\theta'_X \left[
\left(\partial^\mu X^\nu -\partial^\nu X^\mu\right)
 X^\dagger_\mu {\tilde Z}'_{\nu} -
\left(\partial^\mu X^{\nu\dagger} -\partial^\nu X^{\mu\dagger}\right)
 X_\mu {\tilde Z}'_{\nu} \nonumber \right. \\
& \left. + X_\mu X^\dagger_\nu\left(\partial^\mu {\tilde Z}^{\prime\nu} -\partial^\nu {\tilde Z}^{\prime\mu}\right)\right] \\
&-e \varepsilon \cos (\theta_X') \, \tilde{Z}^\prime_\mu \, \bar{f}\gamma^\mu f  + e \varepsilon \sin (\theta_X') \, \tilde{X}_{3 \mu} \,\bar{f}\gamma^\mu f,
\end{aligned}
\label{eq:SIMPL}
  \end{equation}
where $\tan(2\theta'_X) = \frac{2\cos \theta_X \sin \theta_X}{\cos \theta_X^2-\alpha \sin \theta_X^2}$ with $\alpha \equiv 1+q_S^2v_S^2/(I^2v_X^2)$ and the $U(1)_{Z'}$ charge of S $q_S$, and where $\sin \theta_X = \frac{g_{Z'}}{\sqrt{g_X^2 + g_{Z'}^2}}$.  We take $ \theta_X' \ll 1$, therefore $Z'$s are dominantly produced in dark bremsstrahlung events at fixed target experiments. Similarly to the simplified spin-1 DM models, we set $\frac{g_X^2}{4\pi} = 0.5$ such that the dark sector coupling is much larger than the SM coupling $\epsilon e$.

\subsubsection*{Relic Density}
With sufficient couplings between the SM fermions and the dark sector, DM would have once been in thermal equilibrium with the SM bath -- leading to the freeze-out mechanism for DM production. The dominant DM annihilation processes are $X^+ X^- X^+ \to X^+  \tilde{X}_3$ and the so called secluded annihilation $X^+ X^- \to \tilde{X}_3 \tilde{X}_3$, where the former is dominant if $1 < m_{\tilde{X}_3}/m_X < 2$, and the latter if $m_{\tilde{X}_3}/m_X < 1$. This is in contrast to s-channel DM annihilation to SM fermions through $\tilde{X}_3$, which is suppressed since $g_X \gg \epsilon e$. This leaves the relic density contours, or so called \textit{thermal targets} independent of $\epsilon$, only varying with the dark coupling and the masses.

However, we must be careful to check whether DM and the $\tilde{X}_3$ remain in kinetic equilibrium with the SM during freeze-out; otherwise structure formation constraints are not satisfied. DM is easily able to stay in kinetic equilibrium through elastic scattering between DM and $\tilde{X}_3$, given that $\tilde{X}_3$ remains in kinetic equilibrium, since $g_X$ is large. $\tilde{X}_3$, through its decays to SM species, is able to stay in equilibrium with the SM. We impose that,
\begin{align}
n_{\tilde{X}_3,\rm eq}(T_f) \Gamma_{\tilde{X}_3} > H(T_f)n_{X,\rm eq}(T_f)\,,
\label{eq:kin}
\end{align}
where $H(T_f)$ is the Hubble rate at the freeze-out temperature $T_f$, and the equilibrium number density of a species $i$ as a function of temperature $T$ is given by,
\begin{equation}
n_{i, \rm eq}(T) = \frac{\gamma_i m_i^2 T}{2\pi^2} K_2(m_i/T),
\end{equation}
where $\gamma_i$ is the number of internal degrees of freedom and $K_2$ is the modified Bessel function of the second kind. In \eqnref{eq:kin}, $\Gamma_{\tilde{X}_3}$ is the total decay rate of $\tilde{X}_3$, for which the expression is given in~\cite{Choi:2019zeb}.

We consider the particular choice of parameters where $m_{\tilde{X}_3}/m_X \approx 2$, such that the $3 \to 2$ process is dominant. There is a resonance in the $3 \to 2$ cross section when $m_{\tilde{X}_3}/m_X \approx 2$, therefore the relic density drops to zero as the ratio approaches 2 from the left. Since the cross section decreases with increasing DM mass, larger DM masses give larger relic densities. However, as long as the DM mass is sufficient, very close to the resonance the relic density will be in agreement with Planck measurements \cite{Planck:2018vyg}. The relic density is also independent of $\epsilon$ since the annihilation process only contains dark sector particles. Therefore, there is an entire region in $\epsilon$ vs $m_X$ parameter space consistent with the correct relic density, and for lighter masses the DM could be a sub-component.

\subsubsection*{Thermal Targets}

\begin{figure}[h]
    \centering
\includegraphics[width=0.8\linewidth]{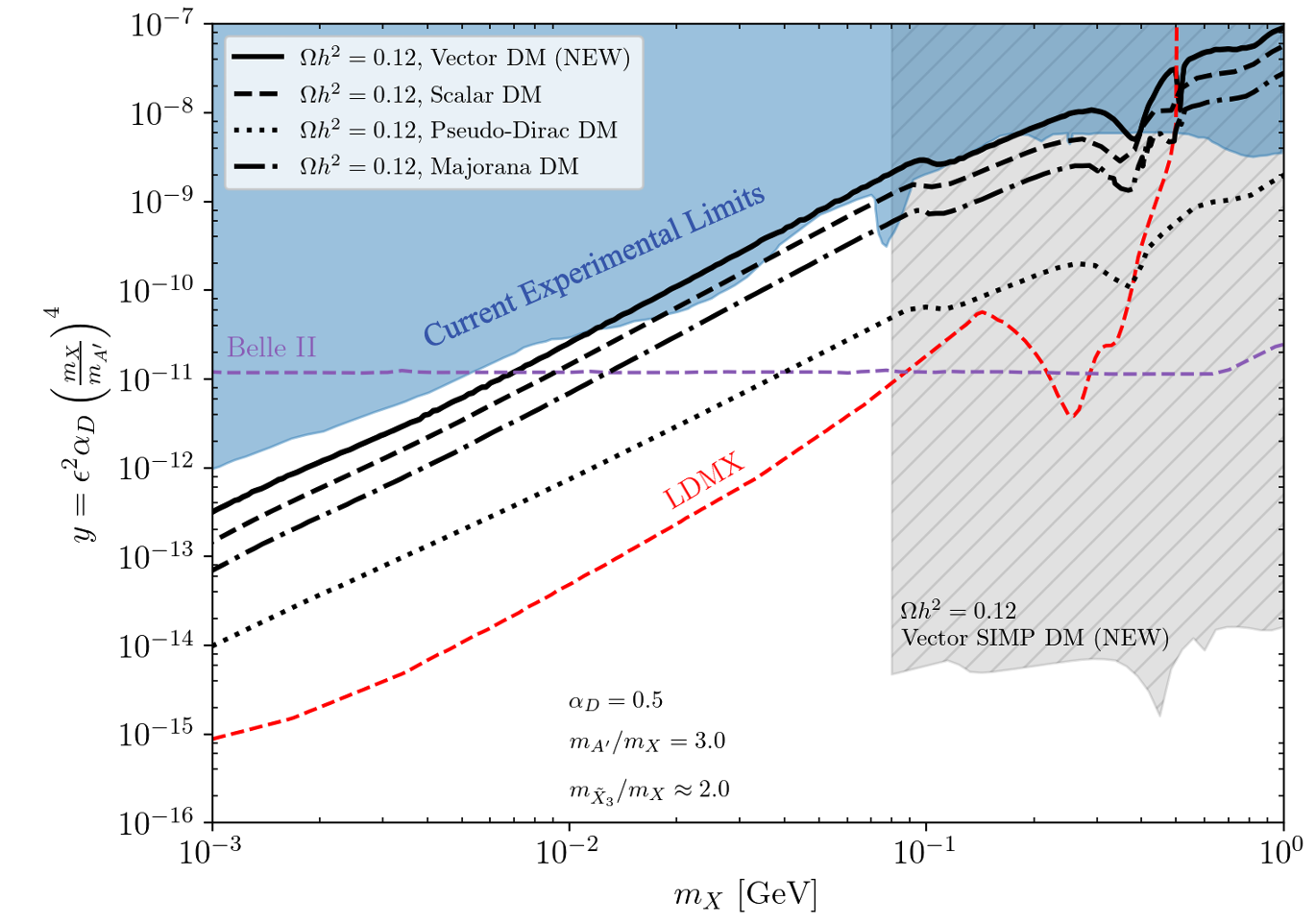}
    \caption{Thermal targets (black lines representing the parameters which reproduce the observed relic density \cite{Planck:2018vyg}) as a function of $y=\epsilon^2 \alpha_D \left(\frac{m_X}{m_{\Ap}} \right)^4$ and $m_X$, for various DM models.}
    \label{fig:thermal_targets}
\end{figure}

We compare the thermal targets of scalar, psuedo-Dirac, majorana, and vector DM (simplified $\Re[b_5]$ and SIMP) in \figref{fig:thermal_targets}. These thermal targets are plotted as a function of $y=\epsilon^2 \alpha_D \left( \frac{m_X}{m_\Ap} \right)^4$ and $m_X$, with approximate current experimental limits\footnote{ Some of these limits are DM model dependent, 
 for example when they involve DM scattering in a detector. For simplicity, we take these limits for complex scalar DM as an approximate benchmark of the current experimental status in this plot. See Paper I for model specific bounds.} in blue and relevant future projections as dashed curves.
As described in detail in Paper I, we find that certain simplified spin-1 DM models are the first models to be probed by upcoming results from NA64 and the future experiment LDMX. Furthermore, the spin-1 SIMP models, having a large viable parameter space, will be probed by LDMX. We have thus shown that the theoretical landscape of future accelerator experiments such as LDMX is rich with viable DM scenarios, including vector DM.

\subsection{Magnetic Dipole Portal}

The phenomenology discussed so far assumes $m_{\rm med} > 2m_{\rm DM}$, so that invisible mediator decay is kinematically open and dominates over visible decays whenever $g_D \gg \epsilon$. We now consider the inverted hierarchy, $m_{\rm med} < 2m_{\rm DM}$, where this channel closes entirely. The phenomenological consequences of this reversed mass hierarchy are (1) the mediator decays exclusively to visible SM final states (on-shell) at accelerator-based experiments, and (2) invisible final states can only be produced off-shell at missing energy/momentum experiments such as LDMX~\cite{LDMX:2025bog} and NA64~\cite{NA64:2023wbi}.
We realize this complementary phenomenological scenario with a light vector DM model featuring a new non-Abelian $SU(2)_D$ symmetry and scalar sector.

We augment the SM gauge group by a new non-Abelian dark gauge symmetry, $SU(2)_D$, under which a scalar doublet $\Phi_D$ and a scalar triplet $\Sigma_D$ are charged, in addition to the dark gauge triplet $X_\mu^a$. The field content
and quantum numbers are summarized in Paper VI. In
contrast to earlier non-Abelian constructions that combine an
$SU(2)\times U(1)$ dark gauge structure
\cite{Choi:2019zeb, Ramos:2021txu}, only a single dark $SU(2)_D$ is introduced here. An Abelian $U(1)_D$ emerges as a residual subgroup once the $SU(2)_D$ symmetry is spontaneously broken.

$SU(2)_D$ breaks spontaneously when the scalar triplet and doublet acquire vevs,
\begin{equation}
    \langle \Sigma_D \rangle = \frac{v_\Sigma}{2}\begin{pmatrix} 1 & 0 \\ 0 & -1 \end{pmatrix} \, ,
    \qquad
    \langle \Phi_D \rangle = \frac{1}{\sqrt{2}}\begin{pmatrix} 0 \\ v_\Phi \end{pmatrix} \, ,
    \label{eq:dark-vevs}
\end{equation}
giving the dark gauge boson masses
\begin{equation}
    m_{X^\pm}^2 = \frac{g_D^2 v_D^2}{4} \, , \qquad
    m_{X^0}^2 = \frac{g_D^2 v_D^2}{4}\sin^2\beta \, ,
    \label{eq:dark-gauge-masses}
\end{equation}
with $v_D \equiv \sqrt{v_\Phi^2 + 2v_\Sigma^2}$ and $\tan\beta \equiv v_\Phi/(\sqrt{2}v_\Sigma)$.

The asymmetry between the two mass terms in \eqnref{eq:dark-gauge-masses} is the structural origin of the inverted hierarchy. The triplet vev contributes to $m_{X^\pm}$ alone, however the doublet vev contributes to both $m_{X^\pm}$ and $m_{X^0}$.  $X_\mu^\pm$ carry charge under a global $U(1)_D$ symmetry generated by $\tau^3$, which is left unbroken by
$\langle\Sigma_D\rangle$ but spontaneously broken by $\langle\Phi_D\rangle$, leaving behind a residual discrete $\mathbb{Z}_2$. The $X_\mu^\pm$ are the lightest states charged under this $\mathbb{Z}_2$ and are therefore stable on cosmological timescales. These are identified as the DM candidates, denoted $m_{\rm DM} \equiv m_{X^\pm}$. Taking $\sin\beta < 1$ then yields $m_{X^0} < m_{\rm DM}$: the neutral dark gauge boson, which will mix into the physical mediator below, is naturally lighter than the DM itself.

The physical scalar spectrum consists of a dark-charged scalar and two neutral scalars, with mass matrices and mixing angles ($\beta$ for the charged sector, $\theta_t$ for the neutral sector) given in the Appendix of Paper VI \cite{Banerjee:2025ejz}. The Higgs-portal couplings between $\Phi_D, \Sigma_D$ and the SM Higgs are neglected, since they are tightly constrained. We take $\cos\beta = \sin\theta_t = 0.7$, and the dark scalars are taken nearly degenerate with the DM, $m_{\Sigma_D^\pm} = m_{\Sigma_D^0} = m_{\Phi_D^0} = 1.2\,m_{\rm DM}$, heavy enough to kinematically forbid DM decay to $\Sigma_D^\pm$, while light enough to remain relevant for the dark Higgs-strahlung signatures.

The dark and visible sectors communicate through a dimension 5 operator\footnote{One can also consider the dimension 6 operator with the scalar doublet investigated in \cite{Foguel:2025hio}, where a similar and complementary vector DM scenario is studied.},
\begin{equation}
    \mathcal{L}_{\text{int}} = -\frac{\sin \zeta}{2v_{\Sigma}} 
    \text{Tr}\left[\Sigma_D X_{\mu \nu}\right] B^{\mu \nu},
    \label{eq:portal}
\end{equation}
where $X_{\mu\nu} \equiv \sigma^a X^a_{\mu\nu}$, $B_{\mu\nu}$ is the hypercharge field strength, and $\zeta$ parametrizes the portal strength. This operator generates KM between the $X^0_\mu$ and the neutral SM gauge bosons,
\begin{equation}
    \mathcal{L}_{\text{mix}} = -\frac{\sin\zeta}{2}X^0_{\mu\nu}B^{\mu\nu} \, .
    \label{eq:kinetic-mixing}
\end{equation}

Beyond the kinetic mixing, the portal simultaneously generates interactions between the charged DM and the SM photon and $Z$ boson, of the magnetic dipole form. The resulting interaction Lagrangian is,
\begin{align}
    \mathcal{L}_{\rm mag}= i C_{VXX} \partial^{[\mu} V^{\nu]} \left( X^-_\mu X^+_\nu - X^+_\mu X^-_\nu \right)\,,
    \label{eq:AXX}
\end{align}
where,
\begin{align}
    C_{VXX} = g_D \sin\zeta \begin{cases} \cos\theta_w\,, &  V_\mu = A_\mu\,, \\
    \left( \sin\theta_w \cos\rho + \tan\zeta \sin\rho \right)\,, & V_\mu = Z_\mu\,, \\
    -\left( \sin\theta_w \sin\rho + \tan\zeta \cos\rho \right)\,, & V_\mu = Z^\prime_\mu\,, \end{cases}   
\end{align}
where $\rho$ is defined in the appendix of Paper VI and we label the new mass eigenstate $Z'$, which behaves analogously to a dark photon at fixed-target experiments.  Below, we consider the relevant constraints on the model parameter space.

\begin{figure}
    \centering
    \includegraphics[width=\linewidth]{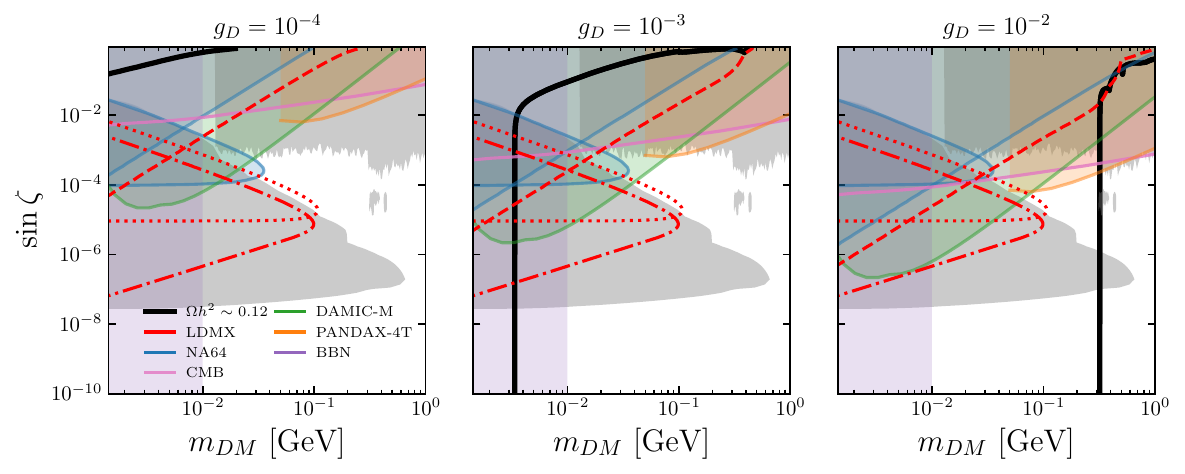}
    \caption{Overview of current and future experimental constraints on the light vector DM model in the $\sin \zeta$ vs $m_{\text{DM}}$ parameter space. The relic density targets, which denote parameter combinations yielding the observed DM abundance \cite{Planck:2018vyg}, are shown as thick solid lines. The subplots illustrate the model’s dependence on the dark gauge coupling, $g_D$, ranging from $10^{-4}$ (left) to $10^{-2}$ (right). Exclusion regions are indicated as follows: grey areas represent bounds from beam dump (CHARM, E774, E141, Orsay, KEK, E137, NuCal) and collider experiments (LHCb, KLOE, BaBar, NA48/2, FASER); pink regions denote CMB constraints \cite{Slatyer:2015jla}; and green and orange lines represent direct detection limits from DAMIC-M \cite{DAMIC-M:2025luv} and PandaX-4T \cite{PandaX:2025rrz}, respectively. Future 90\% C.L. projections are highlighted in red for LDMX Phase II \cite{LDMX:2025bog} (distinguishing between invisible vector meson decays, dashed; and $Z^\prime \to e^+ e^-$ decays outside, dash-dot; or inside, dotted, the detector) and in blue for current NA64 statistics \cite{NA64:2025ddk, NA64:2018lsq}.}
    \label{fig:models_nonAbelian}
\end{figure}

\subsubsection*{Direct Detection Constraints}
In light of recent null results from direct detection collaborations DAMIC-M~\cite{DAMIC-M:2025luv} and PandaX-4T~\cite{PandaX:2025rrz}, we report the corresponding constraints on this light vector DM scenario. These experiments search for DM-electron scattering signatures in the detector material \cite{Catena:2024rym,Mitridate:2022tnv}. DM-electron scattering proceeds via a t-channel diagram through an ordinary photon. We note that scattering through the $Z'$ and $Z$-boson is sub-leading since the momentum transfer squared is $\ll m_{Z'}^2, m_Z^2$. The resulting constraints in the $(\sin \zeta$,$m_\text{DM})$ plane are plotted in \figref{fig:models_nonAbelian} in orange (PandaX-4T) and green (DAMIC-M), alongside the collider and cosmological constraints.

\subsubsection*{Fixed-Target Searches}

\begin{figure}[h]
    \centering
    \includegraphics[width=0.49\linewidth]{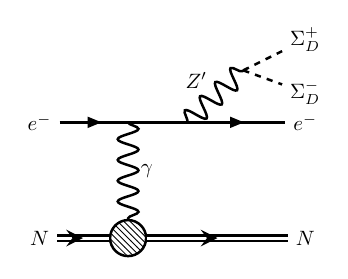}
    \includegraphics[width=0.49\linewidth]{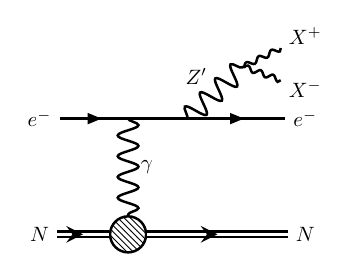}
    \caption{Feynman diagrams of off-shell invisible processes at fixed-target searches for the light vector DM scenario considered in Paper VI \cite{Banerjee:2025ejz}: dark Higgs-strahlung (left) and dark bremsstrahlung (right).}
    \label{fig:models_darkhiggs}
\end{figure}

In the conventional dark photon scenario of \chapref{chap:AcceleratorExperiments}, $m_{A'} > 2m_{\rm DM}$ allows the mediator to be produced on-shell and subsequently decay to a DM pair, so that the event
rate factorizes and depends only on the mediator's properties. Here, by contrast, $m_{Z'} < m_{\rm DM}$ forbids this on-shell channel entirely: DM production must instead proceed through an off-shell $Z'$ (or photon via the magnetic dipole interaction), suppressing the rate relative to the on-shell case and tying it directly to the DM mass and couplings rather than to the mediator alone. The mediator itself, unable to decay invisibly, is left with only one kinematically
available channel: visible decay to SM fermion pairs $Z' \to f^+f^-$. We consider four complementary signatures arising from this picture: (1) dark off-shell bremsstrahlung and (2) dark Higgs-strahlung \cite{Acanfora:2025ikj} (an example process is drawn in \figref{fig:models_darkhiggs}), both giving missing energy/momentum from $X^\pm$ or dark scalar pair production, (3) invisible vector meson decay, described in \secref{section:IVM}, in which bremsstrahlung photons convert to $\rho,\omega,\phi$ (and, at NA64 energies, $J/\psi$) mesons that subsequently decay to $X^+X^-$ through $Z'$–meson mixing, (4) and visible $Z'$
decays, searched for either as displaced energy depositions inside the detector or as missing energy/momentum if the $Z'$ escapes before decaying (discussed in  \secref{sec:VisibleSearches}). 
We find that the off-shell suppression inherent to this hierarchy leads to visible searches and direct detection ultimately setting the strongest constraints overall, rather than invisible searches.

\section{Dark Higher Electromagnetic Moments}
\label{section:DarkMoments}

The kinetic mixing term discussed in \secref{subsection:DarkPhotons} is only one of the ways in which a dark photon can communicate with the SM. higher-order electromagnetic moment interactions can arise as well, each with a distinct momentum dependence and hence a distinct experimental signature \cite{Catena:2025fsl}. Before introducing these interactions in their relativistic, field theory form, it is useful to recall their classical origin as terms in the electromagnetic multipole expansion.

The electromagnetic (or vector) potential of an arbitrary charge (or current) distribution is written as a multipole expansion of arbitrarily high order \cite{Jackson:1998nia}. For a charge distribution $\rho$, the potential as a function of position $\textbf{x}$ is given by,
\begin{equation}
    \Phi(\mathbf{x}) = \frac{1}{4 \pi \epsilon_0} \int \frac{\rho(\mathbf{x}')}{|\mathbf{x}-\mathbf{x}'|} d^3x',
\end{equation}
where $\epsilon_0$ is the permittivity of the vacuum. An analogous expression can be written down for the vector potential due to a current distribution. The potential is expanded in terms of spherical harmonics, where each successive multipole order (monopole, dipole, quadrupole, and so on) captures an additional level of angular structure in the underlying charge (or current) distribution. Progressively higher-orders are therefore required to resolve increasingly complex, non-uniform distributions, such as those that deviate from spherical symmetry or exhibit fine-grained spatial structure.
The coefficients of the multipole expansion of the potential $\Phi$ are called the \textit{moments}. Considering an electrostatic potential of a charge distribution, the coefficient of the first term (the first moment) is called the \textit{monopole} moment (equivalent to the total charge), the second moment is called the \textit{dipole} moment, and so on \cite{Nowakowski:2004cv}. In classical electrodynamics, magnetic monopoles do not exist, thus when considering the magnetic multipole expansion of the vector potential for a current distribution we begin with the \gls{M} moment followed by additional higher-order terms. Beyond the dipole, still higher-order moments exist, such as the quadrupole. However, not every moment appearing in the relativistic treatment discussed below has a counterpart in this classical picture:
the anapole moment, in particular, has no classical electrostatic or magnetostatic analogue, arising instead from parity violation rather than from a static charge or
current distribution.

The relativistic, quantum field theory analogue of this multipole expansion is the parametrization of the fermion-vector vertex in terms of electromagnetic form factors; a complete set of Lorentz invariant operators, whose coefficients generalize the classical electromagnetic moments to the relativistic regime~\cite{Nowakowski:2004cv}.
These effective interactions are generated by vacuum polarization-like loop diagrams, which are inversely proportional to the cutoff scale of order of the loop field masses \cite{Bonciani:2003ai}.
These operators are written down independently of their underlying \gls{uv} origin. Taking the non-relativistic limit of the Hamiltonian operator for the fermion field nevertheless reveals their connection to the static electromagnetic
properties of the fermion, expressed in terms of physical quantities related to the vector field such as the electric field $\mathbf{E}$, magnetic field $\mathbf{B}$, and electromagnetic current $\mathbf{J}$ (at least for the ordinary
photon).

This formalism is well established for the ordinary photon, however an analogous structure arises for a dark photon $A'$, the gauge
boson of a new $U(1)_D$ gauge group, coupled to the SM fermions $f$.
The \gls{KM} term introduced in \secref{subsection:DarkPhotons}, $\frac{\epsilon}{2}F_{\mu\nu}F'^{\mu\nu}$, must be generated by a vacuum polarization-like diagram with \textit{portal matter} fields carrying both SM and dark $U(1)_D$ charges in the loop (where the SM fermions are taken to be singlets under the new $U(1)_D$ group) \cite{Fabbrichesi:2020wbt,Rizzo:2018vlb}. There may also be new scalars which provide the mass generating mechanism of the DM and dark photon, or additional new dark symmetry groups and their corresponding gauge bosons. Crucially, the same portal matter loops responsible for KM can, depending
on the specific loop topology and the quantum numbers of the portal matter content, simultaneously generate these higher-order moment interactions. In certain regions of parameter space these higher-order moments are even expected to dominate over the standard KM term \cite{Rizzo:2021lob}. Since portal matter is expected to lie at or above the TeV scale to evade collider bounds, measurements of these higher-order moment interactions at low-energy fixed-target experiments such as LDMX offer a rare indirect probe of this otherwise inaccessible new physics scale \cite{Rizzo:2024bhn}.

In this Paper IV, we consider five dark electromagnetic moment interactions\footnote{The magnetic and electric dipole terms involve the tensor bilinear $\bar{f}\sigma^{\mu\nu}f$, which connects opposite chirality components and is therefore not $SU(2)_L\times U(1)_Y$ invariant on its own. Gauge invariance requires a compensating scalar insertion carrying the same electroweak quantum numbers as the SM Higgs doublet, after which $\mu_f, d_f \propto v/\Lambda^2$, where $v$ is the \gls{vev} of the scalar and $\Lambda$ is the cutoff scale.}: the ordinary electromagnetic like \gls{KM}, \gls{M}, \gls{E}, \gls{C}, and \gls{A}, given by the following interaction Lagrangian,
\begin{equation}
\begin{aligned}
    \mathscr{L}_{A'f} = & - e \epsilon Q_f \bar{f} \gamma^\mu f A'_\mu  
    -\mu_f \partial_\nu \left( \bar{f} \sigma^{\mu \nu} f \right) A'_\mu -i d_f \partial_\nu \left( \bar{f} \sigma^{\mu \nu} \gamma^5 f \right) A'_\mu \\ 
    &- b_f  \left[ \partial^\nu \partial_\nu \left(\bar{f}\gamma^\mu f \right) - \partial^\mu \partial_\nu \left(\bar{f}\gamma^\nu f \right) \right] A'_\mu \\ 
    &+ a_f  \left[ \partial^\nu \partial_\nu \left(\bar{f}\gamma^\mu \gamma^5 f \right) - \partial^\mu \partial_\nu \left(\bar{f}\gamma^\nu \gamma^5 f \right) \right] A'_\mu,
\label{eq:lagrangian_moments}
\end{aligned}
\end{equation}
where the first term is the standard KM, followed by M, E, C, and A. The key feature of these models is the varying momentum dependence from the derivatives in the Lorentz structures. This causes the recoil electron kinematic distributions at missing momentum experiments to vary between interaction models, when considering each term separately.
The dipole models (M, E) are distinct from the others because they are dimension-5 operators containing a derivative vertex. This introduces an additional momentum dependence that alters the recoil spectra compared to the other interactions.
In the relativistic limit, the pairs (M, E) and (C, A) are degenerate since they only differ by a $\gamma^5$.
The KM model shares an identical recoil shape with C and A due to equivalent energy scaling. The double derivatives of the C/A operators evaluate to $q^2$ at the vertex, which the on-shell condition trivializes into a constant normalization, $m_{A'}^2$.

We compute the relic density for each interaction model separately as a function of the model parameters, drawing the relic target contours. We then simulate dark photon production at LDMX using \madgraph for each model, computing the overall sensitivity and the recoil distributions. We find that LDMX is sensitive to each model for certain model parameters. These expected recoil distributions are then exploited within the likelihood framework presented in \chapref{chap:Likelihood}.

\subsection{Bayesian Model Comparison}
\label{sec:models_bayesian_comparison}

In Paper VII, we quantify the ability of LDMX to distinguish between the dark electromagnetic moment dark photon models.
\begin{figure}
    \centering
    \includegraphics[width=0.75\linewidth]{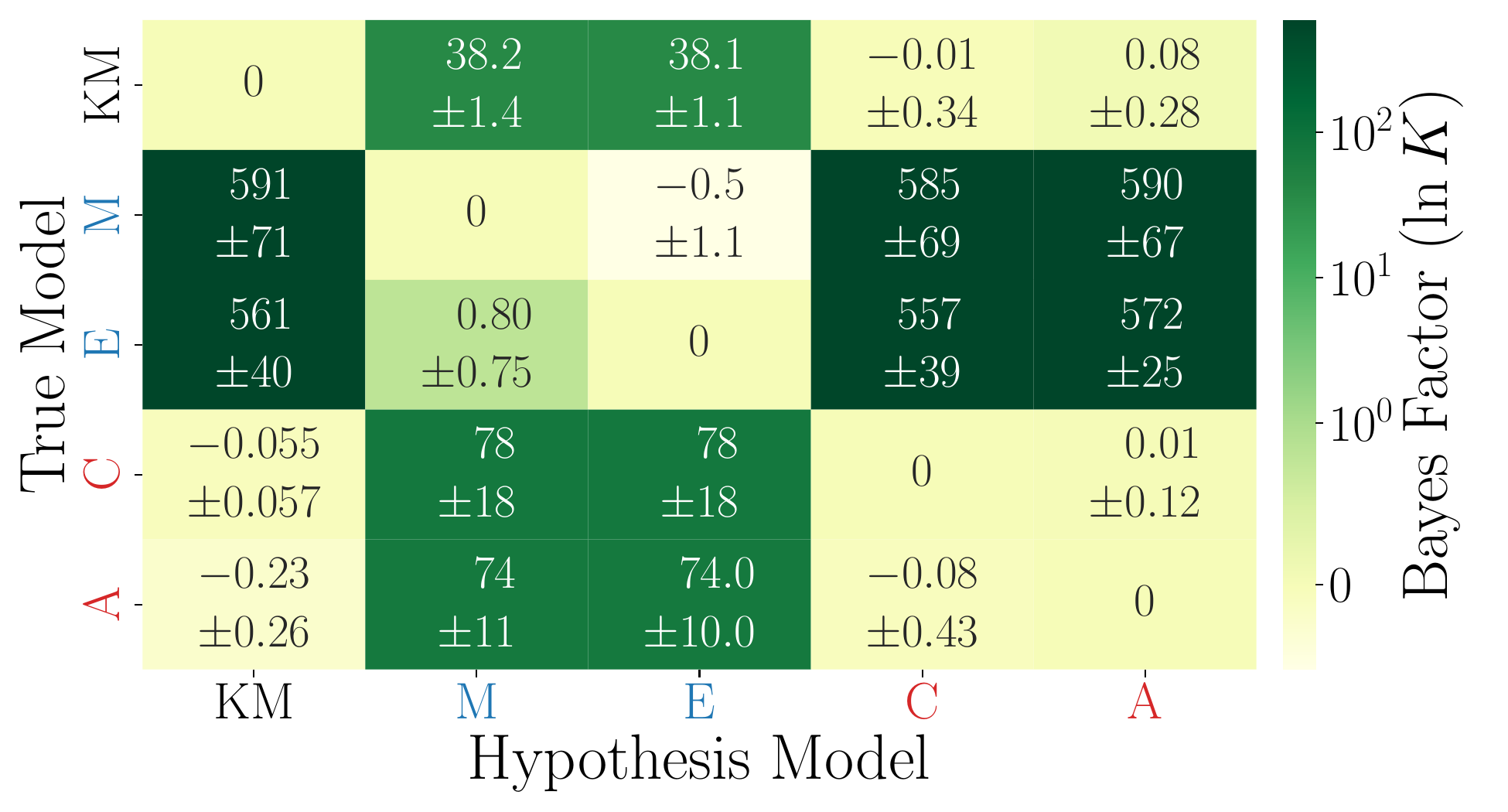}
    \caption{Model discrimination performance using the full two-dimensional recoil electron distribution \cite{Berger}. 
    The matrix displays the Bayes factor, $K$, evaluated using the joint distribution of the final state electron transverse momentum and energy. Rows denote the true generative model used to simulate the pseudo-data, while columns indicate the hypothesis model being evaluated. Each cell reports the median $\log K$ ratio, calculated across an ensemble of ten pseudo-data realizations, with the corresponding ensemble standard deviation reported below it. Large positive values ($\log K \gg 1$) indicate decisive evidence against the hypothesis model in favour of the generative model, whereas values near zero represent indistinguishable model kinematics. All pseudo-data are generated using the benchmark 3 parameters, $m_{A'}=0.1$~GeV, $\mathcal{R}=2.5$.}
    \label{fig:DEM_bayesfactor}
\end{figure}
The model discrimination capability of the LDMX likelihood-based framework is quantified in Fig.~\ref{fig:DEM_bayesfactor}, which displays the median Bayes factor matrix evaluated using the full two-dimensional $(|p_T|_{e^-_f},E_{e^-_f})$ recoil electron distribution for LDMX. The framework exposes the expected degeneracies within the pairs (M, E) and (C, A), which remain mutually indistinguishable ($\log K \approx 0$) due to identical recoil spectral shapes and event yields arising from chiral symmetries in the relativistic limit. Despite a twenty-fold difference in expected signal events at the relic target from KM to C/A, the KM model is nearly indistinguishable from the C/A models because the unconstrained coupling parameter absorbs the normalization discrepancy during evidence evaluation.  This analysis demonstrates that models with fundamentally different recoil electron kinematic shapes can be distinguished at LDMX, for parameter values on the relic target.

\chapter{Ab Initio Modelling of the Target Nucleus}
\label{chap:NuclearPhysics}
\raggedbottom

So far, the computation of the cross section for DM production via electron-nucleus scattering has relied on a rather simple, and quite outdated treatment of the nucleus. The elastic form factor in \eqnref{eq:Bjorken_formfactor} accounts for the atomic electron screening relevant for small momentum transfer and the elastic scattering with the nucleus through a simple parametrization from \cite{ImprovedWW_Tsai,Tsai:1973py}. Similarly, the inelastic form factor in \eqnref{eq:Bjorken_formfactor} includes the inelastic atomic form factor and the elastic proton form factor, assuming the protons are free.

The nucleus, however, is not an elementary particle. It has an internal structure arising from inter-nucleon interactions. This structure is imprinted on the electromagnetic, strong, and weak currents that a probe couples to (e.g., electron-nucleus scattering).
Consequently, a simple analytic parametrization of the nuclear form factor, lacking any microscopic description of nucleon-nucleon correlations, is not sufficient to reliably model the nuclear response relevant for fixed-target light DM searches.
Going beyond these phenomenological models, state-of-the-art \textit{ab initio}\footnote{\textit{Ab initio} methods describe nuclei with a systematically improvable approach, starting from the interactions between nucleons \cite{Ekstrom:2022yea}. Nowadays, \textit{ab initio} computations largely use interactions based on chiral effective field theory \cite{Epelbaum:2008ga}.}
methods rooted in chiral effective field theory interactions aim at a microscopic description of few- and many-body nuclear systems \cite{Ekstrom:2022yea,Hergert:2020bxy}.
Among the many-body methods on the market, the \gls{dDSCGF} method provides a wealth of information about ground and excited states of target nuclei of interest for fixed-target experiments \cite{scalesi2024de,Soma:2020xhv}. Moreover, this method allows for the inclusion of long-range correlations that are captured through rotational symmetry breaking, effectively deforming the system -- allowing for the description of a wide range of nuclei across the nuclear chart.

In Paper VIII, we present the first application of a deformed many-body \textit{ab initio} method to light DM searches at fixed-target experiments. 
We compute the cross section for dark photon bremsstrahlung, drawn in \figref{fig:feynman}, in both the elastic and quasi elastic regimes using the dDSCGF many-body method, quantifying the impact of going beyond standard analytic treatments.
The process for dark photon production is,
\begin{equation}
e^-(p) + N(P_A) \to e^-(p') + N(P_X) + A'(p_{A'}),
\end{equation}
where $N$ is the target nucleus. The final state nucleus can be in its ground state for elastic scattering, or excited into the continuum of $(A-1)+n$ knockout states for quasi elastic scattering, with $P_X$ denoting the four-momentum of the full hadronic final state $X$.
The differential cross section is proportional to,
\begin{equation}
    d\sigma \propto L_{\mu\nu}W^{\mu\nu},
    \label{eq:dsigma}
\end{equation}
where the leptonic tensor $L_{\mu \nu}$ includes the initial and final state beam electrons and the $A'$, while the nuclear tensor $W^{\mu\nu}$ includes the initial and final state nucleus.
The general form of the nuclear tensor is \cite{Rocco:2018vbf,Rocco:2020jlx},
\begin{equation}
W^{\mu\nu}= W_1 \big(-g^{\mu\nu} + \frac{q^\mu q^\nu}{q^2} \big) + \frac{W_2}{m_n^2} \big( P_A^\mu - \frac{P_A q}{q^2} q^\mu \big)\big( P_A^\nu - \frac{P_A q }{q^2} q^\nu \big),
\label{eq:Wmunu}
\end{equation}
where $m_n$ is the mass of the nucleon, $q \equiv (\omega, \vec{q})$ is the momentum transfer to the nucleus via the virtual photon as drawn in \figref{fig:feynman}, and $W_1$ and $W_2$ are the nuclear structure functions defined below. We define the quantity $Q$, which is a scalar that characterizes the momentum transfer scale, satisfying $Q^2 \equiv -q^2$.
The nuclear tensor from \eqnref{eq:dsigma} factorizes from the amplitude squared of the process depicted in \figref{fig:feynman}, allowing for implementation in the Monte Carlo event generator \madgraph.

\section{Elastic Regime}

In the elastic scattering regime, the nuclear structure can be described with a nuclear form factor \cite{Walecka:1995mi,Gnech:2026uan}. From the prescription of \eqnref{eq:Wmunu}, the nuclear form factor relates to the structure functions $W_1$ and $W_2$, where $W_1$ is usually taken to be zero \cite{Tsai:1973py}.
The elastic nuclear form factor\footnote{Strictly, $F$ depends on the direction of $\vec q$ whenever $\rho_{ch}(\vec r\,)$ is not spherically symmetric. For a spherical density the angular integral in Eq.~(\ref{eq:formfactor}) reduces to the familiar scalar form $F(Q)$. For the deformed nuclei considered here, we take $\rho_{ch}(r)$ to be the angle-averaged charge density. Additionally, Eq.~(\ref{eq:formfactor}) is evaluated with $\rho_{ch}(\vec r\,)$ static, implicitly defining $F$ in the frame where the energy transfer vanishes and $|\vec q\,|^2$ coincides with the Lorentz invariant $q^2 = -Q^2 = \omega^2 - |\vec q\,|^2$. Since $F$ depends only on this invariant, we evaluate it at $Q^2$ reconstructed from the generated event kinematics, which correctly accounts for nuclear recoil.}
is a function of the momentum transfer $Q$ and is the Fourier transform of the nuclear charge distribution, $\rho_{ch}$,
\begin{equation}
    F(Q) = \int d\vec{r} \rho_{ch}(\vec{r}) e^{-i \vec{q} \cdot \vec{r}},
    \label{eq:formfactor}
\end{equation}
with $\rho_{ch}$ computed using the dDSCGF many-body method in Paper VIII.

\begin{figure}
    \centering
    \includegraphics[width=0.75\linewidth]{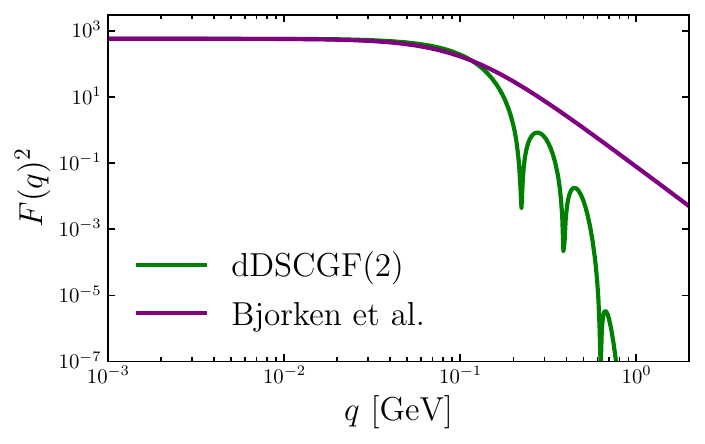}
    \caption{The squared nuclear form factor for $^{52}\text{Cr}$ as a function of the momentum transfer, where $q$ here refers to the scalar quantity $Q$. The analytic form factor from \eqnref{eq:Bjorken_formfactor} is shown in purple, which is a simple parametrization from \cite{Bjorken:2009mm,Tsai:1973py}. Meanwhile, the form factor computed from the \textit{ab initio} dDSCGF method at second order in the many-body expansion is plotted in green. The $\text{N}^3\text{LO}_\text{Texas}$ chiral interaction is used.}
    \label{fig:nuclear_elasticFF}
\end{figure}
Figure~\ref{fig:nuclear_elasticFF} compares two treatments of the elastic nuclear form factor for $^{52}\text{Cr}$. The squared form factor (to which the differential cross section is proportional) is plotted as a function of momentum transfer $q$, for both the form factor from \eqnref{eq:Bjorken_formfactor} (in purple) and from the dDSCGF method (in green). The ``Bjorken et al." curve is the simple analytic parametrization of the nuclear charge distribution used in Refs.~\cite{Bjorken:2009mm,Tsai:1973py}, in which the form factor is modelled as a smooth, monotonically falling function of $q$ built from a phenomenological charge density and does not incorporate any detailed nuclear structure input beyond the overall charge radius. The dDSCGF(2) curve is the corresponding form factor obtained from the \emph{ab initio} many-body approach.
At small momentum transfer the two curves agree well, as expected. In this regime, the form factor is controlled almost entirely by the total charge.
As $Q$ increases, the charge radius\footnote{Charge radii have been shown to be well modelled by chiral effective field theory interactions in many-body nuclear systems~\cite{Demol:2026glc,Hu:2025cjl}. The simple parametrization of the charge radius within \eqnref{eq:Bjorken_formfactor}, however, is not reliable across the nuclear chart, as the nuclear size scale is fixed through the single global relation $d \propto A^{-2/3}$, independent of nuclear structure.} sets the momentum transfer at which the form factor begins to fall off. Here, the two treatments begin to diverge,
since the scattering becomes sensitive to the detailed spatial structure of the nuclear charge distribution. The dDSCGF form factor exhibits pronounced diffraction minima that is absent in the analytic treatment. 
This comparison is directly relevant to the nuclear structure input used in the dark photon bremsstrahlung cross sections. 

$F(Q)^2$ enters as an overall weight on the differential cross section. The nucleus-photon vertex is dressed with the form factor, giving the Feynman rule for a scalar nucleus \cite{Validity_Liu},
\begin{equation}
    \Gamma^{el}_\mu = ieF(Q)\left( P_i + P_f\right)_\mu,
    \label{eq:elastic_vertex}
\end{equation}
which is directly implemented in \madgraph.

The choice of modelling for $F(Q)$ directly impacts the dark photon production cross section, and consequently the predicted signal and background yields at fixed-target experiments. In the $Q \gtrsim 0.1$~GeV region, the elastic form factor obtained from the simplified parametrization begins to disagree appreciably with the one derived from the \textit{ab initio} many-body method. This momentum transfer regime is relevant for $m_{A'} > 0.1$~GeV. Consequently, in this mass range, an \textit{ab initio} treatment of the nuclear form factor can have a non-negligible impact on the predicted signal yield relative to the simplified parametrization conventionally adopted in the fixed-target DM literature.

\section{Quasi Elastic Regime}
\label{sec:quasi-elastic}

At higher momentum transfers, the beam particle scatters quasi-elastically off the individual nucleons within the nucleus, rather than coherently off the nucleus as a whole. This transition is naturally understood through the space resolution of the probe: the electron resolves spatial structures on scales $\sim1/|\vec{q}|$, so once $|\vec{q}|$ becomes comparable to the inverse inter-nucleon spacing, the electron ceases to see the nucleus as a smooth, extended charge distribution and instead couples to individual, quasi-free nucleons. In this regime, the nuclear form factor description that governs elastic scattering breaks down, and the cross section must instead be built from the incoherent sum of scattering processes off the constituent protons and neutrons.
The \textit{quasi elastic} regime is characterized by a broad peak in the energy transfer, $\omega$, spectrum, smeared out by the Fermi motion of the nucleons within the nucleus. Accurately modelling this quasi elastic response is essential for predicting dark photon bremsstrahlung production at fixed-target experiments such as LDMX, since a sizable fraction of the relevant momentum transfer phase space lies in this regime.

We make use of the impulse approximation, where the quasi elastic scattering process reduces to a sum of scatterings with the individual bound nucleons within the nucleus \cite{Benhar:2006wy,Rocco:2020jlx}. We then write the nuclear structure functions, introduced in \eqnref{eq:Wmunu}, in terms of the spectral function \cite{Rocco:2018vbf},
\begin{equation}
    P(\vec{k},E) = \sum_f \big| \big[\, \bra{n_f} \otimes \bra{\Psi_f^{A-1}} \,\big] \ket{\Psi_0^A}  \big|^2 \,
    \delta\!\left(E - E_0^A + E_f^{A-1}\right),
\label{eqn:SF}
\end{equation}
giving the probability of removing a nucleon with excitation energy $E$ and 3-momentum $\vec{k}$ from the initial state nucleus with $A$ nucleons, $\ket{\Psi_0^A}$, resulting in the final state nucleon, $n_f$, and nucleus with $A-1$ nucleons, $\ket{\Psi_f^{A-1}}$.
A key strength of the \gls{dDSCGF} method is its direct access to the spectral function, obtained from the imaginary part of the one-body Green's function.
The nuclear structure functions $W_1$ and $W_2$ are expressed in terms of the nucleon structure functions $w_i^N$ \cite{DeForest:1984qe,Rocco:2020jlx},
\begin{equation}
\begin{aligned}
W_1(|\vec{q}|,\omega) &= \int d^3k dE \Bigg[ ZP_p(\vec{k},E) \frac{m_n}{E_k} \\ & \Big( w_1^p(|\vec{q}|,\tilde{\omega}) + w_2^p(|\vec{q}|,\tilde{\omega})\frac{|\vec{k}\times\vec{q}|^2}{2m_n^2|\vec{q}|^2}\Big) + \text{neutron part}
  \Bigg]
\end{aligned}
\end{equation}
\begin{equation}
\begin{aligned}
W_2(|\vec{q}|,\omega) &= \int d^3k dE \Bigg[  ZP_p(\vec{k},E) \frac{m_n}{E_k} \Big[ w_1^p(|\vec{q}|,\tilde{\omega}) \frac{q^2}{|\vec{q}|^2} \left( \frac{q^2}{\tilde{q}^2}-1 \right) \\ & + \frac{w_2^p(|\vec{q}|,\tilde{\omega})}{m_n^2}\left( \frac{q^4}{|\vec{q}|^4}\left( E_k - \frac{\tilde{\omega} ( E_k \tilde{\omega}-\vec{k}\cdot \vec{q})}{\tilde{q}^2} \right)^2 - \frac{q^2 |\vec{k} \times \vec{q}|^2}{2|\vec{q}|^4}\right) \Big] \\&+ \text{neutron part}
\Bigg],
\end{aligned}
\label{eq:W1W2}
\end{equation}
where $\vec{k}$ ($E_k$) is the initial 3-momentum (energy) of the struck nucleon, the neutron part is proportional to $A-Z$, with the total number of nucleons in the nucleus $A$, and $Z$ the total number of protons. Notice that $\tilde{q}\equiv (\tilde{\omega},\vec{q})$ where $\tilde{\omega} \equiv \omega - E + m_n - E_k$ has been introduced which accounts for the nucleon being bound, thus a fraction of the transferred energy to the nucleus goes into the spectator system.
The nucleon structure functions for $N \in \{p,n\}$ are,
\begin{equation}
w_i^N = \tilde{w}_i^N \delta \left( \tilde{\omega} + \frac{\tilde{q}^2}{2m_n} \right),
\end{equation}
for $i \in \{ 1,2\}$, $\tau = -\tilde{q}^2/4m_n^2$, and,
\begin{equation}
\begin{aligned}
    &\tilde{w}_1^N = \tau G^2_{M,N}(q^2)\\
    &\tilde{w}_2^N = \frac{G^2_{E,N}(q^2)+\tau G^2_{M,N}(q^2)}{1+\tau},
\end{aligned}
\end{equation}
where $G_{E,N}$ and $G_{M,N}$ are the electric and magnetic nucleon form factors \cite{Perdrisat:2006hj},
\begin{equation}
    \begin{aligned}
        &G_{E,p}(q^2) = \left( 1-\frac{q^2}{m_V^2}\right)^{-2} \\
        &G_{M,p}(q^2) = \mu_p \left( 1-\frac{q^2}{m_V^2}\right)^{-2} \\
        &G_{E,n}(q^2) = \frac{\mu_n \tau}{1+5.6 \tau} \left( 1-\frac{q^2}{m_V^2}\right)^{-2}\\
        &G_{M,n}(q^2) = \mu_n \left( 1-\frac{q^2}{m_V^2}\right)^{-2},
    \end{aligned}
\end{equation}
where $m_V^2 = 0.71 \: \text{GeV}^2$ and the proton and neutron magnetic moments are $\mu_p \approx 2.793$ and $\mu_n \approx -1.913$.

\begin{figure}
    \centering
    \includegraphics[width=0.75\linewidth]{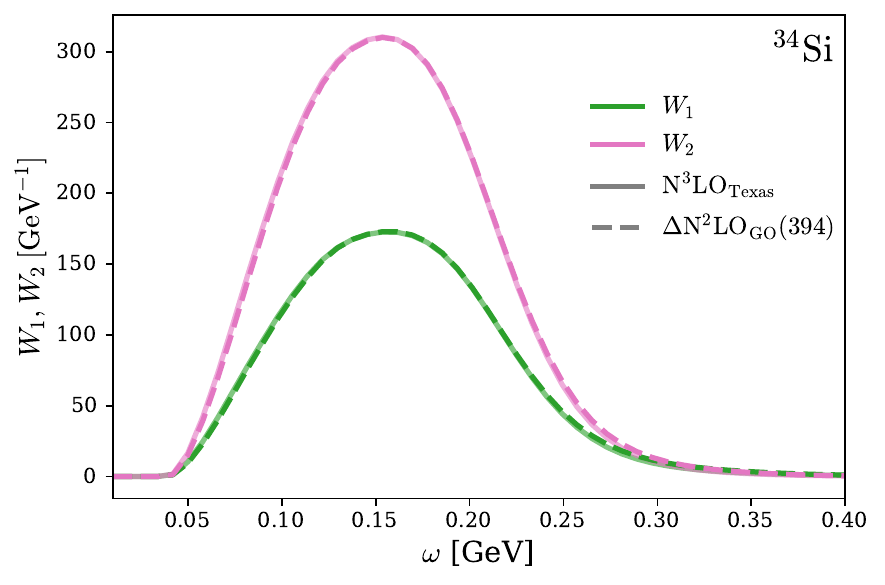}
    \caption{The nuclear structure functions $W_1$ (pink) and $W_2$ (green), given in \eqnref{eq:W1W2}, plotted as a function of the energy transfer $\omega$ for $^{34}$Si at $|\vec{q}|=0.5$ GeV. The structure functions are calculated with the spectral function from the dDSCGF many-body method employing the $\text{N}^3\text{LO}_\text{Texas}$ \cite{Hu:2025cjl} (solid) and the $\Delta \text{N}^2\text{LO}_\text{GO} (394)$ \cite{Jiang:2020the} (dashed) chiral interactions.}
    \label{fig:nuclear_quasielastic}
\end{figure}
For illustration, we plot the nuclear structure functions for $^{34}$Si in \figref{fig:nuclear_quasielastic}, showing both $W_1$ (in pink) and $W_2$ (in green) versus the transfer energy $\omega$. The absolute value of the 3-momentum transfer, $|\vec{q}|$, is set to 0.5 GeV for this example.

Similarly to the vertex in \eqnref{eq:elastic_vertex} for elastic scattering, the vertex for quasi elastic scattering can be represented by,
\begin{equation}
    \Gamma^{QE}_\mu = G_1(|\vec{q}|,\omega) \gamma_\mu + \frac{G_2(|\vec{q}|,\omega)}{2m_n} \left(p_f + p_i \right)_\mu,
\end{equation}
where $p_f$ and $p_i$ are the final and initial state nucleons, and, 
\begin{equation}
    G_1(|\vec{q}|,\omega)^2 = \frac{W_1(|\vec{q}|,\omega)}{2m_n \omega}, \quad G_2(|\vec{q}|,\omega)^2 = \frac{2m_nW_1(|\vec{q}|,\omega)/\omega - W_2(|\vec{q}|,\omega)}{1+\omega/2m_n}.
\end{equation}

We find that the \textit{ab initio} quasi elastic treatment, based on the spectral function formalism, results in a significantly enhanced DM production yield relative to the free proton approximation of \eqnref{eq:Bjorken_formfactor} for dark photons with $m_{A'} \gtrsim 0.1$ GeV. For lighter dark photons, our calculations are suppressed due to the effect of Pauli-blocking.

\part{Conclusion}
\label{part:Conclusion}
\chapter{Summary and Future Directions}

Incontrovertible astrophysical and cosmological data confirm the existence of DM, and deciphering its fundamental particle nature remains a defining imperative of modern physics.
A compelling target in this search is the light DM regime, as sub-GeV candidates can naturally yield the correct thermal relic abundance via the freeze-out mechanism. To realize a thermal relic at sub-GeV scales, a new mediator is required, with simplified DM models containing dark photon mediators serving as the standard paradigm across much of the literature.

This thesis has focused on light DM signatures at intensity-frontier fixed-target experiments -- a powerful search strategy for probing light dark sectors that evade both energy-frontier colliders and traditional nuclear recoil direct detection searches. We have placed particular emphasis on the upcoming experiment LDMX, for which we presented a likelihood-based framework to evaluate discovery prospects, parameter estimation capabilities, model selection power, and projected exclusion limits. Our approach relies on a binned analysis; unbinned likelihood methods, which retain the full event-level information in energy and transverse momentum, offer a promising avenue for future refinement of this framework. We further provided the methodology to test the compatibility of a hypothetical light DM signature at LDMX with a direct detection signal, with the direct detection exposure required to validate it. Ultimately, in the event of an observed excess at a fixed-target search, these frameworks will serve as a useful baseline strategy for interpreting potential excesses and matching tentative experimental signals with underlying dark sector theories, a key step toward constructing a coherent, multi-messenger DM picture.

Beyond electron beam experiments, proton beam dump facilities such as SHiP offer a complementary probe of light DM, and several extensions of our analysis in this setting are left for future work. On the detector side, a dedicated full Monte Carlo re-evaluation of the neutrino-induced background using the updated SHiP design would sharpen the projected sensitivities presented here. On the signal side, our analysis is restricted to the DM-electron scattering channel, owing to the comparatively poor understanding of the background rate for other channels at proton beam dumps; extending this treatment to nuclear recoils would open sensitivity to larger DM masses. Finally, the proton bremsstrahlung production mechanism used throughout follows the improved WW treatment, in which the dark photon is radiated only in the initial state prior to the proton-nucleus interaction. A calculation including final-state radiation off the outgoing proton and its interference with the initial-state contribution remains an avenue for future work.

This thesis has extended beyond the standard paradigm of minimal DM models, exploring vector DM scenarios and richer dark photon interactions through higher-order electromagnetic moments, broadening the theoretical landscape accessible to fixed-target searches and complementary probes such as direct detection via electron recoils. Whereas simplified models offer pragmatic benchmarks, UV-complete scenarios provide theoretical completeness by embedding dark sectors with extended symmetry groups, fields, and mass-generating mechanisms. 
Introducing new gauge symmetries often necessitates extended scalar sectors and dark Higgs mechanisms to generate dark sector masses, inherently yielding richer field content and non-trivial couplings. These extended theories alter physical observables relative to minimal models, introducing additional processes that modify experimental signatures. This thesis has demonstrated the rich resulting phenomenology from UV-complete vector DM scenarios, from dark Higgs-strahlung and invisible vector meson decay at fixed-target experiments, to direct detection via electron scattering. A particularly fruitful next step is to ground effective higher-order dark photon interactions within UV-complete portal matter frameworks, and to consider the corresponding experimental and theoretical consequences. Considering these non-minimal scenarios will be vital for fully leveraging the discovery potential of intensity-frontier facilities.

Orthogonal to the development of theoretical scenarios for light DM, this thesis explored the complementarity of diverse DM searches. We presented a status update of light DM, performing global fits of laboratory, astrophysical, and cosmological constraints. 
With a plethora of next-generation experiments on the horizon, global fits will be vital for transforming disparate data sets into a unified dark sector map. Incorporating full likelihood implementations, such as our framework for LDMX, together with multi-messenger signals like gravitational waves alongside traditional probes, will provide a powerful means for pinpointing the fundamental nature of DM.

The topic of DM requires interdisciplinary input from a wide array of fields.
To further refine signal predictions at fixed-target facilities, we used state-of-the-art \textit{ab initio} nuclear physics methods to model the target nucleus. Our results were in agreement with a simple parametrization for light mediators, but provided a significant enhancement for heavier mediators -- demonstrating the impact of such a nuclear theory informed framework for making DM signal predictions across diverse experimental settings. 
Future extensions of this work will focus on nuclei used in experiments, such as tungsten in LDMX and lead in NA64, to provide improved sensitivities. In addition, extending \textit{ab initio} modelling to proton bremsstrahlung would offer a major theoretical upgrade: whereas current frameworks rely on a simple scaling of the nucleon-level cross section, a nuclear microscopic treatment would be more suited to capture nuclear effects.

This is an exceptionally exciting time for light DM searches, as a new generation of fixed-target, direct detection, and collider experiments comes online to probe the sub-GeV frontier. Realizing this potential will hinge on the rigour of both signal and background modelling, together with the ability to corroborate a candidate signal across independent, complementary probes.

\appendix
\chapter{Weizsäcker-Williams Approximation}
\label{Chapter:Appendix}

The Weizsäcker-Williams (WW) approximation, also known as the equivalent photon approximation, \cite{vonWeizsacker:1934nji,Williams:1935dka} exploits the fact that, the Coulomb field of a fast moving nucleus can be treated as a flux of quasi-real, transverse photons. In the case of fixed-target experiments, in the frame of the relativistic beam electron, the nucleus is in fact fast. This allows the full $2\to3$ process,
\begin{equation}
e^-(p) + N(P_A) \to e^-(p') + N(P_X) + A'(p_{A'}),
\end{equation}
to be factorized into a semiclassical flux of equivalent photons folded with the much simpler $2\to2$ sub-process,
\begin{equation}
    e^-(p) + \gamma(q) \to e^-(p')+A'(p_{A'}),
\end{equation}
simplifying the phase space \cite{Kim:1973he,Bjorken:2009mm}. This appendix reviews the derivation of the approximation following \cite{Kim:1973he,Validity_Liu}.

We will write the squared amplitude in terms of $L_{\mu\nu}W^{\mu\nu}$, where $L_{\mu\nu}$ is the gauge invariant lepton tensor and $W^{\mu\nu}$ is the nuclear tensor \cite{nuclear}. 
\begin{equation}
\begin{aligned}
    L_{\mu\nu} W^{\mu\nu} =& L_{\mu\nu} \left[ W_1  \left( -g^{\mu\nu} + \frac{q^\mu q^\nu}{q^2} \right) \right. \\&\left.+ \frac{W_2}{M^2} \left(P_A^{\mu} - \frac{q^\mu (q \cdot P_A)}{q^2}\right)\left(P_A^{\nu} - \frac{q^\nu (q \cdot P_A)}{q^2}\right)\right],
\end{aligned}
\end{equation}
with the nucleus mass $M$ (not the nucleon mass from the quasi-elasticc regime of \secref{sec:quasi-elastic}), and the momentum transfer $q$.
Due to gauge invariance, $L_{\mu\nu}q^\mu = 0$, therefore,
\begin{equation}
     L_{\mu\nu} W^{\mu\nu} = L_{\mu\nu} \left[ -W_1  g^{\mu\nu} + \frac{W_2}{M^2} P_A^{\mu} P_A^{\nu} \right]
\end{equation}
Considering the $W_2$ term, we can write it as follows \cite{ImprovedWW_Tsai},
\begin{equation}
    \frac{L_{\mu\nu}P_A^{\mu}P_A^{\nu}}{M^2} = \frac{L_{\mu \nu}}{M^2} \left[ P_A^{\mu} + \frac{q^\mu M}{q^z - q^0}\right] \left[ P_A^{\nu} + \frac{q^\nu M}{q^z - q^0}\right].
\end{equation}
Expanding this out gives \cite{ImprovedWW_Tsai},
\begin{equation}
    \frac{L_{\mu\nu}P_A^{\mu}P_A^{\nu}}{M^2} = \big(L_{00} 
    + L_{zz} - 2L_{0z}\big) \frac{(q^z)^2}{(q^z - q^0)^2}  +  \big(L_{xx}(q^x)^2 + L_{yy}(q^y)^2 \big)\frac{1}{(q^z - q^0)^2}.
\end{equation}
The longitudinal component, $L_{00} + L_{zz} - 2L_{0z}$, is negligible compared to the transverse component since the energy of the incident electron is much greater than $m_\Ap$ and $m_e$, therefore we can neglect it. Since $q_x = q_\perp \cos\phi$ and $q_y = q_\perp \sin\phi$, with the azimuthal angle $\phi$, we now have,
\begin{equation}
    \frac{L_{\mu\nu}P_A^{\mu}P_A^{\nu}}{M^2} \approx \frac{q_\perp^2}{(q_z - q_0)^2} (\cos^2 \phi L_{xx} + \sin^2 \phi L_{yy}).
    \label{eq:LPP}
\end{equation}

We define the squared momentum transfer,
\begin{equation}
    t = -q^2  = -\omega^2 + |\vec{q}|^2,
\end{equation}
with the energy transfer $\omega$.
We make the assumption that near $t_{min}$, $q_0 \approx 0$, so we can write $t_{min}$ as \cite{appendix},
\begin{equation}
    t_{min} \approx \left( \frac{U}{2E(1-x)}\right)^2
    \label{eq:tmin_U}
\end{equation}
where,
\begin{equation}
    U = E^2 \theta^2 x + \frac{m_\Ap(1-x)}{x} + m_e^2 x,
\end{equation}
 $x \equiv \frac{E_k}{E}$, $E$ is the initial electron energy, $E_k$ is the dark photon energy, and $\theta$ being the angle of the dark photon from the beam axis.

Near the minimum momentum transfer $|\vec{p}-\vec{k}|$ is parallel to $|\vec{\pp}|$, and since $\vec{q}=(\vec{p}-\vec{k})-\vec{\pp}$, $\vec{q}$ is also parallel to $\vec{p}-\vec{k}$ and $\vec{\pp}$. We will define $\vec{\pp}$ to point in the $z$ direction, therefore near minimum momentum transfer $\vec{q}$ also points in the $z$ direction. Hence $q_\perp \approx 0$ which leads to,
\begin{equation}
    t_{min} \approx q_z^2 - q_0^2,
    \label{eq:tmin_q}
\end{equation}
and
\begin{equation}
    t - t_{min} = q_x^2 + q_y^2 = q_\perp^2 \cos^2\phi + q_\perp^2 \sin^2\phi = q_\perp^2.
    \label{eq:tmin_qxy}
\end{equation}
We define,
\begin{equation}
    t_{min} = t^\prime_{min} + 2q_0\sqrt{t^\prime_{min}}
\end{equation}
where
\begin{equation}
    t^\prime_{min} = \frac{(p\cdot k - m_\Ap^2/2)^2}{(E-E_k)^2} = (q_z - q_0)^2.
    \label{eq:tminp}
\end{equation}
Substituting eqs. \ref{eq:tmin_qxy} and \ref{eq:tminp} into eq. \ref{eq:LPP},
\begin{equation}
     \frac{L_{\mu\nu}P_A^{\mu}P_A^{\nu}}{M^2} \approx \frac{t-t_{min}}{t^\prime_{min}} \left(  \cos^2\phi L_{xx} + \sin^2\phi L_{yy}\right) 
\end{equation}
Due to the photon propagator having a $1/q^4 = 1/t^2$ dependence we can evaluate $L_{xx}$ and $L_{yy}$ at $t = t_{min}$, since small momentum transfers dominate this cross section. Near $t=t_{min}$, $t'_{min}\approx t_{min}$.
$L_{xx}$ and $L_{yy}$ are independent of $\phi$ when $t\approx t_{min}$ since $q_{min}$ only has a $z$ component (as explained above eq. \ref{eq:tmin_q}), and $\phi$ comes from the $x$ and $y$ components of $q$. This allows for easy integration,
\begin{equation}
    \int_0^{2\pi} \frac{L_{\mu\nu}P_A^{\mu}P_A^{\nu}}{M^2} d\phi \approx  \pi \frac{t-t_{min}}{t_{min}} \left. \left( L_{xx}+L_{yy}  \right) \right|_{t=t_{min}}.
\end{equation}
We can write $ L_{xx}+L_{yy}$ as $-L_{\mu\nu}g^{\mu\nu}$, since $L_{00}$ and $L_{zz}$ are safely neglected,
\begin{equation}
    \int_0^{2\pi} \frac{L_{\mu\nu}P_A^{\mu}P_A^{\nu}}{M^2} d\phi \approx \pi \frac{t-t_{min}}{t_{min}} \left.\left( -L_{\mu \nu}g^{\mu\nu}\right) \right|_{t=t_{min}}.
\end{equation}
Now we combine the $W_1$ and $W_2$ terms,
\begin{equation}
     \int_0^{2\pi} L_{\mu\nu}W^{\mu\nu} d\phi \approx  \left[ 2 \pi W_1 +\frac{t-t_{min}}{t_{min}} \pi W_2  \right] \left.\left(  -L_{\mu \nu}g^{\mu\nu} \right) \right|_{t=t_{min}}.
\end{equation}
$W_1$ is often neglected in the literature \cite{Tsai:1973py}.
This allows us to write the cross section of the $2\to 3$ process in terms of the $2\to 2$ process \cite{Tsai:1973py,Validity_Liu},
\begin{equation}
    \frac{1}{8M^2}\int\frac{d\phi}{2\pi}\mathcal{A}^{23} \approx \frac{t-t_{min}}{2t_{min}}\mathcal{A}^{22}_{t=t_{min}},
\end{equation}
where $e$ is the elementary electric charge, $\mathcal{A}^{23}=\bar{|M^{23}|^2}\frac{t^2}{e^4 \epsilon^2 F(q)^2}$, $\mathcal{A}^{22} = \frac{\bar{|M^{22}|^2}}{e^2 \epsilon^2}$, and $F(q)^2 = \frac{W_2^{el}(q)}{2M}$ is the elastic nuclear form factor introduced in \chapref{chap:NuclearPhysics}.
The differential cross section in the lab frame of the $2\to 3$ process is then \cite{Bjorken:2009mm},
\begin{equation}
    \frac{d\sigma(p+P_i \to p^{\prime}+k+P_f)}{dE_k d\cos\theta_\Ap} = \left. \frac{d\sigma(p+q \to p^{\prime}+k)}{d(p\cdot k)} \right|_{t=t_{min}} \frac{\alpha \chi}{\pi} \frac{E x \beta_\Ap}{1-x},
\end{equation}
with,
\begin{equation}
    \chi \equiv \int_{t_{min}}^{t_{max}} \frac{(t-t_{min})}{t^2}W_2 dt,
\end{equation}
\begin{equation}
    \beta_\Ap \equiv \sqrt{1-\frac{m_\Ap^2}{E_k^2}},
\end{equation}
$\alpha = \frac{e^2}{4\pi}$, and $t_{max} \approx m_\Ap^2$.

\newpage
\thispagestyle{empty} 
\cleardoublepage\addcontentsline{toc}{chapter}{Bibliography}

\printbibliography



%
\cleardoublepage

\addcontentsline{toc}{chapter}{Appendix}

\end{document}